\documentclass[11pt]{article}

\usepackage{jheppub} 

\usepackage{amsmath}
\usepackage{amssymb}
\usepackage{braket}
\usepackage{mathrsfs}
\usepackage{bm}
\usepackage{color}
\usepackage[normalem]{ulem}

\def\nn{\nonumber \\}

\def\vev#1{\left\langle #1 \right\rangle }
\def\abs#1{\left| #1 \right| }
\def\tr{ \text{Tr}\, }

\def\rd{ {\rm d} }
\def\ip#1#2{\left\langle #1 , #2 \right\rangle }

\def\m{ \mathcal{M} }
\def\h{{ \cal H}}
\def\g{{\cal G}}
\def\C_A{\mathsf{C_A}}
\def\n_G{N_\varphi}
\def\fcs[#1#2]#3{ f^{\phantom{#1#2}#3}_{#1#2} }
\def\rpi{ \mathbf{R}^{(\pi)} }

\def\bpi{{\overline \pi}}
\def\api{\abs{{\overline \pi}}}
\def\avphi{\abs{\overline \pi}}
\def\bvphi{{\overline \pi}}

\newcommand{\Lie}{{\mathscr{L}}}

{\count255=\time\divide\count255 by 60 \xdef\hourmin{\number\count255}
  \multiply\count255 by-60\advance\count255 by\time
  \xdef\hourmin{\hourmin:\ifnum\count255<10 0\fi\the\count255}}

\begin{document}

\preprint{CERN-TH-2016-116}

\title{Geometry of the Scalar Sector}

\author{Rodrigo Alonso,$^1$}
\author{Elizabeth E.~Jenkins,$^{1,2}$}
\author{Aneesh V.~Manohar$^{1,2}$}

\affiliation{1. Department of Physics, University of California at San Diego, La Jolla, CA 92093, USA}
\affiliation{2. CERN TH Division, CH-1211 Geneva 23, Switzerland}

\abstract{
The $S$-matrix of a quantum field theory is unchanged by field redefinitions, and so only depends on geometric quantities such as the curvature of  field space. Whether the Higgs  multiplet transforms linearly or non-linearly under electroweak symmetry is a subtle question since one can make a coordinate change to convert a field that transforms linearly into one that transforms non-linearly. Renormalizability of the Standard Model (SM) does not depend on the choice of scalar fields or whether the scalar fields transform linearly or non-linearly under the gauge group, but only on the geometric requirement that the scalar field manifold ${\m}$ is flat. We explicitly compute the one-loop correction to scalar scattering in the SM written in non-linear Callan-Coleman-Wess-Zumino (CCWZ) form, where it has an infinite series of higher dimensional operators, and show that the $S$-matrix is finite. 

Standard Model Effective Field Theory (SMEFT) and Higgs Effective Field Theory (HEFT) have curved ${\m}$, since they parametrize deviations from the flat SM case. We show that the HEFT Lagrangian can be written in SMEFT form if and only if ${\cal M}$ has a $SU(2)_L \times U(1)_Y$ invariant fixed point. Experimental observables in HEFT depend on local geometric invariants of ${\m}$ such as sectional curvatures, which are of order $1/\Lambda^2$, where  $\Lambda$ is the EFT scale.  We give explicit expressions for these quantities in terms of the structure constants for a general $\g \to \h$ symmetry breaking pattern.  The one-loop radiative correction in HEFT is determined using a covariant expansion which preserves manifest invariance of ${\m}$ under coordinate redefinitions.  The formula for the radiative correction is simple when written in terms of the curvature of ${\m}$ and the gauge curvature field strengths.  We also extend the CCWZ formalism to non-compact groups, and generalize the HEFT curvature computation to the case of multiple singlet scalar fields.

}
\date{\today\quad\hourmin}

\maketitle

\section{Introduction}

Current experimental data is consistent with the predictions of the Standard Model (SM) with a light Higgs boson of mass $\sim 125$~GeV. The measured properties of the Higgs boson agree with SM predictions, but the current experimental accuracy of measured single-Higgs boson couplings is only at the level of  $\sim10\%$, and no multi-Higgs boson couplings have been measured directly.  It is important to consider generalizations of the SM with additional parameters in order to quantify the accuracy to which the SM is valid or to detect deviations from SM predictions.  

Over the past 40 years, many theoretical ideas have been proposed for the underlying mechanism of electroweak symmetry breaking. Theories that survive must be consistent with the currently observed pattern of electroweak symmetry breaking, which is well-described by the SM.   A general model-independent analysis of electroweak symmetry breaking can be performed using effective field theory (EFT) techniques.  Assuming there are no additional light particles beyond those of the SM at the electroweak scale $v \sim 246$\,GeV, the EFT has the same field content as the SM.  There are two main EFTs used in the literature, the Standard Model Effective Field Theory (SMEFT) and Higgs Effective Field Theory (HEFT). In this paper, we make the relationship between these two theories precise.

The Higgs boson $h$ of the SM is a neutral $0^+$ scalar particle.  In the SM Lagrangian, it appears in a complex scalar field $H$, which transforms as $\mathbf{2}_{1/2}$ under the $SU(2)_L \times U(1)_Y$ electroweak gauge symmetry.  An oft-stated goal of the precision Higgs physics program is to test whether (a) the Higgs boson transforms as part of a complex scalar doublet which mixes linearly under $SU(2)_L \times U(1)_Y$ with the three ``eaten" Goldstone bosons $\bm{\varphi}$, or (b) whether the Higgs field is a singlet radial direction which does not transform under the electroweak symmetry.  In case (b), the three Goldstone modes $\bm{\varphi}$ transform non-linearly amongst themselves under the electroweak symmetry, in direct analogy to pions in QCD chiral perturbation theory, and do not mix with the singlet Higgs field.   In case (a), there are relations between Higgs boson and Goldstone boson (i.e.\ longitudinal gauge boson) interactions, whereas in case (b), no relations are expected in general.  An objective of this paper is to explore the distinction between these two pictures for Higgs boson physics.

The properties of the scalar sector of the SM and its EFT generalizations can be clarified by studying it from a geometrical point of view~\cite{Alonso:2015fsp}. The scalar fields define coordinates on a scalar manifold ${\m}$.  The geometry of ${\m}$ is invariant under coordinate transformations, which are scalar field redefinitions.  The quantum field theory $S$-matrix also is invariant under scalar field redefinitions, so it depends only on coordinate-independent properties of ${\m}$.  Consequently, experimentally measured quantities depend only on the geometric invariants of ${\m}$, such as the curvature.   Formulating physical observables geometrically avoids arguments based on a particular choice of fields.  It also allows us to correctly pose and answer the question of whether the Higgs boson transforms linearly or non-linearly under the electroweak gauge symmetry.   Further, a geometric analysis gives a better understanding of the structure of the theory and its coordinate-invariant properties.

The UV theory can have additional states, such as massive meson excitations in the case of theories with strong dynamics. At low energies, the EFT interactions in the electroweak symmetry breaking sector are described by a Lagrangian with scalar degrees of freedom on some manifold $\cal M$, with the Lagrangian expanded in gradients of the scalar fields. The geometric description captures the features of the UV dynamics needed to make predictions for experiments at energies below the scale of new physics.

The geometrical structure of non-linear sigma models has been worked out over many years, mainly in the context of supersymmetric sigma models (see e.g.~\cite{Meetz:1969as,Honerkamp:1996va,Honerkamp:1971sh,Ecker:1972bm,AlvarezGaume:1981hn,AlvarezGaume:1981hm,Boulware:1981ns,Salam:1981xd,Friedan:1980jm,Gaillard:1985uh,AlvarezGaume:1987kj}). The applications to the SM Higgs sector presented here are new, and they provide a better understanding of the structure of HEFT and the search for signals of new physics through the couplings of the Higgs boson.

Some of the results in this paper have already been given in Ref.~\cite{Alonso:2015fsp}.  Here we provide more explanation of the results presented there, as well as details of explicit calculations in that work.  These calculations include the proof of renormalizability of the SM written in non-linear form, and the derivation of the one-loop effective action for a curved scalar manifold ${\cal M}$.  For most of the paper, we will assume that the scalar sector has an enlarged global symmetry, known as custodial symmetry.  Also note that we will usually treat the scalar sector in the ungauged case, referring to the scalar fields as Higgs and Goldstone bosons.  The gauged version of the theory follows immediately by replacing ordinary derivatives by gauge covariant derivatives.  In the gauged case, the Goldstone bosons are eaten via the Higgs mechanism, becoming the longitudinal polarization states of the massive electroweak gauge bosons. Thus, the Higgs-Goldstone boson relations we refer to are in fact relations between the couplings of the Higgs boson and the three longitudinal gauge boson states $W^\pm_L$ and 
$Z_L$~\cite{cornwall,vayonakis,leequiggthacker}.

The organization of the paper is as follows.  The relationship between the SM, SMEFT and HEFT is discussed in Sec.~\ref{sec:eft} from a geometrical point of view.  It is shown that SMEFT is a special case of HEFT when ${\m}$ is expanded about an $O(4)$ invariant fixed point.  Further, it is shown that the existence of such an $O(4)$ invariant fixed point is a necessary and sufficient condition for the existence of a choice of scalar fields such that the Higgs field transforms linearly under the electroweak gauge symmetry. In Sec.~\ref{sec:O(N)model}, a scalar field redefinition is performed on the SM Lagrangian to write it in terms of the non-linear exponential scalar field parametrization of Callan, Coleman, Wess and Zumino (CCWZ)~\cite{Coleman:1969sm,Callan:1969sn}.   In this non-linear parametrization, the SM contains an infinite series of terms with arbitrarily high dimension, but it nonetheless remains renormalizable.  We demonstrate  renormalizability of the CCWZ form of the Lagrangian by an explicit calculation of the one-loop correction to $\phi \phi \to \phi \phi$.  The $S$-matrix is finite, even though Green's functions are divergent.  The one-loop calculations in the linear and non-linear parameterizations only differ by equation-of-motion terms.  Both parameterizations have a divergence-free $S$-matrix at one loop after including the usual counterterms computed in the unbroken phase. Sec.~\ref{sec:curved} presents the covariant formalism for curved scalar field space.  We discuss global and gauge symmetries in terms of Killing vectors of the scalar manifold, and we derive the one-loop correction to the effective action for curved ${\m}$. In Sec.~\ref{sec:CCWZ}, the geometric formulation of $\g/\h$ theories is connected with the standard coordinates of CCWZ.  We give formul\ae\ for the curvature tensor in terms of field strengths for a general sigma model.  We also discuss the extension of the CCWZ standard coordinates to non-compact groups.  As shown in  Ref.~\cite{Alonso:2015fsp}, the sign of deviations from SM values of Higgs boson-longitudinal gauge boson scattering amplitudes is controlled by sectional curvatures in HEFT.  For $\g/\h$ theories based on compact groups, these sectional curvatures are typically positive.  We compute the sectional curvature, and show that in certain cases, it can be negative. In Sec.~\ref{sec:custodial}, we briefly discuss the SM and custodial symmetry violation, and the relation between the SM scalar manifold and the configuration space of a rigid rotator. Sec.~\ref{sec:multiHiggs} generalizes HEFT to the case of multiple singlet Higgs bosons.  Finally, Sec.~\ref{sec:conclusions} provides our conclusions.  Additional formulae are provided in the appendices, including intermediate steps in the computation of the one-loop correction to HEFT given in Refs.~\cite{Guo:2015isa,Alonso:2015fsp}, and discussion of the complications for non-reductive cosets.

\section{ SM $\subset$ SMEFT $\subset$ HEFT}\label{sec:eft}

In this section, we discuss the scalar sector of the SM and its EFT generalizations, SMEFT and HEFT, as well as the relationship between these three theories. We begin with a summary of the scalar sector of the SM.  

The SM scalar Lagrangian (with the gauge fields turned off) is
\begin{align}
L &= \partial_\mu H^\dagger \partial_\mu H - \lambda \left( H^\dagger H - \frac{v^2}{2} \right)^2 \,.
\label{2.2a}
\end{align}
This scalar Lagrangian is the most general $SU(2)_L \times U(1)_Y$ invariant Lagrangian with terms of dimension $\le 4$ built out of a Higgs doublet $H$ that transforms as  $\mathbf{2}_{1/2}$ under $SU(2)_L \times U(1)_Y$.  As is well-known, the SM scalar sector has an enhanced global custodial symmetry group $O(4) \sim SU(2)_L \times SU(2)_R$.  This global symmetry can be made manifest by writing the SM complex scalar doublet field $H$ in terms of four real scalar fields,
\begin{align}
H & \equiv\frac{1}{\sqrt 2} \left[ \begin{array}{cc}  \phi^2 + i\phi^1 \\ \phi^4 - i \phi^3  \end{array}\right]\,.
\label{2.1}
\end{align}
Substitution in Eq.~(\ref{2.2a}) yields the Lagrangian
\begin{align}
L &
=\frac12 \partial_\mu \bm{\phi} \cdot \partial_\mu \bm{\phi} - \frac{\lambda}{4}  \left( \bm{\phi} \cdot \bm{\phi} - v^2\right)^2\, ,
\label{2.2}
\end{align}
where $\bm{\phi}=(\phi^1,\phi^2,\phi^3,\phi^4)$.  Lagrangian Eq.~(\ref{2.2}) is invariant under ${\cal G}= O(4)$ global symmetry transformations
\begin{align}
\bm{\phi} &\to O \bm{\phi}, & O^T O &=\mathbf{1}.
\label{2.3}
\end{align}
The scalar field $\bm{\phi}$ transforms {\it linearly} as the four-dimensional vector representation of the global symmetry group $\g=O(4)$.  The minimum of the potential is the three-sphere $S^3$ of radius $v$,
\begin{align}
\langle\bm{\phi} \cdot \bm{\phi} \rangle &=v^2\,,
\label{2.4}
\end{align}
which is the Goldstone boson vacuum submanifold of the SM. The radius of the sphere, $v\sim 246$ GeV, is fixed by the gauge boson masses.  It is conventional to choose the vacuum expectation value 
\begin{eqnarray}
\vev{\bm{\phi} }= v \left[ \begin{array}{c} 0 \\ 0 \\ 0 \\ 1 \end{array} \right],
\end{eqnarray}
and expand the Lagrangian about this vacuum state in the shifted fields $\phi^4 \equiv v+ \mathsf{h}$ and $\phi^a \equiv \varphi^a$, $a=1,2,3$,
\begin{align}
\bm{\phi}&= \left[ \begin{array}{c} \varphi^1 \\ \varphi^2 \\ \varphi^3 \\ v + \mathsf{h} \end{array} \right],
\qquad &H=\frac{1}{\sqrt 2} \left[ \begin{array}{cc}  \varphi^2 + i\varphi^1 \\ v+ \mathsf{h} - i \varphi^3  \end{array}\right]\,.
\label{2.5}
\end{align}
The vacuum expectation value $\vev{\bm{\phi} }$ spontaneously breaks the global symmetry group $\g = O(4)$ to the unbroken global symmetry group $\h = O(3)$.  The Goldstone bosons $\bm{\varphi}^a$, $a=1,2,3$, transform as a triplet under the unbroken global symmetry, whereas $\mathsf{h}$ transforms as a singlet.  We will refer to both the enlarged global symmetries $\g=O(4) \sim SU(2)_L \times SU(2)_R$ and $\h=O(3) \sim SU(2)_V$ as custodial symmetries.  The unbroken global symmetry group $\h$ leads to the relation $M_W=M_Z \cos \theta_W$, which is a successful prediction of the SM.  The experimental success of this gauge boson mass relation implies that custodial symmetry is a good approximate symmetry of the SM.

The Lagrangian Eq.~(\ref{2.2}) in terms of shifted fields Eq.~(\ref{2.5}) becomes 
\begin{align}
L & =  \frac12 \partial_\mu \bm{\varphi} \cdot \partial_\mu \bm{\varphi} + \frac12  \, \left(\partial_\mu \mathsf{h}\right)^2 - \frac{\lambda}{4}  \left( \mathsf{h}^2 + 2 \mathsf{h} v + \bm{\varphi \cdot \varphi} \right)^2 .
\label{2.6}
\end{align}
The singlet $\mathsf{h}$ is the physical Higgs field with mass
\begin{align}
m_{\mathsf{h}}^2 &= 2 \lambda v^2\,,
\label{2.7}
\end{align}
whereas the Goldstone bosons are strictly massless. In the gauged theory, the three Goldstone bosons $\bm{\varphi}^a$ of the $\g \to \h$ global symmetry breakdown are ``eaten" via the Higgs mechanism, becoming the longitudinal polarization states of the massive $W^\pm$ and $Z$ gauge bosons.  Note that the $O(4)$-invariant potential $V(\mathsf{h}, \bm{\varphi})$ depends on an $O(4)$-invariant combination of both $\mathsf{h}$ and $\bm{\varphi}$.

Equating the scalar kinetic energy term in Eq.~(\ref{2.6}) with
\begin{eqnarray}
L_{\rm KE} &=& \frac 12 g_{ij}\left(\phi \right) \left( \partial_\mu \phi^i \right) \left( \partial^\mu \phi^j \right), \qquad i,j=1,2,3,4, 
\label{scalarmetric}
\end{eqnarray}
defines the scalar metric $g^{\rm SM}_{ij}(\phi) = \delta_{ij}$ for the SM scalar manifold ${\cal M}$ with coordinates given by the scalar fields $\phi^i$.  Distances on ${\cal M}$ are determined by $ds^2 = g_{ij}\left( \phi \right) d\phi^i d\phi^j$.

\begin{figure}
\centering
\includegraphics[bb=315 240 485 375,width=7cm]{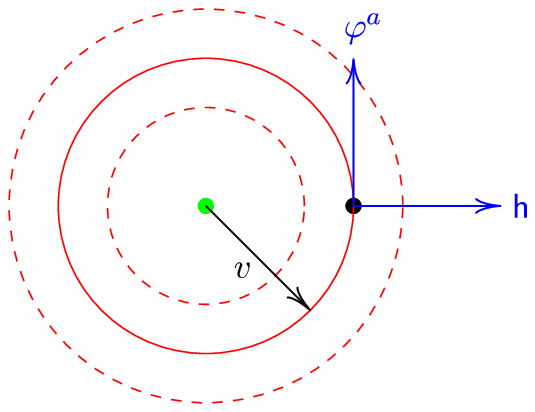}\hspace{1cm}
\raise0.1cm\hbox{\includegraphics[bb=320 575 490 710,width=7cm]{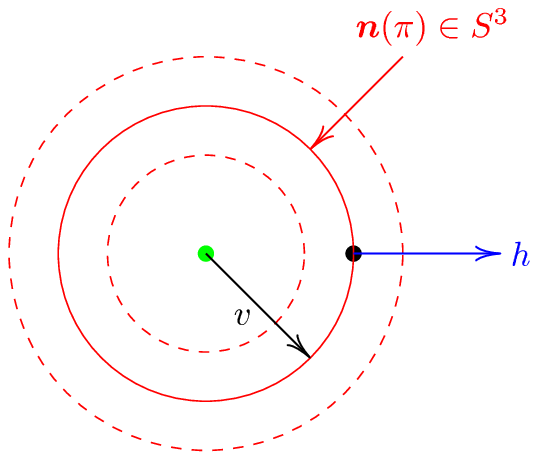}}
\caption{\label{fig:1} Two-dimensional depiction of the four-dimensional scalar manifold ${\m} = \mathbb{R}^4$ of the SM.  The SM vacuum is the black dot shown in the figure. The origin (green dot) is an $O(4)$ invariant fixed point.  The left and right diagrams show the fields in Cartesian and polar coordinates, respectively. $O(4)$ symmetry acts linearly on the Cartesian coordinates. In polar coordinates, $h$ is $O(4)$-invariant, and the angular coordinates $\bm{n}(\pi)$ transform non-linearly under the $O(4)$ symmetry.  The scalar manifold ${\m}$ is flat, so the scale $\Lambda$ setting the curvature is formally infinite. }
\end{figure}

The four-dimensional SM scalar manifold ${\m}= \mathbb{R}^4$ is shown in Fig.~\ref{fig:1}.  The $O(4)$ symmetry acts by rotations. The minimum of the potential is the solid red curve, and forms the three-dimensional Goldstone boson submanifold $S^3$ of radius $v$. The parameterization Eq.~(\ref{2.5}) is a Cartesian coordinate system for ${\m}$ centered on the vacuum (black dot), where $\mathsf{h}$ is the horizontal direction, and $\bm{\varphi}^a$, $a=1,2,3$, are the three other directions orthogonal to $\mathsf{h}$.  The angular coordinates of $S^3$ are $\varphi^a/v$.  The $O(4)$ symmetry acts \emph{linearly} on $(\varphi^1,\varphi^2,\varphi^3,v+\mathsf{h})$. 

In Cartesian coordinates, it seems intuitively clear that $\varphi^a$ and $\mathsf{h}$ interactions are related, given that the four scalar fields belong to the same Higgs doublet Eq.~(\ref{2.1}).  However, the precise relation is subtle. In order to understand this point better, it is instructive to express the SM Lagrangian Eq.~(\ref{2.2}) in polar coordinates as well.

In polar coordinates,\footnote{We use $\mathsf{h},\varphi$ for the fields in Cartesian coordinates, and $h,\pi$ (or $h,\mathbf{n}$) in polar coordinates.}
\begin{align}
\bm{\phi} &= (v+h) \bm{n}(\pi)\,, & \bm{n \cdot n} &=1\,,
\label{2.8}
\end{align}
where $(v+h)$ is the magnitude of $\bm{\phi}$, and $\bm{n}(\pi) \in S^3$ is a four-dimensional unit vector.   The four shifted scalar fields consist of the three dimensionless angular coordinates $\bpi^a=\pi^a/v$ (the direction of $\bm{n}(\pi)$ on $S^3$), and the radial coordinate $h$. The SM Lagrangian in polar coordinates is
\begin{align}
L &= \frac12 (v+h)^2 \left(\partial_\mu \bm{n} \right)^2 + \frac12 \left(\partial_\mu h \right)^2  - \frac{\lambda}{4}
\left(h^2 + 2 v h\right)^2\,.
\label{2.9}
\end{align}
An advantage of expressing the SM Lagrangian in polar coordinates is that the three Goldstone boson fields of $\bm{n}(\pi)$ are derivatively coupled.   In addition, the scalar potential in polar coordinates only depends on the radial coordinate $h$, whereas in Cartesian coordinates it depends on all four scalar fields.

The $O(4)$ symmetry transformations of $\m$ in polar coordinates are
\begin{align}
h &\to h, & \bm{n} &\to O\, \bm{n},
\label{2.10}
\end{align}
so the Higgs field $h$ is invariant under $O(4)$ transformations, and $\bm{n}$ transforms linearly by an orthogonal transformation that preserves the constraint  $\bm{n \cdot n}=1$.   Due to the constraint, however, only three of the four components of $\bm{n}$ are independent.  Without loss of generality, one can take the first three components of  $\bm{n}$ to be the independent components.  Then, the fourth component $n^4$ is a non-linear function of the independent components $n^{1,2,3}$.  The non-linear constraint  $\bm{n \cdot n}=1$ turns the linear $O(4)$ transformation on $\bm{n}$ into a {\it non-linear} transformation when written in terms of unconstrained fields.  Thus, the $O(4)$ transformation on the three independent angular coordinates $\pi^a/v$ is a {\it non-linear} transformation.

Many different parameterizations of $\bm{n}(\pi)$ in terms of the independent unconstrained coordinates $\pi^a/v$ are possible.  Two natural non-linear parameterizations are the square root parameterization and the exponential parameterization, which are defined by
\begin{align}
\bm{n}(\pi) &=\frac{1}{v} \left[ \begin{array}{cc}  \pi^1 \\ \pi^2 \\ \pi^3 \\ \sqrt{v^2-\bm{\pi} \cdot \bm{\pi}}  \end{array}\right]\, ,
\label{2.11}
\end{align}
and
\begin{align}
\bm{n}(\pi) &
= {\rm{exp}} \left(\frac{1}{v} \left[ \begin{array}{cccc} 0 & 0 & 0 & \pi^1 \\ 
0 & 0 & 0 & \pi^2 \\ 
0  & 0 & 0 & \pi^3 \\ 
-\pi^1 & -\pi^2 & -\pi^3 & 0 \end{array}\right] \right) \left[ \begin{array}{c} 0 \\ 0 \\ 0 \\ 1 \end{array} \right],
\label{nexp}
\end{align}
respectively. For most of this paper, we use the exponential parameterization for $\bm{n}(\pi)$ since it corresponds to the standard coordinates of CCWZ.  

Rotations in the $12$, $13$ and $23$ planes act linearly on $(n^1,n^2,n^3)$, and leave $n^4$ invariant.  However, rotations in the $14$, $24$ and $34$ planes mix $(n^1, n^2, n^3)$ and $n^4$.  For example, a $14$ rotation gives
\begin{align}
\delta n^1 &= \delta \theta \, n^4 ,  &
\delta n^2 &= 0, &
\delta n^3 &=0, &
\delta n^4 &= -\delta \theta \, n^1 .
\end{align}
In terms of the independent unconstrained coordinates $\pi^a$ of the square root parameterization, $12$, $13$ and $23$ rotations act linearly, but a 14 rotation gives
\begin{align}
\delta \pi^1 &= \delta \theta \, \sqrt{v^2-\bm{\pi}\cdot \bm{\pi} },  &
\delta \pi^2 &= 0, &
\delta \pi^3 &=0. 
\label{2.12}
\end{align}
The $O(4)$ transformation Eq.~(\ref{2.12}) is \emph{non-linear}.  Consequently, Eq.~(\ref{2.10}) is called a \emph{non-linear} transformation, since it is non-linear when written in terms of unconstrained coordinates $(\pi^1, \pi^2, \pi^3)$.

In polar coordinates, $\bm{n}$ and $h$ are very different objects, and it is not at all obvious that $\bm{n}$ and $h$ interactions are related. Nevertheless, all we have done is switch from Cartesian coordinates $\{\varphi^a, \mathsf{h}\}$ to polar coordinates $\{ \pi^a, h \}$ while keeping the Lagrangian fixed. This change of coordinates does not affect physical observables such as $S$-matrix elements. Any relations that exist amongst physical observables must be present irrespective of the choice of coordinates.

We have summarized the standard analysis of the SM in Cartesian and polar coordinates. In Cartesian coordinates, the Higgs field $\mathsf{h}$ and the three Goldstone fields ${\varphi}^a$ form a four-dimensional representation which transforms \emph{linearly} under $O(4)$. In polar coordinates, the Higgs field $h$ is an $O(4)$ singlet or invariant, and the three Goldstone bosons $\pi^a$ parameterizing the $S^3$ unit vector $\bm{n}(\pi)$ transform among themselves under the \emph{non-linear} $O(4)$ transformation law Eq.~(\ref{2.10}).  The Higgs boson field $h$ in polar coordinates is not the same field as the Higgs boson field $\mathsf{h}$ in Cartesian coordinates.  The relation between the two Higgs boson fields is
\begin{align}
(v+h)^2 &= (v+\mathsf{h})^2 + \bm{\varphi} \cdot \bm{\varphi},
\label{2.13}
\end{align}
so that
\begin{align}
h &= \mathsf{h} + \frac{\bm{\varphi} \cdot \bm{\varphi}}{2 v}   - \frac 12 \frac{\mathsf{h}\,\bm{\varphi} \cdot \bm{\varphi}}{v^2} + \ldots
\label{2.14}
\end{align}
By the Lehmann-Symanzik-Zimmermann (LSZ) reduction formula, $h$ and $\mathsf{h}$ give the \emph{same} $S$-matrix, and both are perfectly acceptable choices for the Higgs boson field.\footnote{The nomenclature ``the Higgs field'' is misleading, since there is no unique choice for the Higgs field.}

\subsection{$O(4)$ Fixed Point}

We now return to the question of whether the Higgs field transforms linearly or non-linearly under the electroweak gauge symmetry, and whether interactions of the Higgs boson and the three Goldstone bosons (i.e.\ longitudinal gauge boson polarizations) are related. As we have just seen, this question is not well-posed in the SM, since the answer depends on the choice of coordinates.
However, it is intuitively clear that there is an underlying relationship between the couplings of the Higgs and Goldstone bosons in the SM that does not remain valid in the general context of HEFT.  We need to formulate any coupling relations in a coordinate-invariant way.  There are two conditions which make the SM special --- (i) there is a point $\bm{\phi}= \bm{0}$ (or $H=\bm{0}$) of $\m$ which is an $O(4)$ invariant fixed point, and (ii) the scalar manifold $\m$ is flat, i.e.\ it has a vanishing Riemann curvature tensor.\footnote{In Cartesian coordinates, $g_{ij}^{\rm SM}(\phi) = \delta_{ij}$, and it trivially follows that the Riemann curvature tensor vanishes.  Since the curvature is coordinate independent, it also vanishes in polar coordinates, even though the metric is more complicated.} As we now see, relations in the SM between the couplings of the Higgs boson and the three Goldstone bosons arise from these two conditions which are no longer true in HEFT in general.
 
We first analyze whether the Higgs field is part of a multiplet that transforms \emph{linearly} under the $O(4)$ symmetry.  Even in the SM, the answer to this question depends on the choice of coordinates.  The coordinate-invariant formulation of the question is: Does there exist a choice of coordinates for $\m$ such that the Higgs field is part of a multiplet that transforms \emph{linearly} under the $O(4)$ symmetry? We now show that the answer is yes if and only if $\m$ has an $O(4)$ invariant fixed point.\footnote{In theories without custodial symmetry, the fixed point is $SU(2)_L \times U(1)_Y$ invariant.}

It is clear from the $O(4)$ transformation law Eq.~(\ref{2.3}) for $\bm{\phi}$ that the origin $\bm{\phi}=\bm{0}$ is an $O(4)$ invariant fixed point.  Any other theory that can be formulated using fields $\bm{\phi}$ which transform \emph{linearly} under the $O(4)$ symmetry also must have an $O(4)$ invariant fixed point at $\bm{\phi}=\bm{0}$.  Thus, if there exists a choice of coordinates $\bm{\phi}$ which transform linearly under the $O(4)$ symmetry, then the scalar manifold ${\cal M}$ has an $O(4)$ invariant fixed point.

Now, we prove the converse statement.  Consider a general scalar manifold $\m$, which is described by coordinates which transform under $O(4)$ transformations and which contains an $O(4)$ invariant fixed point $P$.  Is there a choice of coordinates such that the scalar fields transform linearly under the $O(4)$ symmetry?  The key result we need for the proof in this direction is the linearization lemma of Coleman, Wess and Zumino~\cite{Coleman:1969sm}, which states that if $P$ is an $O(4)$ invariant fixed point, there exists a set of coordinates in a neighborhood of $P$ which transform \emph{linearly} under $O(4)$ transformations in some (possibly reducible) representation of $O(4)$.
If this $O(4)$ representation contains the four-dimensional vector representation of $O(4)$, then the four coordinates  $\phi^i$, $i=1,2,3,4$, which transform as a vector, can be combined into a Higgs doublet $H$, as in Eq.~(\ref{2.1}).  Thus, the Higgs field is part of a linear representation $H$ if and only if there is an $O(4)$ invariant fixed point whose tangent space transforms under $O(4)$ in a representation that contains the vector representation.  In most of our examples, the scalar manifold is four-dimensional, and the tangent space of $P$ automatically transforms as the vector representation, so we will omit the condition that the tangent space transforms as the vector representation.

The condition that ${\cal M}$ contains an $O(4)$ fixed point divides theories into those which can and cannot be written in a form where the Higgs boson is part of a multiplet that transforms \emph{linearly} under the electroweak gauge symmetry group ${\g}_{\rm gauge}= SU(2)_L \times U(1)_Y$ (or the larger global custodial symmetry group ${\g}= O(4) = SU(2)_L \times SU(2)_R$). There are theories which satisfy the condition that ${\cal M}$ contains an $O(4)$ invariant fixed point, but which do not have relations between the couplings of the Higgs boson and the Goldstone bosons.  To understand this point better, we now introduce SMEFT and HEFT.

\subsection{SMEFT}

SMEFT is an effective theory with the most general Lagrangian written in terms of SM fields, including all independent higher dimension operators with dimension greater than four, suppressed by an EFT power counting scale $\Lambda$.  The independent operators at dimension six, and their renormalization~\cite{Buchmuller:1985jz,Grzadkowski:2010es}, has been worked out in detail~
\cite{Grojean:2013kd,Elias-Miro:2013gya,Elias-Miro:2013mua,Jenkins:2013zja,Jenkins:2013wua,Alonso:2013hga,Alonso:2014rga,Alonso:2014zka}.

In SMEFT, all operators involving scalar fields are written in terms of the Higgs doublet field $H$.   For simplicity, at present we assume that the custodial symmetry group of SMEFT is ${\g} = O(4)$.  The SMEFT scalar kinetic energy term, which consists of all operators built out of Higgs doublet fields with two derivatives, is
\begin{eqnarray}
L_{\rm KE} &=& \partial_\mu H^\dagger \partial^\mu H + \frac{1}{\Lambda^{d-4}} \sum_i C_i O_i^{(d)} \nn
&=& \partial_\mu H^\dagger \partial^\mu H + \frac{1}{\Lambda^2} C_{H D} \left( H^\dagger \partial_\mu H \right)^* \left( H^\dagger \partial^\mu H \right) + \cdots ,
\end{eqnarray}
where the sum in the first line is over all independent mass dimension $d$ operators built out of two derivatives and Higgs doublet fields $H^\dagger$ and $H$, and the second line gives the explicit expression including the leading $d=6$ operator.
Using Eq.~(\ref{2.1}) to write the Higgs doublet $H$ in terms of four real scalars $\bm{\phi}$, yields a scalar kinetic energy term of the form  
\begin{align}
L_{\rm KE} &=  \frac12  \left[ A\left( \frac{\bm{\phi \cdot \phi}}{\Lambda^2} \right)\ \partial_\mu \bm{\phi} \cdot \partial^\mu \bm{\phi}
+  B\left( \frac{\bm{\phi \cdot \phi}}{\Lambda^2} \right)\ \frac{ \left(\bm{\phi}\cdot  \partial_\mu \bm{\phi} \right)^2 }{\Lambda^2} \right] ,
\label{2.15}
\end{align}
where the arbitrary functions $A(z)$ and $B(z)$ are defined by power series expansions in their argument $z \equiv \bm{\phi} \cdot \bm{\phi}/\Lambda^2$.  In the $\Lambda \to \infty$ limit, the kinetic energy term of SMEFT reduces to the SM kinetic energy term, so the functions $A(z)$ and $B(z)$ satisfy $A(0)=1$ and $B(0)=0$.  Comparison of Eq.~(\ref{2.15}) with Eq.~(\ref{scalarmetric}) yields the SMEFT scalar metric
\begin{align}
g_{ij}(\phi) &=  A\left( \frac{\bm{\phi \cdot \phi}}{\Lambda^2} \right)\  \delta_{ij} 
+  B\left( \frac{\bm{\phi \cdot \phi}}{\Lambda^2} \right)\ \frac{\phi_i \phi_j }{\Lambda^2}\,.
\label{2.15a}
\end{align}
The Riemann curvature tensor $R_{ijkl}(\phi)$ of the curved scalar manifold $\m$ in SMEFT can be calculated from the above metric. The SM is a special case of the SMEFT in which all higher dimension operators with $d >4$ are set to zero, or equivalently, one takes the limit $\Lambda \to \infty$.  From Eq.~(\ref{2.15a}), we see that in this limit the SMEFT metric yields the SM scalar metric $g^{\rm SM}_{ij}(\phi) = \delta_{ij}$ in Cartesian coordinates, and $\m \to \mathbb{R}^4$ becomes \emph{flat} with vanishing Riemann curvature tensor.

Most composite Higgs models~\cite{Kaplan:1983fs,Dugan:1984hq} can be written in SMEFT form. A simple example is the $SO(5) \to SO(4)$ composite Higgs model~\cite{Agashe:2004rs}. The  symmetry breaking field lives on a sphere of radius $f$ in five dimensions, and can be written as
\begin{align}
\left[ \begin{array}{c} \bm{\phi} \\[5pt] \sqrt{f^2-\bm{\phi \cdot \phi}} \end{array}\right]\,.
\label{o5}
\end{align}
$\bm{\phi}$ is the SMEFT field, and the Lagrangian can be written in SMEFT form. In general, composite Higgs theories solve the hierarchy problem by vacuum misalignment. There is a field configuration where the vacuum is ``aligned,'' so that the electroweak symmetry is unbroken. This is the point $\bm{\phi}=\bm{0}$ of SMEFT, and $\bm{\phi}$ measures deviations from this point, as in Eq.~(\ref{o5}). In the neighborhood of $\bm{\phi}=\bm{0}$, $\bm{\phi}$ gives a linear representation of $O(4)$. For HEFT to reduce to SMEFT form, this representation must transform as the vector of $O(4)$. Composite Higgs models which are consistent with experimental data are of this type~\cite{Alonso:2014wta,Hierro:2015nna}.

The SMEFT is the EFT generalization of the SM where the scalar manifold has an $O(4)$ invariant fixed point, so that the Lagrangian can be written in terms of the Higgs doublet field $H$ or the four-dimensional vector field $\bm{\phi}$ on which the $O(4)$ symmetry acts linearly.  This restriction is not enough to give the same scattering amplitudes of Higgs bosons and Goldstone bosons (longitudinal gauge bosons) as the SM, which can be verified by explicit computation using Eq.~(\ref{2.15}).  In Refs.~\cite{Alonso:2015fsp,Alonso:2016btr}, it was shown that the high energy behavior of the cross sections for $W_LW_L \to W_L W_L$ and $W_L W_L \to h h$ scattering depend on two sectional curvatures which can be obtained from the Riemann curvature tensor $R_{ijkl}(\phi)$. The one-loop radiative correction in the scalar sector also depends on the Riemann curvature tensor $R_{ijkl}(\phi)$~\cite{Alonso:2015fsp}. The details of these calculations are presented later in this paper.  The important point is that the $\phi \phi \to \phi \phi$ scattering cross sections and the one-loop radiative correction in SMEFT are equal to the SM values if and only if $\m$ is \emph{flat}, i.e. the Riemann curvature tensor of SMEFT vanishes.  This statement is a coordinate-independent condition, which is true in the SM using either Cartesian or polar coordinates.  Thus, the intuitive idea that the Goldstone boson and Higgs boson directions in Fig.~\ref{fig:1} are related in the SM can be formulated precisely as the condition that $\m$ in the SM is a four-dimensional flat Euclidean space.

\subsection{HEFT}

\begin{figure}
\centering
\includegraphics[bb=85 610 260 738,width=7cm]{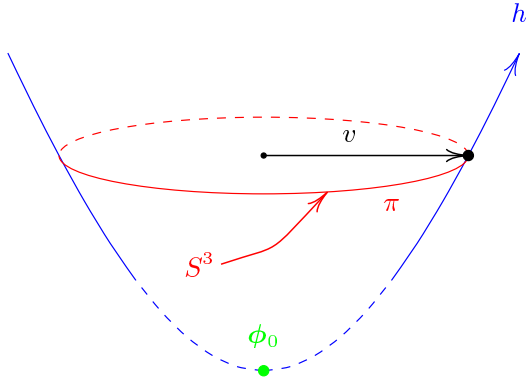}
\caption{\label{fig:heft} The HEFT scalar manifold. There is $S^3$ for each value of $h$. An $O(4)$ invariant fixed point exists if there is a value of $h$ for which the radius of $S^3$ vanishes. The fixed point $\bm{\phi}_0$ at $h=h_*$ is shown in a dotted region of $\m$ since it need not exist. There is no boundary at the transition between the solid and dotted regions, if  the dotted region does not exist. Instead, the manifold can extend to infinity, or is smoothly connected without a point where $F(h)=0$. SMEFT has a scalar manifold where $\bm{\phi}=0$ is an $O(4)$ invariant fixed point that always exists, and are like the HEFT manifold including the dotted section.}
\end{figure}
HEFT is a generalization of the SM using the polar coordinate form of the SM Lagrangian, Eq.~(\ref{2.9}).  The theory is written in terms of three angular coordinates $\pi^a/v$ that parametrize a unit vector $\bm{n}({\pi}) \in S^3$, and one or more  coordinates $\{h_i\}$. As in the SM, the unit vector $\bm{n}$ parametrizes the Goldstone bosons directions~\cite{Longhitano:1980iz,Longhitano:1980tm,Appelquist:1980vg,Appelquist:1980ae,Feruglio:1992wf}. Here we restrict to one additional $h$ field. The case of multiple $\{h_i\}$ is considered in Sec.~\ref{sec:multiHiggs}.
The coordinate $h$ is chosen so that $h=0$ is the ground state. 
The HEFT Lagrangian is
\begin{align}
L &= \frac12 v^2 F(h)^2 \left(\partial_\mu \bm{n} \right)^2  + \frac12 \left(\partial_\mu h \right)^2 -V(h) + \ldots
\label{heft}
\end{align}
where $F(h)$ is an arbitrary dimensionless function with a power series expansion in $h/v$~\cite{Grinstein:2007iv}, normalized so that 
\begin{align}
F(0)=1\,,
\label{f0}
\end{align}
since the radius of $S^3$ in the vacuum is fixed to be $v$ by the gauge boson masses. The HEFT manifold is shown schematically in Fig.~\ref{fig:heft}.  $\m$ has a coordinate $h$, with an $S^3$ fiber at each value of $h$.  While $h$ is often called the radial direction by analogy with the polar coordinate form of the SM, in HEFT, $h$ is simply a scalar field, and need not be the radius of anything. HEFT power counting is discussed in \cite{Gavela:2016bzc}, and is a combination of chiral power counting~\cite{Weinberg:1978kz,Gasser:1983yg} and naive dimensional analysis~\cite{Manohar:1983md}.  The terms omitted in Eq.~(\ref{heft}) are the NLO operators~\cite{Alonso:2012px,Alonso:2012pz,Buchalla:2013rka,Gavela:2014vra,Brivio:2016fzo}.

The $O(4)$ transformation laws for $h$ and $\bm{n}$ are given in Eq.~(\ref{2.10}), so $h$ is invariant and $\bm{n}$ transforms non-linearly. The SM and SMEFT are both special cases of HEFT. In the SM, the radial function is
\begin{align}
F^{\rm SM}(h) &= \left(1 + \frac{h}{v} \right)\,.
\label{fsm}
\end{align}
The SMEFT kinetic energy term Eq.~(\ref{2.15}) yields the polar coordinate kinetic energy term
\begin{align}
L &= 
 \frac12  (v+h)^2  A\left(z \right) \left(\partial_\mu \bm{n} \right)^2 + \frac12 \left[  A\left(z\right)+ z\, B\left(z\right)  \right] \left(\partial_\mu h \right)^2, & z&= \frac{(v+h)^2}{\Lambda^2}.
\label{2.19}
\end{align}
This kinetic energy term can be put into the standard form of HEFT by performing a field redefinition on $h$ to make the coefficient of the  $(\partial_\mu h)^2$ term equal to $1/2$.  Thus, the HEFT scalar metric for one singlet Higgs field is
\begin{eqnarray}
g_{ij}(\phi) &=& \left[  \begin{array}{cc} F(h)^2 g_{ab}(\pi) & 0 \\ 0 & 1 \end{array} \right],
\end{eqnarray}
where the function $F(h)$ is parametrized by coefficients $c_n$, $n \ge 1$,
\begin{eqnarray} 
F(h) &=& 1 + c_1 \left( \frac{h}{v} \right) + \frac 12 c_2 \left( \frac{h}{v} \right)^2 + \cdots .
\end{eqnarray}
The coefficient $c_1$ is already constrained by experiment to be equal to its SM value $c_1=1$ to a precision of about $10\%$.  The coefficient $c_2$ is not constrained at present. The HEFT scalar metric reduces to the SM scalar metric when $F(h) = F^{\rm SM}(h) = 1 + h/v$.
 
In the SMEFT, the functions $A$ and $B$ in Eq.~(\ref{2.15a}) are expanded out in powers of $\bm{\phi\cdot\phi}$, whereas in the HEFT literature, they are treated as arbitrary (unexpanded) functions.

When is it possible to rewrite HEFT in SMEFT form?  We have seen that a necessary and sufficient condition is that there must exist an $O(4)$ invariant fixed point $P$ on $\m$. One can then define $\bm{\phi}$ as coordinates around $P$ and write the Lagrangian in terms of $\bm{\phi}$. The general HEFT manifold consists of $h$ and a sequence of spheres of radius $v F(h)$ fibered over each point of $h$. The HEFT manifold is depicted in Fig.~\ref{fig:heft}. $O(4)$ acts on the point $\bm{n}$ on the surface of $S^3$ by rotation, so that $O(4)$ maps points on the the red curve onto itself. No point of $S^3$ is invariant under the full $O(4)$ group, so the only way to have an $O(4)$ invariant fixed point is if the sphere has zero radius, i.e. if $F(h_*)=0$ for some $h_*$.  Such a point may not exist; its existence depends on the structure of the HEFT manifold. For example, if $F(h) = e^{h/v} \cosh (1+h/f)$ the HEFT manifold has no $O(4)$ invariant fixed point. In the SM, $F(h)$ is given by Eq.~(\ref{fsm}), and $F(h_*)=0$ at $h_*=-v$. If there is an $O(4)$ fixed point, the HEFT can be written as a SMEFT. Some examples are given in Refs.~\cite{Brivio:2013pma,Brivio:2014pfa,Brivio:2016fzo}.

To summarize, HEFT with no $O(4)$ invariant point, i.e.\ no point where $F(h)=0$, cannot be written in SMEFT form, and hence cannot be written using a doublet field $H$ (or equivalently, a four-dimensional vector field $\bm{\phi})$ which transforms linearly under the electroweak gauge symmetry.  This statement answers the question posed in the introduction: when do the scalar fields of HEFT transform linearly or non-linearly under the gauge symmetry?  They transform linearly if and only if $F(h_*)=0$ for some $h_*$, so that there is a $O(4)$ fixed point.   

Thus, we have shown that the relationship of the SM, SMEFT and HEFT is described by the hierarchy SM $\subseteq$ SMEFT $\subseteq$ HEFT.  SMEFT is a special case of HEFT when there is a value of the Higgs field $h_*$ where $F(h_*)=0$. The SM is the special case of SMEFT (and HEFT) when there are no higher dimension operators in the theory, and so $\m$ is flat.

One can convert the SMEFT Lagrangian to HEFT form using Eq.~(\ref{2.8}) to switch from Cartesian and polar coordinates. One can attempt to convert from HEFT to SMEFT form using
\begin{align}
\frac{ \bm{\phi} }{ (\bm{\phi \cdot \phi})^{1/2}} &=  \bm{n}
\end{align}
with $(\bm{\phi \cdot \phi})^{1/2}$ some function of $h$.
This substitution gives a Lagrangian $L(\bm{\phi})$ that need not be analytic in $\bm{\phi}$. However, if there is an $O(4)$ fixed point, then there is a suitable change of variables such that the resulting Lagrangian is analytic in $\bm{\phi}$.

Scattering amplitudes are evaluated in perturbation theory by expanding the action in small fluctuations about the vacuum (the black dot) in Fig.~\ref{fig:heft}.  The curvature of $\m$ is a \emph{local} quantity, given by the metric and its derivatives up to second order, evaluated at the vacuum state. Scattering amplitudes, and hence experimentally measurable cross sections depend directly on the curvature~\cite{Alonso:2015fsp,Alonso:2016btr}, so the curvature of the EFT scalar manifold can be determined experimentally.

Whether there is an $O(4)$ invariant fixed point where $F(h_*)=0$ is a non-perturbative question, since $F(0)=1$ in the ground state.  One has to move a distance of at least $h \sim v$ away from the ground state to probe the existence of a fixed point where $F(h)$ vanishes.

\section{Renormalization of the $O(N)$ Model}\label{sec:O(N)model}

One of the main points of Refs.~\cite{Alonso:2015fsp,Alonso:2016btr} and this paper is that the scalar sector can be studied in a coordinate-invariant way.  Thus, the SM written in the linear Cartesian coordinates Eq.~(\ref{2.6}), and the SM written in non-linear polar coordinates Eq.~(\ref{2.9}), are completely equivalent formulations of the same theory. In particular, even though Eq.~(\ref{2.9}) is a non-linear formulation of the SM, where the Lagrangian contains operators of arbitrarily high dimension, it is still renormalizable. In this section, we demonstrate this result by explicit computation of the one-loop $\phi \phi \to \phi \phi$ scattering amplitude. It is instructive to see how the theory is renormalizable even when written in non-linear form --- we find that Green's functions can be divergent but the $S$-matrix is finite. We compute the scattering amplitude in the $O(N)$ theory in the linear and non-linear formulations.  The SM is the special case $N=4$.  Our results are related to the well-known calculations by Longhitano~\cite{Longhitano:1980iz,Longhitano:1980tm} and by Appelquist and Bernard~\cite{Appelquist:1980vg,Appelquist:1980ae} in the non-linear sigma model with no Higgs field, and by Gavela et al.~\cite{Gavela:2014uta} in HEFT.

\subsection{Preliminaries}\label{sec:prelim}

The $O(N)$ sigma model has an $N$-component real scalar field $\bm{\phi}^i = (\phi^1,\ldots,\phi^N)$ with Lagrangian
\begin{align}
L &= \frac12 \partial_\mu \bm{\phi} \cdot \partial^\mu  \bm{\phi} -\frac14 \lambda\left( \bm{\phi} \cdot  \bm{\phi}- v^2\right)^2\, ,
\label{LagON}
\end{align}
which is invariant under transformations 
\begin{align}
\bm{\phi} &\to O \bm{\phi}, & O^T O &=1,\label{phiON}
\end{align}
where $O$ is a real $N \times N$ orthogonal matrix.  The global symmetry group of the theory is ${\g}=O(N)$, which has $N(N-1)/2$ generators.  We are mainly interested in the broken phase $v^2>0$. The minimum of the potential in Eq.~(\ref{LagON}) is at $\langle \bm{\phi} \cdot  \bm{\phi} \rangle = v^2$, so the set of minima form the surface $S^{N-1}$, the sphere in $N$-dimensions, with radius $v$.  All points on $S^{N-1}$ are equivalent vacua.  One can make an $O(N)$ transformation so that
\begin{align}
\vev{\bm{\phi}} &\equiv \bm{\phi_0} = v\,  \bm{\chi}_0 , & \bm{\chi}_0 &= \left[ \begin{array}{c} 0 \\ \vdots \\ 0 \\ 1 \end{array}\right],
\label{3.3}
\end{align}
where $\bm{\chi}_0$ is a unit vector pointing to the North pole of the sphere. The global symmetry group ${\g}=O(N)$ of the theory is spontaneously broken to the subgroup ${\h}=O(N-1)$, the rotations that leave $\bm{\phi}_0$ invariant.  The vacuum manifold is ${\g}/{\h}=S^{N-1}$. The number of broken generators is $\n_G= (N-1)$, so the theory has $\n_G$ Goldstone bosons. 

The generators of $O(N)$ are
\begin{align}
\left[ M_{ab} \right]^i{}_j &= -i \left({\delta^i}_{a} \delta_{jb} - \delta_{ja} {\delta^i}_{b}\right),
& 1 \le a < b \le N,
\label{3.4}
\end{align}
where the non-zero entries of $M_{ab}$ have $-i$ in row $a$, column $b$, and $i$ in row $b$, column $a$. It is often convenient to consider $M_{ab}$ without the restriction $a<b$, which includes each unbroken generator twice, since $M_{ab} = - M_{ba}$. The matrices have been normalized so that
\begin{align}
\tr M_{ab} M_{cd} &= 2\left(\delta_{ac}\delta_{bd}-\delta_{ad} \delta_{bc} \right) .
\label{3.5}
\end{align}
The broken $O(N)$ generators are
\begin{align}
\left[ X_{a} \right]^i{}_j \equiv \left[ M_{aN} \right]^i{}_j &= -i \left({\delta^i}_{a} \delta_{jN} - \delta_{ja} {\delta^i}_{N}\right)
=
-i
\left(
\begin{array}{ccccc}
0 & & \cdots &  & 0 \\
 \vdots & & \vdots & & \vdots \\
0 & & \cdots & & 1 \\
\vdots & & \vdots & & \vdots \\
0 & \cdots & -1 & \cdots & 0 \\
\end{array}\right),
& a=1,\ldots, N-1, 
\label{3.6}
\end{align}
The unbroken $O(N)$ generators are $M_{ab}$, $1 \le a < b \le N-1$,  which are the generators of the $O(N-1)$ subgroup.

The unbroken transformations with the vacuum choice $\bm{\phi}_0$ are $O(N-1)$ rotations that leave the North pole fixed, i.e.\ rotations among the first $(N-1)$ components of $\bm{\phi}$. Of course, one could have picked any other vacuum state $\bm{\phi}_{\bm{n}}$, a vector of length $v$ pointing in some direction $\bm{n}$, which is invariant under  ${\h}_{\bm{n}}$, $O(N-1)$ transformations that leave $\bm{n}$ fixed. Since $\bm{\phi}_0$ can be rotated to $\bm{\phi}_{\bm{n}}$ by a ${\g}=O(N)$ transformation, the two vacua are equivalent and ${\h}_{\bm{n}}$ is conjugate to ${\h}_0$, ${\h}_{\bm{n}}=g {\cal H}_0 g^{-1}$, where $g \in O(N)$ is the transformation that maps $\bm{\phi}_0$ to $\bm{\phi}_{\bm{n}}$, $\bm{\phi}_{\bm{n}}=g \bm{\phi}_0$.

In the linear realization, one expands about the classical vacuum $\bm{\phi_0}$ in Cartesian coordinates
\begin{align}
\bm{\phi}(x) &= \left[ \begin{array}{c} \varphi^1(x) \\ \vdots \\ \varphi^{\n_G}(x) \\ v + \mathsf{h}(x) \end{array}\right].
\label{3.7}
\end{align}
The Lagrangian Eq.~(\ref{LagON}) with this field parametrization is
\begin{align}
L 
&=\frac12 \left(\partial_\mu \mathsf{h}\right) \left(\partial^\mu \mathsf{h}\right) + \frac12 \partial_\mu \bm{\varphi} \cdot \partial^\mu \bm{\varphi} -\frac14 \lambda\bigl(\mathsf{h}^4 + 2\mathsf{h}^2\, \bm{\varphi \cdot \varphi}  + (\bm{\varphi \cdot \varphi})^2 + 4v \mathsf{h}^3 +  4 v \mathsf{h} \,  \bm{\varphi \cdot \varphi}  + 4\mathsf{h}^2v^2 \bigr)\,.
\label{3.8}
\end{align}
The unbroken global symmetry subgroup ${\cal H}=O(N-1)$ under which $\bm{\varphi}$ is a vector is manifest in this coordinate system, but the original global symmetry group ${\cal G}=O(N)$ of the underlying theory is not obvious.  From Eq.~(\ref{3.8}), we see immediately that all $\bm{\varphi}$ are massless, and $\mathsf{h}$ is massive with
\begin{align}
m_{\mathsf{h}}^2 &= 2\lambda v^2 \,.
\label{3.9}
\end{align}
The masses and couplings in Eq.~(\ref{3.8}) are given in terms of two parameters $\lambda$ and $v$, which is a reflection of the hidden $O(N)$ invariance of the theory.

We now parameterize the $O(N)$ model in a different way, following the non-linear realization of CCWZ.  Let
\begin{align}
\bm{\phi}(x) &=\left[v+h(x)\right] \xi(x) \bm{\chi}_0 \, ,
\label{3.10}
\end{align}
where
\begin{align}
\xi(x) &\equiv \exp \left(\bm{\Pi}\right) = \exp\,  \frac{1}{v} \left[ \begin{array}{cccc} 0 & \ldots & 0 & \pi^1 \\ 
0 & \ldots & 0 & \pi^2 \\
\vdots & & \vdots & \vdots \\ 0  & \ldots & 0 & \pi^{\n_G} \\ -\pi^1 & \ldots & -\pi^{\n_G} & 0 \end{array}\right] , & \bm{\Pi} &\equiv \frac{i\pi^a X^a}{v}.
\label{3.11}
\end{align}
Eq.~(\ref{3.10}) is a polar coordinate system in field space with radial coordinate $(v+h)$ and $(N-1)$ dimensionless angular coordinates $\bm{\pi}/v$ of the sphere $S^{N-1}$. The field $\xi(x)$ is a real orthogonal matrix, so
\begin{align}
\bm{\phi \cdot \phi} &= (v+h)^2\,.
\end{align}
The Lagrangian Eq.~(\ref{LagON}) with this field parameterization is
\begin{align}
L &= \frac12(v + h)^2 \bm{\chi_0}^T \left( \partial_\mu \xi\right)^T \left(\partial^\mu \xi \right) \bm{\chi}_0 + \frac12 \left(\partial_\mu h \right) \left( \partial^\mu h \right)-\frac14 \lambda\left(h^2+2hv\right)^2.
\label{3.13}
\end{align}
The potential only depends on the radial coordinate $h$; it is independent of the Goldstone boson fields $\bm{\pi}$, which are massless and derivatively coupled. Expanding the exponential $\xi(x)$ in a power series gives the leading terms
\begin{align}
L 
&= 
\frac12 \left(1+\frac{h}{v}\right)^2 \left[ \partial_\mu \bm{\pi} \cdot \partial^\mu \bm{\pi}  \right] 
+\frac1{6v^2} \left(1+\frac{h}{v}\right)^2 \left[  (\bm{\pi}\cdot \partial_\mu \bm{\pi})^2-
 (\bm{\pi}\cdot \bm{\pi})(\partial_\mu \bm{\pi} \cdot \partial^\mu \bm{\pi}) \right]+\ldots \nn
&+ \frac12 \left( \partial_\mu h \right)  \left( \partial^\mu h \right) -\frac14 \lambda\left(h^2+2hv\right)^2 
 \label{3.14}
\end{align}
The full expression is given in Appendix~\ref{app:full}. The Lagrangian Eq.~(\ref{3.14}) naively looks like a non-renormalizable theory with an infinite set of higher dimension operators. However, it is simply the renormalizable Lagrangian Eq.~(\ref{LagON}) written using a different parametrization of the fields.  

The Lagrangians Eq.~(\ref{LagON}) and Eq.~(\ref{3.14}) correspond to different choices of coordinates for the scalar manifold $\mathcal M$, and they are related by a field redefinition.  Since the $S$-matrix is invariant under a field redefinition, the two theories have the same $S$-matrix.  Renormalizability of Lagrangian Eq.~(\ref{3.14}) is hidden, as is $O(N)$ invariance.  Treating Eq.~(\ref{3.14}) as an EFT with the usual power counting rules (for a pedagogical review, see \cite{Manohar:1996cq}) gives the same $S$-matrix as Eq.~(\ref{LagON}).  In particular, Eq.~(\ref{3.14}) is a renormalizable theory with a \emph{finite} number of renormalization counterterms even though it looks superficially non-renormalizable.

\subsection{Renormalization}

The linear $O(N)$ model including renormalization counterterms is
\begin{align}
L &= \frac12 Z_\phi \partial_\mu \bm{\phi} \cdot \partial^\mu  \bm{\phi} -\frac14 Z_\lambda \lambda  \mu^{2\epsilon} \left(
Z_\phi \bm{\phi} \cdot  \bm{\phi}- Z_v^2 v^2 \mu^{-2\epsilon}\right)^2\,.
\label{3.21}
\end{align}
In dimensional regularization in $4-2\epsilon$ dimensions, the one-loop counterterms $Z_\phi$, $Z_\lambda$ and $Z_v$ are given by
\begin{align}
Z_i &= 1 + \frac{\delta_i}{16\pi^2\epsilon}, &
\delta_\phi &= 0, &
\delta_\lambda &= \lambda(N+8), &
\delta_v &=-3\lambda .
\label{3.22}
\end{align}
These renormalization counterterms can be computed using perturbation theory in the unbroken phase, where $v^2<0$.  The combinations $Z_\lambda$ and $Z_\lambda Z_v^2$ are the counterterm renormalizations of the $O(N)$ invariant operators $( \bm{\phi} \cdot  \bm{\phi})^2$ and $\bm{\phi} \cdot  \bm{\phi}$, and they are gauge independent. 

Field theory divergences arise from the short distance structure of the theory. Thus the renormalization counterterms do not depend on whether the symmetry is unbroken or spontaneously broken; the \emph{same} counterterms Eq.~(\ref{3.21}) also renormalize the broken theory.\footnote{There are subtleties in the gauged case, which are discussed later.} In the broken phase, one uses Eq.~(\ref{3.7}) with the replacement $v \to \left(v + \Delta v \right) \mu^{- \epsilon}$, 
\begin{align}
\phi(x) &= \left[ \begin{array}{c} \varphi^1(x) \\ \vdots \\ \varphi^{\n_G}(x) \\ \left( v + \Delta v \right) \mu^{-\epsilon} + \mathsf{h}(x) \end{array}\right],
& \n_G &=N-1\,.
\label{3.23}
\end{align}
The tadpole shift $\Delta v$ has a perturbative expansion in powers of $\lambda$, and it is computed order by order in perturbation theory by cancelling the tadpole graphs to maintain $\vev{h}=0$.   At tree-level, $\Delta v=0$.  The Lagrangian Eq.~(\ref{3.8}) including renormalization counterterms is
\begin{align}
L &=  \frac12 Z_\phi \,\partial_\mu \bm{\varphi} \cdot \partial^\mu \bm{\varphi}  + \frac12 Z_\phi \left( \partial_\mu \mathsf{h} \right) \left( \partial^\mu \mathsf{h} \right) \nn
&-\frac14 Z_\lambda \lambda  \mu^{2\epsilon} \biggl( Z_\phi  \bm{\varphi \cdot \varphi} + Z_\phi \left[\left( v + \Delta v \right) \mu^{-\epsilon} + \mathsf{h} \right]^2
 - Z_v^2 v^2 \mu^{-2\epsilon}\biggr)^2\, 
\label{3.24}
\end{align}
The Lagrangian Eq.~(\ref{3.8}) gives finite Green's functions and finite $S$-matrix elements in the broken phase. The underlying ${\cal G}$-symmetry of the theory ensures that the counterterms in Eq.~(\ref{3.24}) are given in terms of $Z_{\phi}$, $Z_\lambda$ and $Z_v$ of the unbroken theory Eq.~(\ref{3.22}), plus a tadpole shift $\Delta v$. The $\bm{\varphi \cdot \varphi}$ term in Eq.~(\ref{3.24}) is a pure counterterm, and keeps the Goldstone bosons massless in the presence of radiative corrections. The Higgs mass is $m_{\mathsf{h}}^2=2\lambda v^2$.

In the non-linear realization, one uses
\begin{align}
\bm{\phi} &= \left[\left( v + \Delta v \right) \mu^{-\epsilon}+h(x)\right]\, \xi(x) \bm{\chi}_0 \,
\label{3.25}
\end{align}
with $\xi(x)$ given by Eq.~(\ref{3.11}). Since Eq.~(\ref{3.25}) is simply a different choice of field coordinates in comparison to  Eq.~(\ref{3.23}), the renormalization constants and tadpole shift $\Delta v$ are the same. Note that no $Z$ factor is needed in the exponent of $\xi(x)$. The $\pi^a/v$ in Eq.~(\ref{3.11}) are periodic variables, since a $2\pi$ rotation about some axis is equivalent to the identity transformation, and cannot be multiplicatively renormalized.

The renormalized Lagrangian in the non-linear parameterization is 
\begin{align}
L &= \frac12 \left[\left( v + \Delta v \right) \mu^{-\epsilon}+h(x)\right]^2 Z_\phi^2 \,\bm{\chi_0}^T \left(\partial_\mu \xi \right)^T \left(\partial^\mu \xi \right) \bm{\chi}_0 
+\frac12 Z_\phi^2  \left(\partial_\mu h \right)  \left(\partial^\mu h \right) \nn
&-\frac14 Z_\lambda \lambda\left(Z_\phi \left[\left( v + \Delta v \right) \mu^{-\epsilon}+h(x)\right]^2 - v^2 \mu^{-2\epsilon} \right)^2 ,
\label{3.26}
\end{align}
with $Z_{\phi,\lambda,v}$ given by Eq.~(\ref{3.22}).  This Lagrangian can be expanded in a power series in $\pi$ and used in perturbation theory. The claim which we wish to prove is that Eq.~(\ref{3.26}) gives finite $S$-matrix elements (but not necessarily Green's functions), since it is a field redefinition of Eq.~(\ref{3.24}).

\subsection{$\pi \pi$ Scattering}\label{sec:pipi}

\begin{figure}
\begin{center}
\includegraphics[scale=0.35]{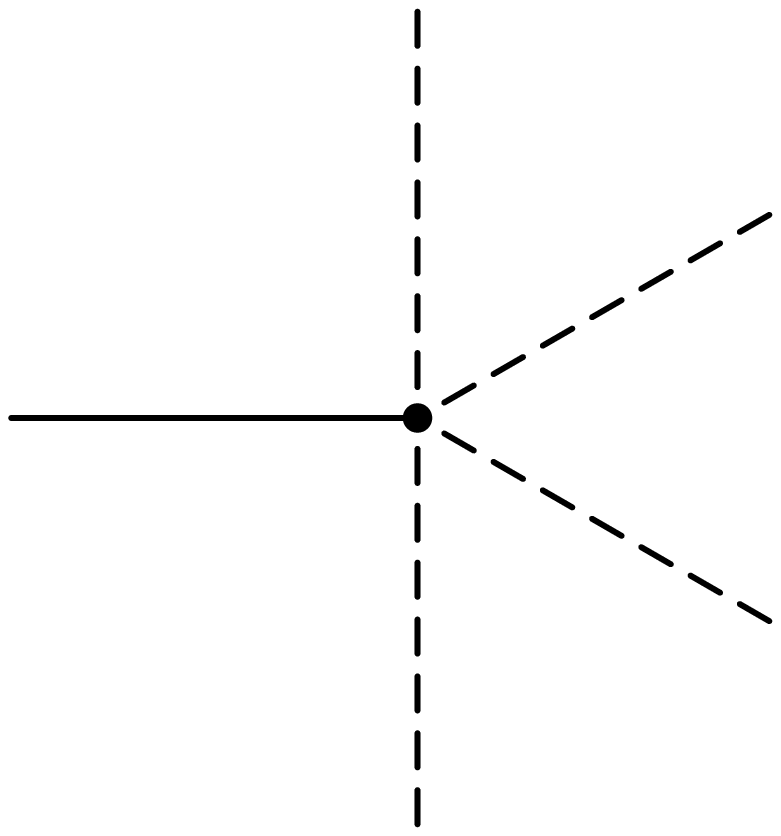}\hspace{2cm}
\includegraphics[scale=0.35]{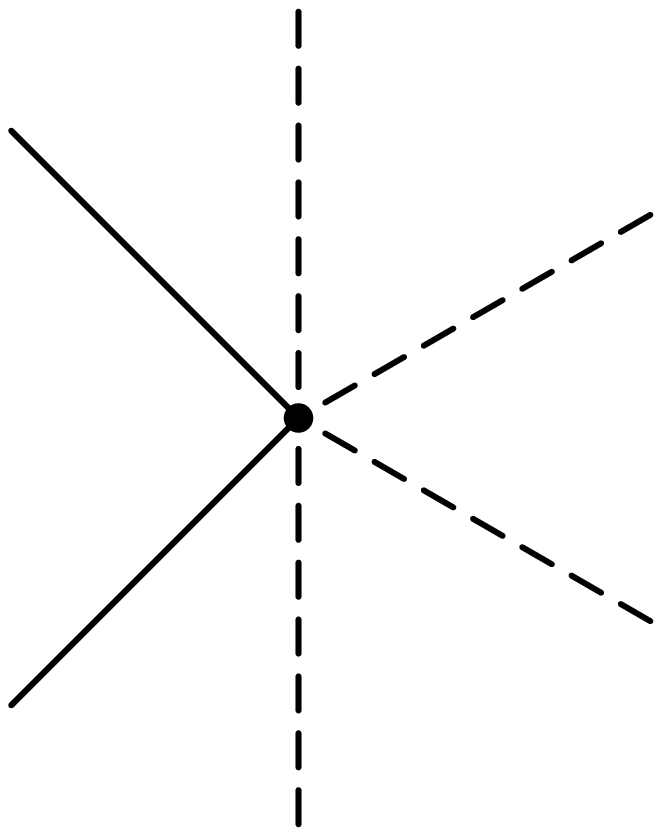}
\end{center}
\caption{ \label{fig:4} Some vertices in the Lagrangian Eq.~(\ref{3.26}). Solid lines are $h$ and dashed lines are $\pi$. }
\end{figure}

\begin{figure}

\begin{align*}
\renewcommand{\arraycolsep}{1cm}
\begin{array}{cc}
\includegraphics[scale=0.3]{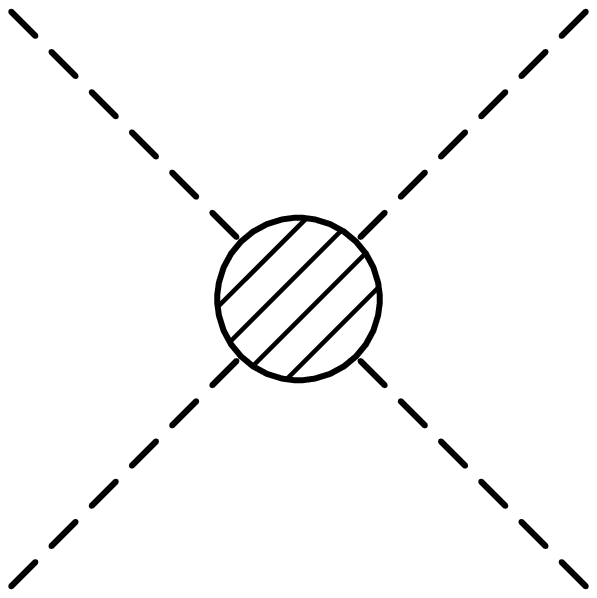} &
\includegraphics[scale=0.3]{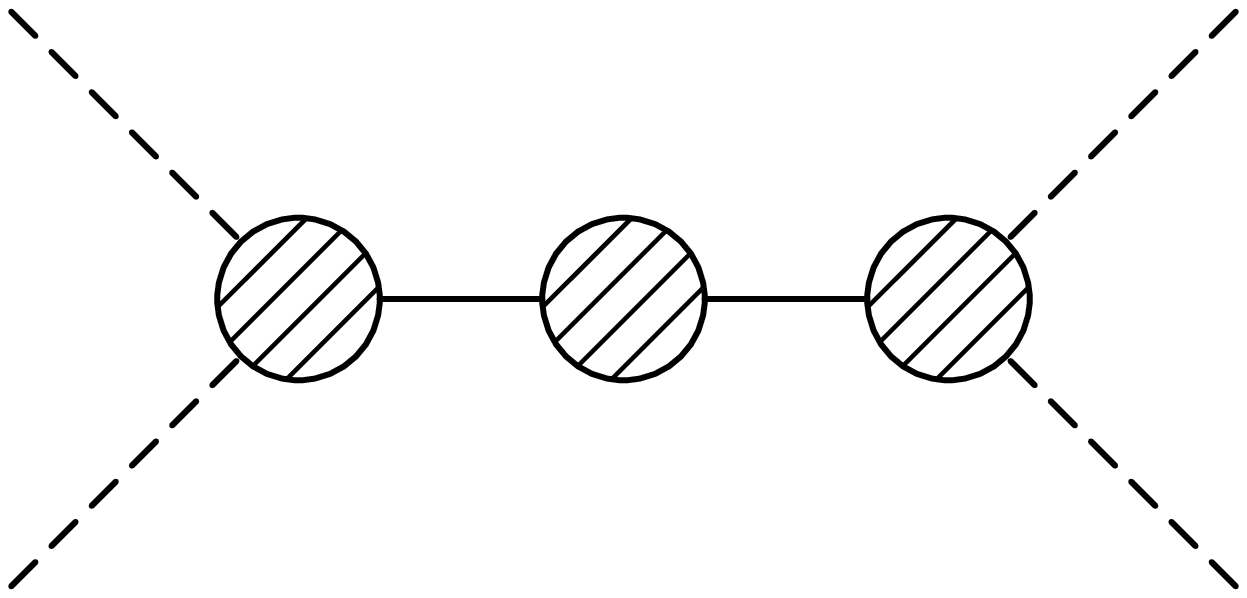} \\[0.2cm]
(a) & (b) 
\end{array}
\end{align*}
\caption{\label{fig:S} Skeleton graphs for the $\pi \pi \to \pi \pi$ scattering $S$-matrix. The shaded blobs are irreducible vertices and two-point functions. There is also a pion wavefunction correction to the amplitude.}
\end{figure}

The finiteness of the $S$-matrix using Lagrangian Eq.~(\ref{3.26}) seems surprising, and is worth explaining in some detail. The Lagrangian Eq.~(\ref{3.26}) contains vertices with an arbitrary number of fields. For example, it contains the vertices in Fig.~\ref{fig:4} which involve five and six scalar fields.
%
%
%
%
We will use Eq.~(\ref{3.26}) to compute the infinite part of $\pi \pi$ scattering to one loop.\footnote{In our notation, for the linear case, we compute $\varphi \varphi \to \varphi \varphi$, and for the non-linear case $\pi \pi \to \pi \pi$.}
In the non-linear case, we will only give the explicit results for the amplitude to $\mathcal{O}(p^4)$, but we have checked that the $S$-matrix is finite to all orders in $p$. The skeleton graphs that contribute to the $S$-matrix for $\pi \pi \to \pi \pi$ are shown in Fig.~\ref{fig:S}. The tree-level amplitude is given by the skeleton graphs in Fig.~\ref{fig:S} with the blobs replaced by tree vertices, and the one-loop correction to the amplitude is given by using the one-loop irreducible vertex for one blob in each graph, and tree vertices for the rest.
We will give the results of the various contributions using the linear parameterization, Eq.~(\ref{3.24}), and the non-linear one, Eq.~(\ref{3.26}), which will be denoted by subscripts $L$ and $N$, respectively. In this subsection, we only compute the infinite parts of the graphs, and  omit an overall factor of $i/(16\pi^2\epsilon)$.
%
%
%
%

\begin{figure}
\begin{align*}
\renewcommand{\arraycolsep}{1cm}
\begin{array}{ccc}
\includegraphics[scale=0.35]{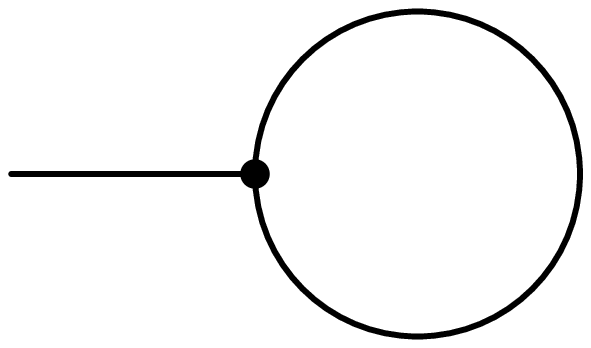} &
\includegraphics[scale=0.35]{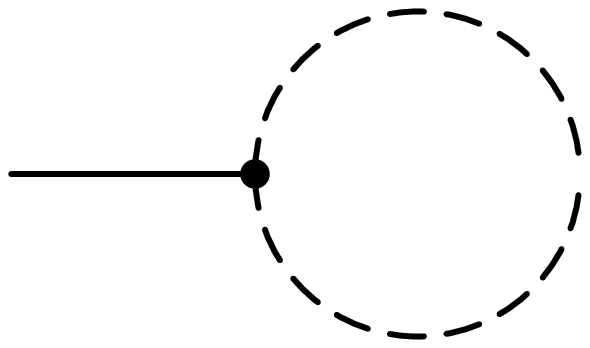} &
\raise0.45cm\hbox{\includegraphics[scale=0.35]{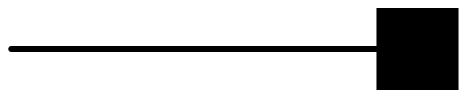}} \\[0.2cm]
(a) & (b) & (c)
\end{array}
\end{align*}
\caption{\label{fig:tadpole} $h$ tadpole graphs.  Graph $(c)$ includes counterterm and tadpole vertices.}
\end{figure}

The $\pi$ tadpole vanishes by $O(N-1)$ invariance. The $h$ tadpole graphs are shown in Fig.~\ref{fig:tadpole} and give the $h$ one-point function
%
%
%
\begin{align}
\Gamma^{(\mathsf{h})}_L &= \Gamma^{(h)}_N = 3 \lambda v m_h^2 + 0 + \left[-2 \lambda v^2 \Delta v -\lambda v^3(\delta_\phi - 2 \delta_v)\right]
\label{3.27}
\end{align}
where the three terms are the infinite contributions from the three diagrams.  The linear and non-linear parameterizations give the same result. Using the counterterms from Eq.~(\ref{3.22}), $m_h^2 = 2 \lambda v^2$, and requiring that $\Gamma^{(h)}$ vanishes gives
\begin{align}
\Delta v  &=  0 .
\label{3.28}
\end{align}
Note that the tadpole shift $\Delta v$ is finite in the non-gauged case, but it develops an infinite piece when gauge interactions are turned on.

The infinite contribution to the $h$ two-point function is
%
%
\begin{figure}
\begin{align*}
\renewcommand{\arraycolsep}{1cm}
\begin{array}{ccc}
\includegraphics[scale=0.35]{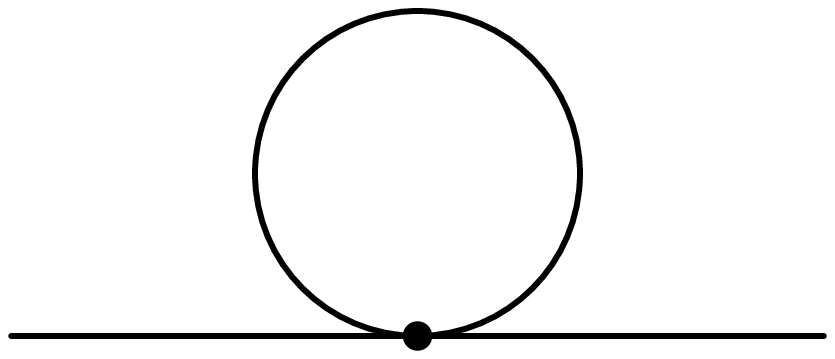} &
\includegraphics[scale=0.35]{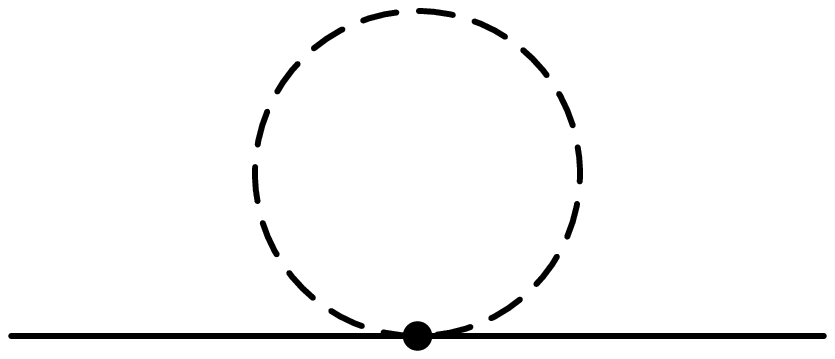} &
\lower0.45cm\hbox{\includegraphics[scale=0.35]{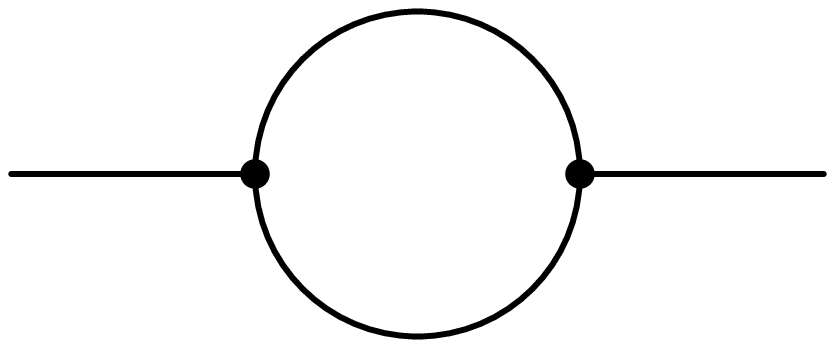}} \\[0.5cm]
(a) & (b) & (c) 
\end{array}
\end{align*}

\begin{align*}
\renewcommand{\arraycolsep}{1cm}
\begin{array}{cc}
\lower0.45cm\hbox{\includegraphics[scale=0.35]{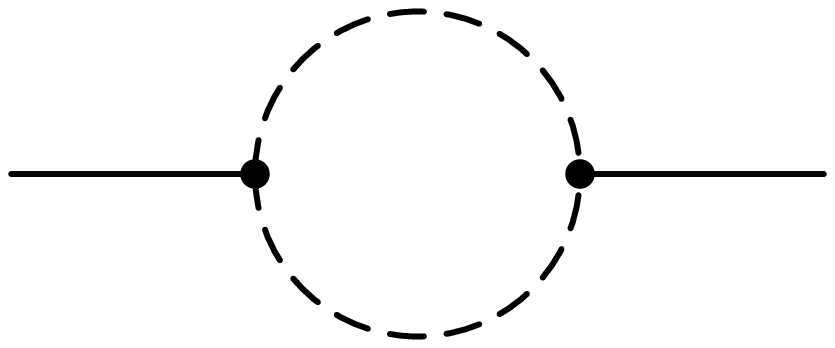}} &
\includegraphics[scale=0.35]{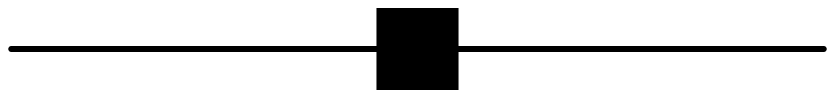} \\[0.5cm]
(d) & (e)
\end{array}
\end{align*}
\caption{\label{fig:h} $h$ propagator graphs.  Graph $(e)$ includes counterterm and tadpole vertices.}
\end{figure}
%
%
\begin{align}
\Gamma^{(\mathsf{hh})}_L &= 3 \lambda m_h^2 + 0 + 18 \lambda^2 v^2 + 2 \lambda^2 v^2 \n_G +\left[-6 \lambda v \Delta v - \lambda v^2  \left(2 \delta_\lambda +5 \delta_\phi -2 \delta_v \right) \right] =0 \nn
\Gamma^{(hh)}_N &= 3 \lambda m_h^2 + 0 + 18 \lambda^2 v^2 + \frac{p^4}{2v^2} \n_G +\left[-6 \lambda v \Delta v - \lambda v^2  \left(2 \delta_\lambda +5 \delta_\phi -2 \delta_v \right) \right] \nn
&=\frac{p^4}{2v^2} \n_G-2 \lambda^2 v^2 \n_G = \frac{1}{2v^2} \n_G \left(p^2-m_h^2\right)\left(p^2+m_h^2\right)
\label{3.29}
\end{align}
from the individual graphs in Fig.~\ref{fig:h}. The two forms of the Lagrangian give a different result. In the non-linear parameterization,  $\pi$ is derivatively coupled, so graph (d) is $\mathcal{O}(p^4)$; in the linear parameterization, the graph is $\mathcal{O}(p^0)$ since there is a $\mathsf{h} \bm{\varphi} \cdot \bm{\varphi}$ coupling in the potential.  In the non-linear parameterization, the one-loop corrected $h$ propagator in Fig.~\ref{fig:S}(b) is
\begin{align}
\frac{1}{p^2-m_h^2} \left[ 1 -  \frac{1}{16\pi^2\epsilon}\frac{1}{2v^2} \n_G\left(p^2+m_h^2\right) \right]
\label{hprop}
\end{align}
on expanding out the correction Eq.~(\ref{3.29}), and does not have a double pole in $(p^2-m_h^2)$ because $\Gamma^{(hh)}_N \propto \left(p^2-m_h^2\right)$. This feature is important for the cancellation of divergences.

The $\pi$ propagator graphs give the infinite contribution to the two-point function
%
%
\begin{figure}
\begin{align*}
\renewcommand{\arraycolsep}{0.5cm}
\begin{array}{cccc}
\includegraphics[scale=0.35]{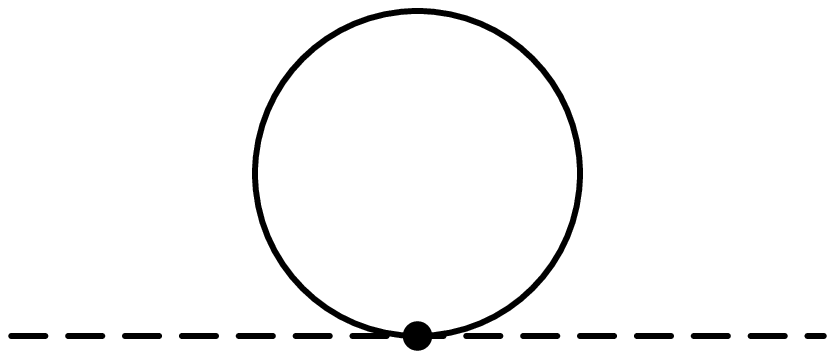} &
\includegraphics[scale=0.35]{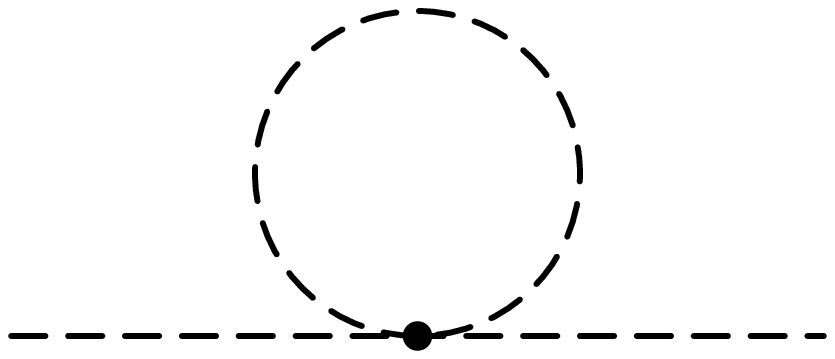} &
\lower0.45cm\hbox{\includegraphics[scale=0.35]{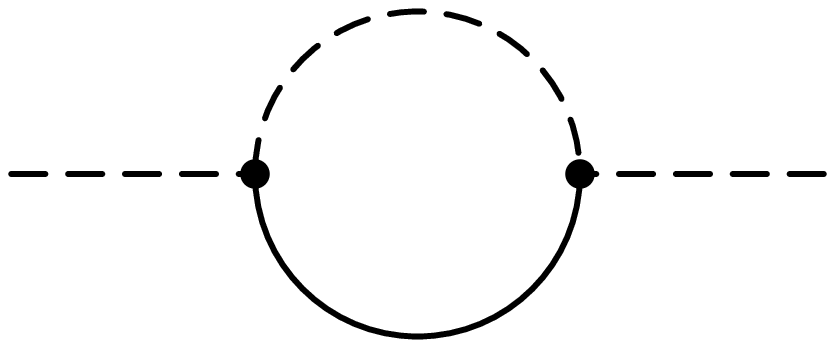}} &
\includegraphics[scale=0.35]{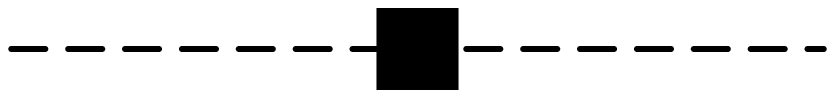} \\[0.5cm]
(a) & (b) & (c) & (d)
\end{array}
\end{align*}
\caption{\label{fig:pi} $\pi$ propagator graphs.  Graph $(d)$ includes counterterm and tadpole vertices.}
\end{figure}
%
%
\begin{align}
\Gamma^{(\varphi\varphi)}_L &= \lambda m_h^2 \delta_{ab} + 0 + 4 \lambda^2 v^2 \delta_{ab} +\left[-2 \lambda v \Delta v - \lambda v^2  \left(\delta_\phi -2 \delta_v \right) \right] \delta_{ab} =0,\nn
\Gamma^{(\pi\pi)}_N &= -\frac{1}{v^2}  m_h^2 p^2\delta_{ab} + 0 + \frac{1}{v^2}  (m_h^2+p^2) p^2 \delta_{ab} + \frac{2 \Delta v}{v} p^2 \delta_{ab} = \frac{1}{v^2}p^4 \delta_{ab},
\end{align}
where $a,b$ are $\pi$ flavor indices.  The loop contribution to the Goldstone boson mass is cancelled by the counterterm and tadpole vertices in the linear parameterization. In the non-linear parameterization, the one-loop correction to the Goldstone boson propagator is $\mathcal{O}(p^4)$, by the chiral counting rules.

\begin{figure}
\begin{align*}
\renewcommand{\arraycolsep}{0.5cm}
\begin{array}{cccc}
\includegraphics[scale=0.35]{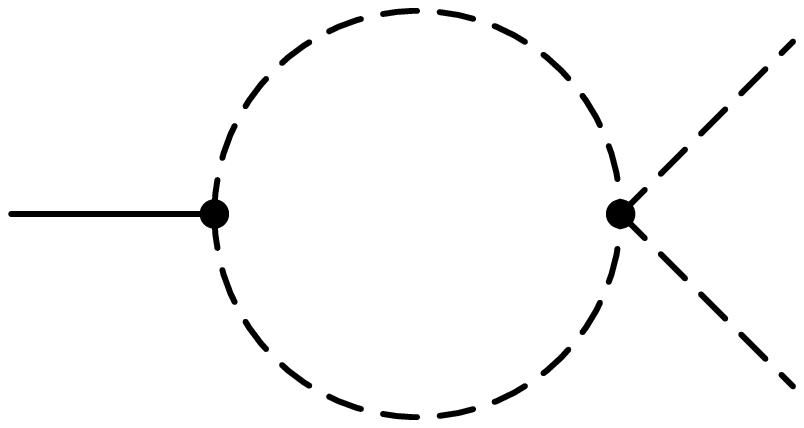} &
\includegraphics[scale=0.35]{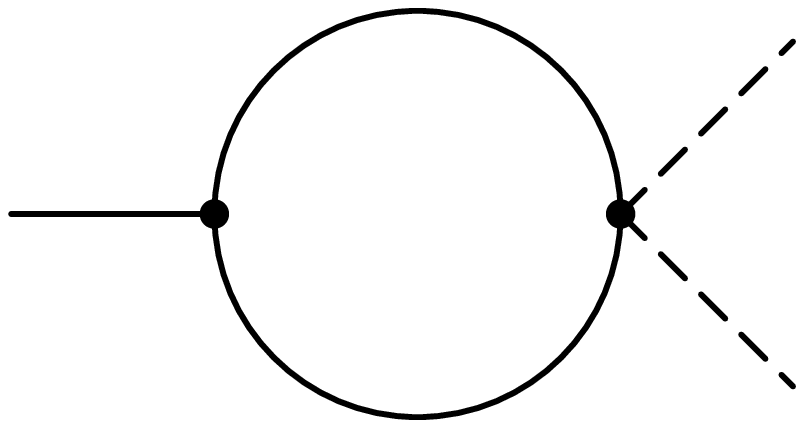} &
\lower0.5cm\hbox{\includegraphics[scale=0.35]{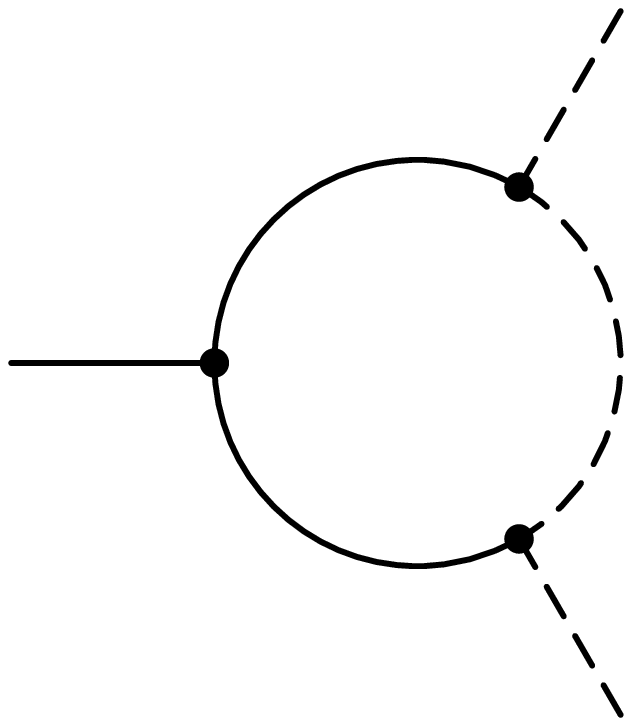}} &
\lower0.5cm\hbox{\includegraphics[scale=0.35]{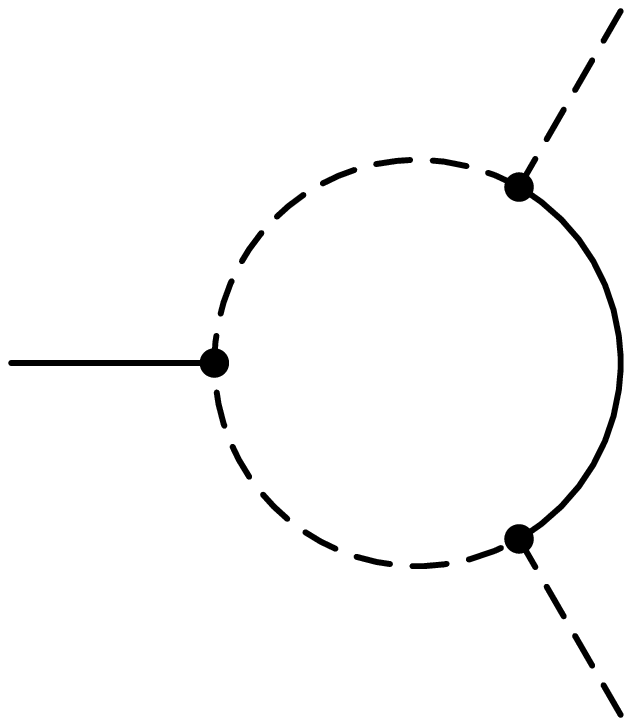}}\\[0.5cm]
(a) & (b) & (c) & (d)
\end{array}
\end{align*}

\begin{align*}
\renewcommand{\arraycolsep}{1cm}
\begin{array}{ccc}
\includegraphics[scale=0.35]{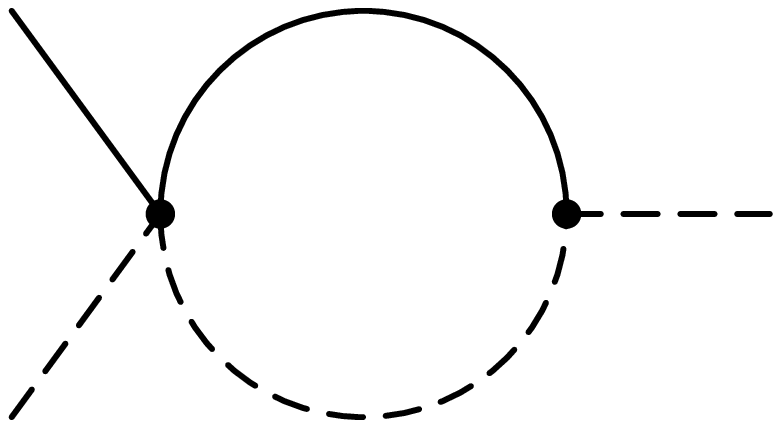} &
\includegraphics[scale=0.35]{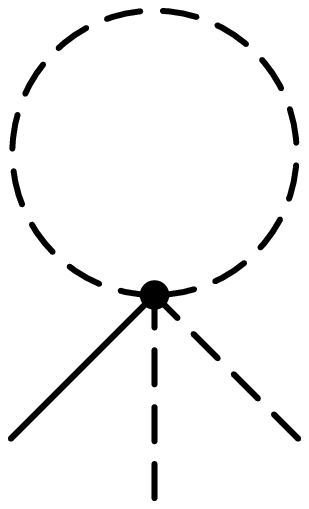} &
\includegraphics[scale=0.35]{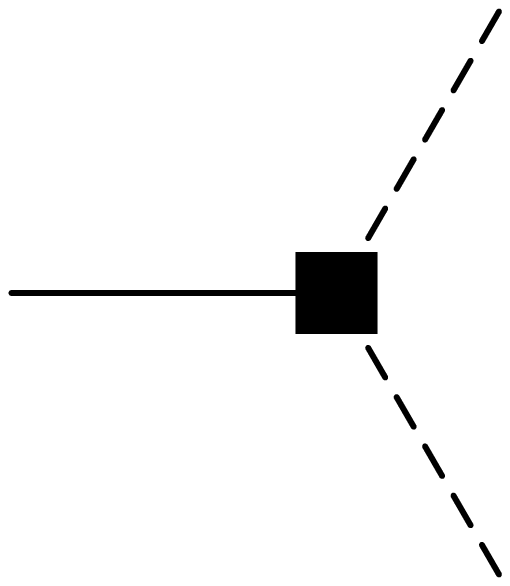} \\[0.5cm]
(e) & (f) & (g)
\end{array}
\end{align*}
\caption{\label{fig:hpp} $h\pi \pi$  graphs.  Graph $(g)$ includes counterterm and tadpole vertices.  Graph $(f)$ is not present for the linear case.}
\end{figure}

The $h \pi \pi$ vertex correction from the graphs in Fig.~\ref{fig:hpp} is
%
%
%
%
\begin{align}
\Gamma^{(\mathsf{h}\varphi\varphi)}_L &= 2 \lambda^2 v (\n_G+2) \delta_{ab} + 6 \lambda^2 v \delta_{ab} + 0 + 0 + 8\lambda^2 v \delta_{ab} +0+ \left[2 \lambda (-2 v \delta_\phi - v \delta_\lambda - \Delta v )\right]\delta_{ab} =0,\nn
\Gamma^{(h\pi\pi)}_N &= \left\{-\frac{1}{3v^3}(\n_G-1) (p_1+p_2)^2 \left[p_1^2+p_2^2+3 p_1 \cdot p_2\right]\right\}\delta_{ab}
 + 6\frac{\lambda}{v} (p_1 \cdot p_2)\delta_{ab} - \frac{6}{v}\lambda (p_1 \cdot p_2)\delta_{ab} \nn
&+\left\{ \frac{2}{v^3}\left( m_h^2 p_1 \cdot p_2-2 p_1^2 p_2^2 - (p_1 \cdot p_2)^2 -( p_1^2 + p_2^2) p_1 \cdot p_2\right) \right\} \delta_{ab}\nn
& +\left\{ -\frac{4}{v^3}\left[\frac12 m_h^2 p_1 \cdot p_2 +\frac1{4} p_1 \cdot p_2 (p_1^2 + p_2^2) \right] \right\}\delta_{ab}
 + 0  -\frac{2\Delta v}{v^2} (p_1 \cdot p_2) \delta_{ab} \nn
&=\frac{1}{3v^3}\biggl[-6 \n_G (p_1 \cdot p_2)^2-(5 \n_G+4)(p_1 \cdot p_2) (p_1^2+p_2^2) \nn
&-(\n_G-1)((p_1^2)^2+(p_2^2)^2)-(2\n_G+10) p_1^2p_2^2 \biggr]\delta_{ab},
\label{3.32}
\end{align}
where the pions have incoming momentum and flavor $p_1,a$ and $p_2,b$, respectively. Graph $(f)$ does not exist for the linear case.

One can already see non-trivial evidence for finiteness of the $S$-matrix for $h \to \pi \pi$ in Eq.~(\ref{3.32}). On-shell, only the first term in $\Gamma^{(h\pi\pi)}_N$ is non-zero, and is precisely cancelled by the $h$ propagator correction in Eq.~(\ref{hprop}).

\begin{figure}
\begin{align*}
\renewcommand{\arraycolsep}{0.5cm}
\begin{array}{ccccc}
\includegraphics[scale=0.35]{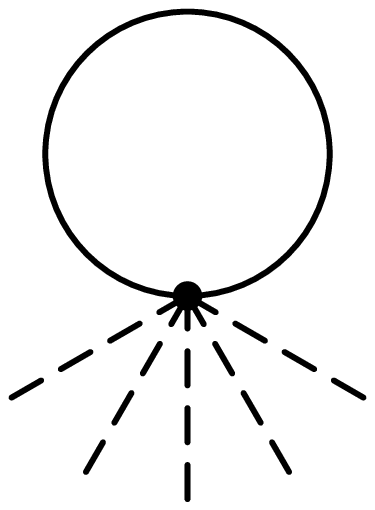} &
\includegraphics[scale=0.35]{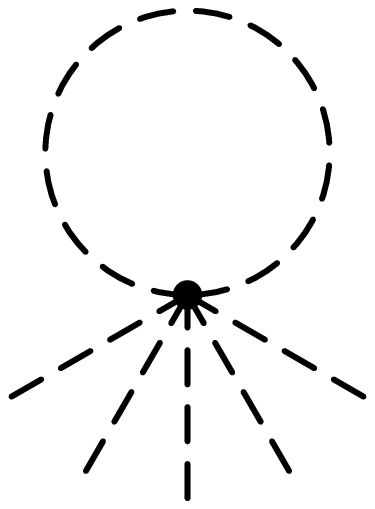} &
\includegraphics[scale=0.35]{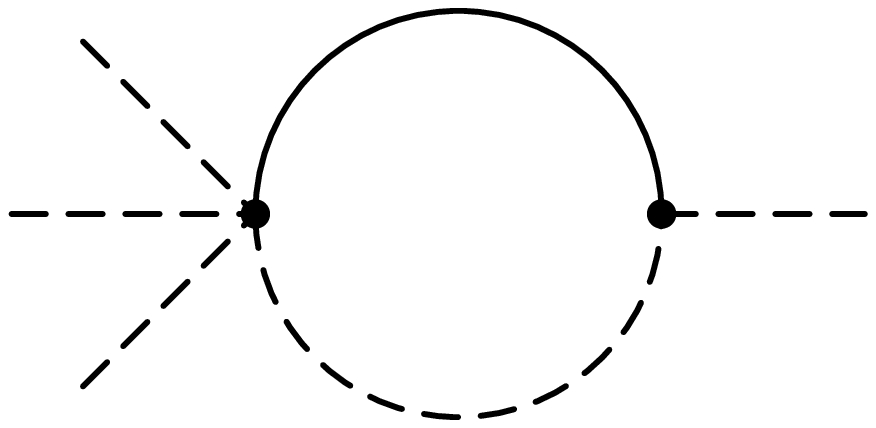} &
\includegraphics[scale=0.35]{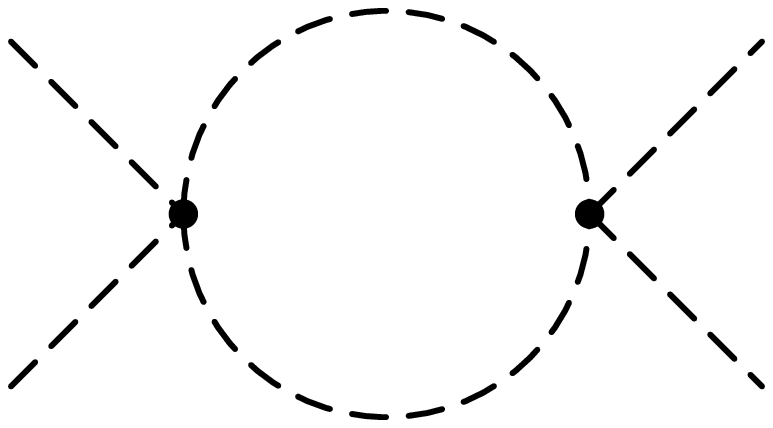} &
\includegraphics[scale=0.35]{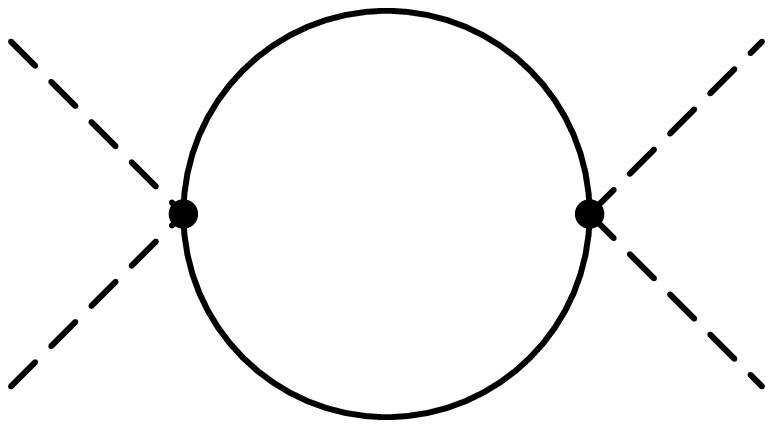} \\[0.5cm]
(a) & (b) & (c) & (d) & (e)
\end{array}
\end{align*}

\begin{align*}
\renewcommand{\arraycolsep}{0.75cm}
\begin{array}{cccccc}
\lower0.5cm\hbox{\includegraphics[scale=0.35]{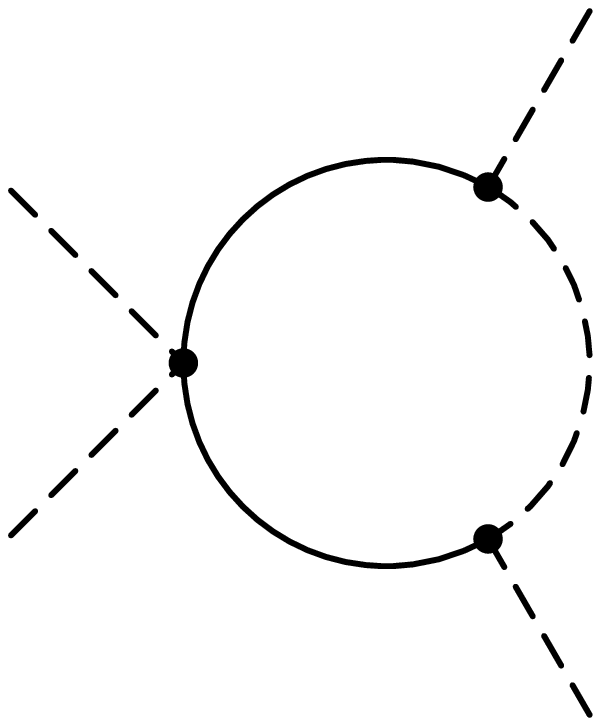}} &
\lower0.5cm\hbox{\includegraphics[scale=0.35]{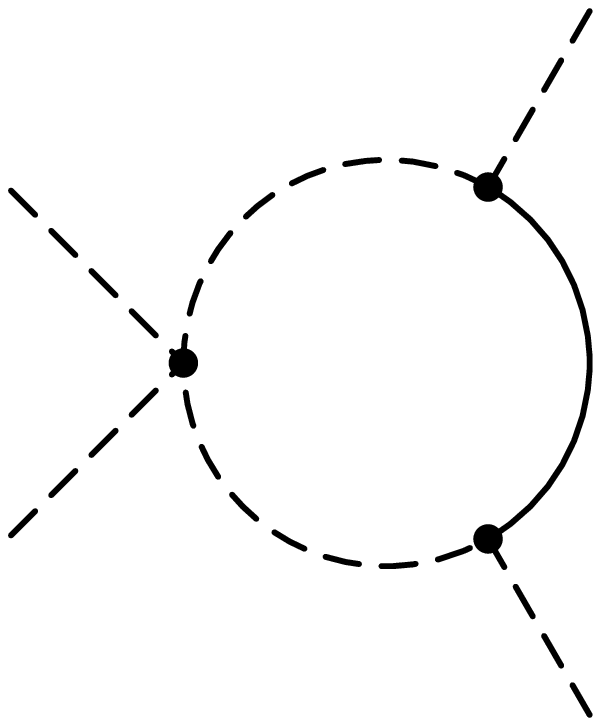}} &
\lower0.5cm\hbox{\includegraphics[scale=0.35]{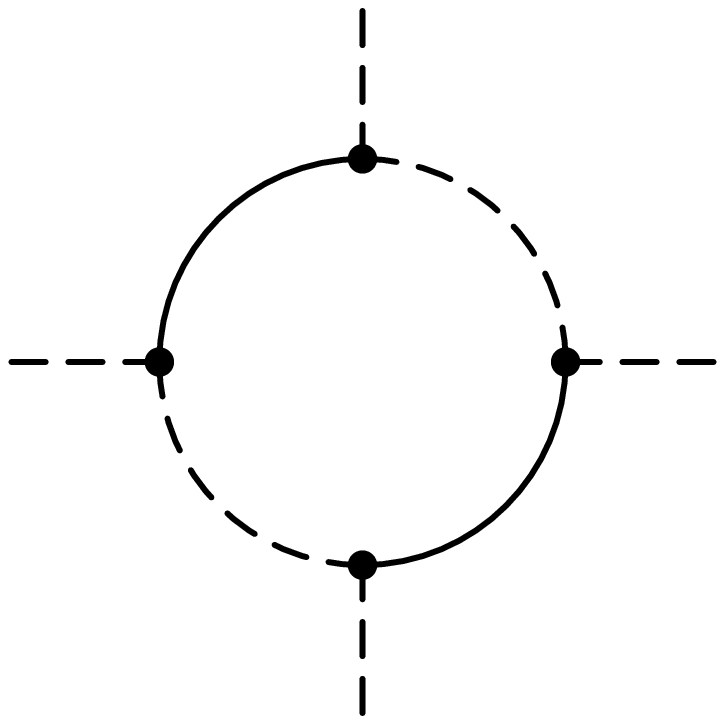}} &
\includegraphics[scale=0.35]{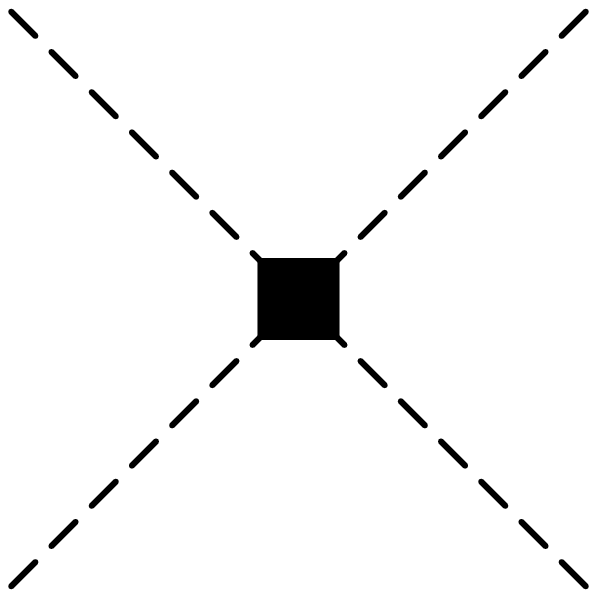} \\[0.5cm]
(f) & (g) & (h) & (i)
\end{array}
\end{align*}
\caption{\label{fig:4pi} $\pi^4$  graphs.  Graph $(i)$ includes counterterm and tadpole vertices.}
\end{figure}

The $\pi^4$ graphs are shown in Fig.~\ref{fig:4pi}. The pions have incoming momentum and flavor $p_i,a_i$, $i=1,2,3,4$.%
%
%
%
Let $\mathcal{I}$ denote the flavor structure
\begin{align}
\mathcal{I} &= \delta_{a_1a_2}\delta_{a_3a_4} 
+\delta_{a_1a_3}\delta_{a_2 a_4} + \delta_{a_1a_4}\delta_{a_2 a_3}.
\end{align}
Graphs $(a,b,c)$ do not exist for the linear case. Then, in the linear parameterization, the graphs give
\begin{align}
\Gamma^{(\varphi\varphi\varphi\varphi)}_L &= 0 + 0 + 0 + 2 \lambda^2 (\n_G+8) \mathcal{I}+ 2\lambda^2 \mathcal{I} +  0 + 0 + 0 +\left[-
2 \lambda (2\delta_\phi+\delta_\lambda)\right]\mathcal{I} =0,
\label{3.34}
\end{align}
using $N_\varphi = (N-1)$.
The non-linear parameterization has a much more complicated tensor structure. Since we are only computing the infinite parts of the amplitude, we can write them as the matrix elements of local operators, which provides a more compact form for the results. Using the operators
\begin{align}
O_1&=\left( \partial_\mu \bm{\pi} \cdot \partial^\mu \bm{\pi}  \right) \left( \partial_\nu \bm{\pi} \cdot \partial^\nu \bm{\pi}  \right) , & 
O_2&=\left( \partial_\mu \bm{\pi} \cdot \partial_\nu \bm{\pi}  \right) \left( \partial^\mu \bm{\pi} \cdot \partial^\nu \bm{\pi}  \right) ,\nn
O_3&=\left( \partial^2 \bm{\pi} \cdot  \bm{\pi}  \right)\left( \partial_\mu \bm{\pi} \cdot \partial^\mu \bm{\pi}  \right) , &
O_4&=\left( \partial^2 \bm{\pi} \cdot  \partial_\mu \bm{\pi}  \right)\left( \bm{\pi} \cdot \partial^\mu \bm{\pi}  \right), \nn
O_{5}&=\left( \partial^2 \bm{\pi} \cdot \bm{\pi}  \right)\left( \partial^2 \bm{\pi} \cdot  \bm{\pi}  \right), &
O_{6}&=\left( \partial^2 \bm{\pi} \cdot \partial^2  \bm{\pi}  \right)\left( \bm{\pi} \cdot  \bm{\pi}  \right), \nn
O_{7}&=-m_h^2\left(  \partial^2 \bm{\pi} \cdot   \bm{\pi}  \right)\left( \bm{\pi} \cdot \bm{\pi}  \right), &
O_{8}&=-m_h^2\left(  \partial_\mu \bm{\pi} \cdot    \partial^\mu\bm{\pi}  \right)\left( \bm{\pi} \cdot \bm{\pi}  \right),
\end{align}
the contribution from the loop graphs to $\Gamma^{(\pi\pi\pi\pi)}_N$ is
\begin{align}
\renewcommand{\arraystretch}{1.25}
\renewcommand{\arraycolsep}{0.2cm}
\begin{array}{c|cccccccc|c}
& (a) & (b) & (c) & (d) & (e) & (f) & (g) & (h) & \text{Total} \\
\hline
O_1 & 0 & 0 & 0 & 3 \n_G-7 & 3 & -6 & 8 & 2 & 3 \n_G \\
O_2 & 0 & 0 & 0 & 4 & 0 & 0 & -8 & 4 & 0 \\
O_3 & 0 & 0 &12 & 4 \n_G-8 & 0 & 0 & 0 & 0 & 4 \n_G+4\\
O_4 & 0 & 0 & -12 & 4 & 0 & 0 & 0 & 0 & -8 \\
O_{5} & 0 & 0 & 4 & \frac{4\n_G}{3}-2 & 0 & 0 & -2 & 0 & \frac{4}{3}\n_G \\
O_{6} & 0 & 0 & -4 & \frac23 & 0 & 0 & 2 & 0 &  -\frac43\\
O_{7} & -1 & 0 & 2 & 0 & 0 & 0 & -1 & 0 & 0\\
O_{8} & -3 & 0 & 6 & 0 & 0 & 0 & -3 & 0 & 0
\end{array}
\label{table}
\end{align}
times $1/(12 v^4)$. Graph $(i)$ is proportional to $\Delta v$ and vanishes by Eq.~(\ref{3.28}).

We can now study the finiteness of the $S$-matrix. In the linear case, all the irreducible vertices are finite, and so is the $S$-matrix. The counterterms were chosen to render the irreducible vertices finite in the unbroken sector. Symmetry breaking does not affect the short distance behavior of the theory, and the same counterterms of the unbroken theory also make the broken theory finite.

More interesting is the divergence structure of the $S$-matrix in the non-linear parameterization. The infinite part of the total amplitude $iA$ can be written as the matrix element of local operators,
\begin{align}
A &= \frac{\n_G}{3v^4}O_3 - \frac{4}{3v^4}O_4 + \frac{2\n_G-3}{18v^4}O_{5} + \frac{1}{18v^4}O_{6}\,,
\label{3.37}
\end{align}
which vanishes on-shell since $\partial^2 \bm{\pi} = 0$, so the $\pi\pi$ scattering $S$-matrix is finite at one-loop, as claimed. Some interesting cancellations are necessary for the on-shell $S$-matrix to be finite. The operators $O_{1,2}$ do not vanish on-shell. Eq.~(\ref{table}) shows that there is a non-zero contribution to
$O_1$ in $\Gamma^{(\pi\pi\pi\pi)}_N$, but not to $O_2$. The $O_1$ contribution is cancelled by the Higgs propagator correction Eq.~(\ref{hprop}) in Fig.~\ref{fig:S}(b). The Fig.~\ref{fig:S}(b) amplitude is proportional to the square of the tree-level $h \pi \pi$ vertex, and only produces the tensor structure $O_1$. A bit more algebra shows that the one-loop $S$-matrix is finite to all orders in $p$. 
\begin{figure}
\begin{center}
\renewcommand{\arraycolsep}{1cm}
\begin{align*}
\begin{array}{cc}
\includegraphics[scale=0.4]{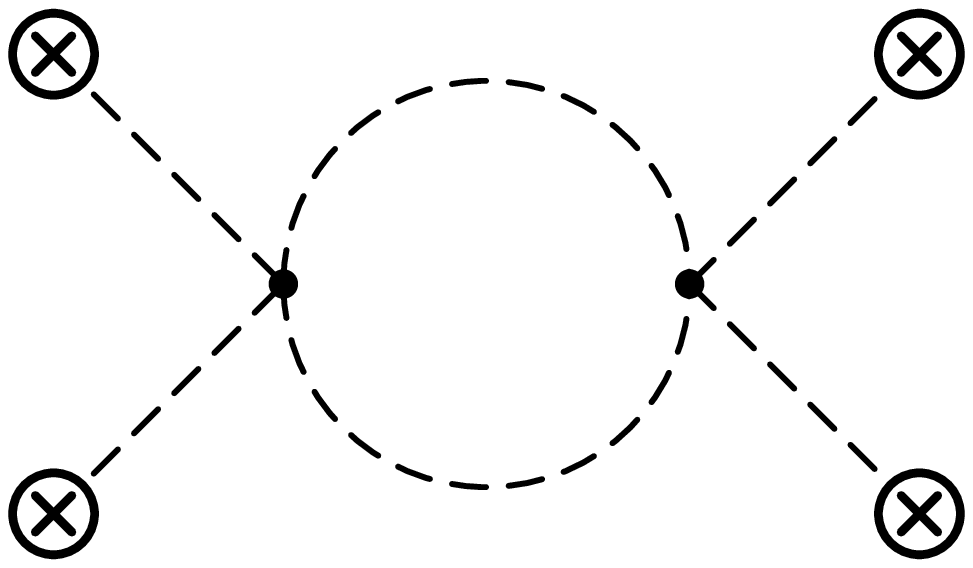} &
\includegraphics[scale=0.4]{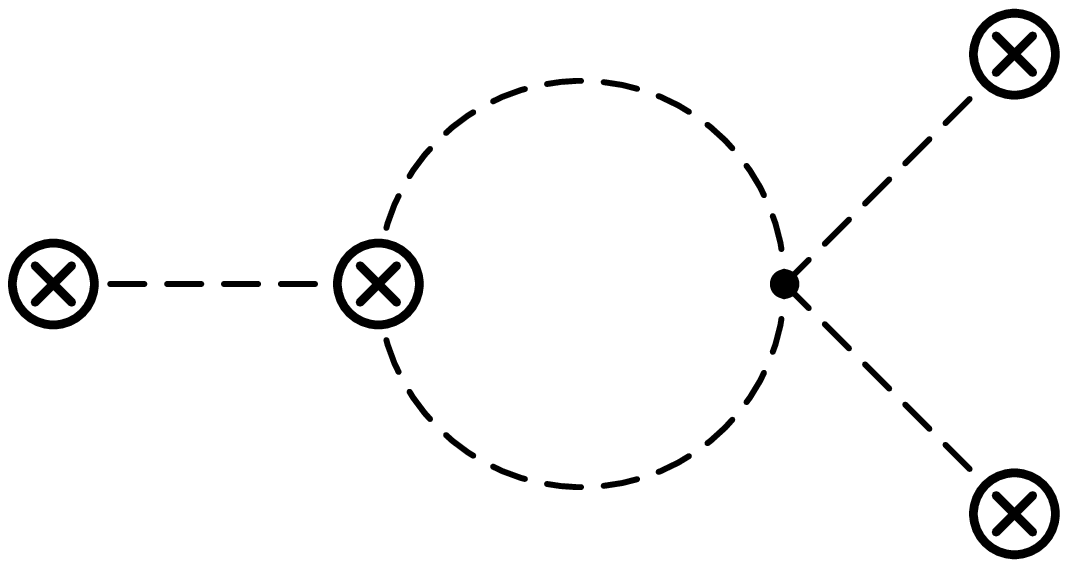} \\
(a) & (b)
\end{array}
\end{align*}
\end{center}
\caption{\label{fig:11}}
\end{figure}

The non-covariant terms in the scattering amplitude that vanish on-shell arise from field redefinitions in the functional integral.  A coordinate transformation is equivalent to a redefinition of the source, i.e.\ to a field redefinition in the generators of 1PI graphs $\Gamma(\Phi)$. Start with an $O(N)$ covariant source term
\begin{align}
L_J &= \bm{J \cdot \phi} =(v+h) \left( \bm{J} \cdot \xi \bm{\chi}_0 \right),
\label{3.38}
\end{align}
where
\begin{align}
\xi \bm{\chi}_0 &= \left[ \begin{array}{c} \frac{ \sin \abs{\overline \pi} } {\abs{\overline \pi}} \frac{\bm{\pi^a}}{v}  
\\ \cos \abs{\overline \pi} \end{array} \right] , \qquad \abs{\overline \pi}^2 \equiv \frac{\pi^a \pi^a}{v^2} .
\end{align}
$W(J)$ with source term $\bm{J \cdot \phi}$ is covariant, and $n$-point Green's functions are given by expanding $W(J)$ in powers of $J$. This expansion is still covariant. On the other hand, the Feynman graph computation we have done is equivalent to computing with a source $\bm{J \cdot \pi}$ and then Legendre transforming in $\bm{\pi}$. This procedure is not covariant, as explained in detail in the next section. The covariant version uses the coupling
\begin{align}
\left(1+\frac{h}{v}\right)\left[\bm{J \cdot \pi} - \frac1{6v^2} (\bm{J \cdot \pi})( \bm{\pi \cdot \pi})+ \ldots \right] .
\label{Jsource}
\end{align}
The fourth-order contribution to $W(J)$ has contributions from the graphs shown in Fig.~\ref{fig:11}, where graph (b) involves the $J\pi^3$ term in Eq.~(\ref{Jsource}). Both graphs are the same order in $p$, since graph (a) has an extra $p^2$ from the $\pi^4$ interaction, and an extra $1/p^2$ from the extra propagator relative to graph (b). The two graphs add to give a covariant contribution if one uses Eq.~(\ref{3.38}) for the source term.

Note the following remarks.
\begin{itemize}

\item Green's functions are finite in the linear parameterization, but not in the non-linear parameterization. The 4-point function $\Gamma^{(\pi\pi\pi\pi)}_N$ has the divergence Eq.~(\ref{3.37}) which is not cancelled by any counterterm.

\item The divergence Eq.~(\ref{3.37}) is not chirally invariant, i.e.\ it cannot be written in a ${\cal G}=O(N)$ invariant way.  Although it explicitly breaks the $O(N)$ symmetry, the breaking is unphysical since it does not enter measurable quantities such as on-shell $S$-matrix elements.

\item It is not necessary to add a counterterm to cancel Eq.~(\ref{3.37}). One can ignore it, or make a field redefinition (which does not change the $S$-matrix) to remove it. It is possible to remove it by making a field redefinition because the operator vanishes on-shell.

\item In the usual computation of hadronic weak decays, one replaces penguin operators $\left( \bar \psi T^A \gamma^\nu P_L \psi \right) D^\mu G^A_{\mu \nu}$ by four-fermion operators  using the QCD equations of motion.  In this case, penguin graphs are infinite, since there is no counterterm to cancel them, but the weak decay $S$-matrix is finite. This situation is analogous to the situation we are finding in the $O(N)$ model. In both cases, the field redefinitions that eliminate the equation of motion terms involve divergent $1/\epsilon$ terms.

\item In the $O(N)$ non-linear sigma model, i.e.\ the $O(N)$ theory with the $h$ field set to zero, Appelquist and Bernard~\cite{Appelquist:1980vg,Appelquist:1980ae} showed that there were $O(N)$ non-invariant counterterms which vanished on-shell. The reason for these terms is explained in the next section. In the Appelquist-Bernard calculation, there were, in addition, $\mathcal{O}(p^4)$ divergences proportional to $O_1$ and $O_2$ at one-loop, as expected in chiral perturbation theory.
In our calculation, these higher order in $p$ divergences do not occur because of extra graphs involving $h$ which cancel the divergences.

\item A similar situation occurs in renormalization of HEFT. Ref.~\cite{Gavela:2014uta} found non-covariant terms
\begin{align*}
\frac{1}{32\pi^2\epsilon}\!\left\{\!\left(\frac32+10\eta +18\eta^2\right)\!\frac{\left({\bm\varphi}\Box {\bm\varphi}\right)^2}{v^4}-c_1\!\left(3+10\eta\right)\!\frac{{\bm\varphi}\Box {\bm\varphi}\Box h}{v^3} \right\},
\end{align*}
which vanish on-shell. Ref.~\cite{Alonso:2015fsp} showed that these terms arose from the use of non-covariant perturbation theory due to the non-covariant term in Eq.~(\ref{4.6}).

\end{itemize}

Finally, we return to an important subtlety in the gauged case. As mentioned earlier, the gauged $O(N)$ sigma model has different counterterms in the unbroken and broken phases, even though symmetry breaking is an infrared effect and does not change the short distance structure of the theory. The reason that the counterterms differ is that the $O(N)$ theory in the unbroken phase is quantized using the gauge fixing term
\begin{align}
L_{\text{gf}} &=  -\frac{1}{2\xi} \left(\partial^\mu A_\mu^a\right)^2
\end{align}
whereas the broken theory is quantized using the gauge fixing term
\begin{align}
L_{\text{gf}} &= 
 -\frac{1}{2 \xi}
 \left[ \partial^\mu  {A}^{a}_\mu + i \xi^\prime g v\left(\bm{\chi}_0^{T} T^a \bm{\phi}-  \bm{\phi}^{T}.
 T^a \bm{\chi}_0\right)\right]^2 \,.
\end{align}
Gauge theory counterterms do depend on the gauge fixing term, so the two theories can have different counterterms if they are quantized using different gauge fixing terms. If one uses the \emph{same} gauge fixing term for both theories, then the renormalization counterterms are the same.

The renormalizability of the $O(N)$ theory in non-linear coordinates will hold to arbitrary loop order, as is obvious from the general arguments given earlier.

\section{Covariant Formalism for Curved Scalar Field Space}\label{sec:curved}

In this section, we review the well-known geometric formulation of non-linear sigma models~\cite{Gaillard:1985uh,Gasser:1983yg,Honerkamp:1996va,Honerkamp:1971sh,Jackiw:1974cv}. The use of functional methods for quantum corrections, combined with a covariant formalism sheds light on a number of technical issues identified in Refs~\cite{Appelquist:1980ae,Gavela:2014uta}. This covariant formalism has wide applicability --- the CCWZ phenomenological Lagrangian is a special case of the geometric approach in a particular choice of coordinates, as discussed in Sec.~\ref{sec:CCWZ}.

\subsection{Scalar Fields on a Curved Manifold ${\cal M}$}\label{sec:DE}

Consider $N$ real scalar fields $\phi^{i}$ which are the coordinates of a curved scalar manifold $\mathcal{M}$.  The scalar action for the $O(p^2)$ Lagrangian (with no gauge fields) containing all operators with up to two derivatives is
\begin{align}
S=\int d^4x \ \mathcal L\left[\phi(x) \right] =\int d^4x \left( \frac12 \, g_{ij}\left(\phi \right) \, \left( \partial_\mu\phi \right)^{i} \left(\partial^\mu \phi \right)^{j}
+{\cal I}\left(\phi \right)\right) \,,
\label{p2Lag}
\end{align}
where ${\cal I}(\phi)$ is an invariant scalar density on $\m$.  The two-derivative terms define the scalar metric $g_{ij}(\phi)$ of ${\cal M}$.  Under a scalar field redefinition or change of scalar coordinates $\phi^\prime\left(\phi \right)$, 
the derivative $\left( \partial_\mu \phi^{i} \right)$ transforms as a contravariant vector
\begin{align}
\partial_\mu \phi^{\prime \, i}&=\left( \frac{\partial \phi^{\prime \, i}}{\partial \phi^j}\right) \partial_\mu\phi^j\,,
\end{align}
and the metric $g_{ij}\left(\phi \right)$ transforms as a tensor with two lower indices,
\begin{align}
g_{ij}^\prime &= \left( \frac{\partial \phi^k}{\partial \phi^{\prime\, i}}\right)  \left( \frac{\partial \phi^l}{\partial \phi^{\prime\, j}}\right) g_{kl}\, .
\end{align}
Thus, the Lagrangian also is an invariant scalar density.  The potential ${\cal I}(\phi)$ is non-zero, in general.  It is a constant if all the fields $\phi^i$ are exact Goldstone bosons of an enlarged global symmetry.

The first variation of the action yields the equation of motion for the field $\phi$.  Under an infinitesimal variation $\phi \to \phi + \eta$, the linear in $\eta$ variation of the action is
\begin{align}
\delta S &= \int d^4x \ \left( - g_{ij}  \left(\mathcal{D}_\mu (\partial^\mu \phi)\right)^i +  {\cal I}_{,\,j}\right) \eta^j\, ,
\label{4.3}
\end{align}
where
\begin{eqnarray}
\left(\mathcal{D}_\mu \eta \right)^i \equiv \partial_\mu \eta^i +  \Gamma^i_{kj}\,\left( \partial_\mu \phi \right)^k\, \eta^j
\label{4.4}
\end{eqnarray}
is the covariant derivative on a vector field $\eta^i$ and $\Gamma^i_{jk}(\phi)$ is the Christoffel symbol.  From Eq.~(\ref{4.3}), one obtains the classical equation of motion
\begin{align}
E_j  &= g_{ij}  \left( \mathcal{D}_\mu \left( \partial^\mu \phi \right)\right)^i - {\cal I}_{,\,j} = g_{ij} \left( \partial^2 \phi^i + \Gamma^i_{kj}  \left( \partial_\mu \phi \right)^k  \left( \partial_\mu \phi \right)^j \right) - {\cal I}_{,\,j} = 0\,,
\label{EOMphi}
\end{align}
which is the wave equation for $\phi$ on the curved manifold ${\cal M}$. 

The second variation of the action under an infinitesimal variation $\phi \to \phi + \eta$ is
\begin{align}
\delta^2 S &=  \frac12 \int d^4x \biggl[ g_{ij} \left( \mathcal{D}_\mu \eta \right)^i \left(\mathcal{D}_\mu \eta \right)^j
-R_{ijkl} \, \eta^i (\partial_\mu \phi)^j \eta^k (\partial_\mu \phi)^l
- E_j \, \Gamma^j_{kl} \eta^k \eta^l + \mathcal{I}_{;\, ij}\, \eta^i \eta^j \biggr]\,,
\label{4.6}
\end{align}
where $R_{ijkl}$ is the  Riemann curvature tensor and
\begin{eqnarray}
\mathcal{I}_{;\, ij} &=& \nabla_i \nabla_j \mathcal{I} = \frac{\partial^2 \mathcal{I}}{\partial \phi^i \partial \phi^j} - \Gamma_{ij}^k \frac{\partial \mathcal{I}}{\partial \phi^k}.
\end{eqnarray}
Eq.~(\ref{4.6}) is not covariant because of the third term which depends explicitly on the connection $\Gamma^i_{jk}$.  This term, however, vanishes on shell since it is proportional to the equation of motion $E_i$. The non-covariant term leads to non-covariant divergences in Green's functions which vanish in $S$-matrix elements.  Non-covariant terms arose in the explicit calculation in Sec.~\ref{sec:O(N)model} of the $O(N)$ model in flat space. Even though they have no physical consequences, the appearance of non-covariant terms is puzzling since the original theory is covariant. The non-covariant terms occur because the infinitesimal variation $\phi \to \phi + \eta$ is not a covariant parameterization of fluctuations to second order in $\eta$, as was explained in Ref.~\cite{Honerkamp:1996va,Honerkamp:1971sh}.

The variation of the scalar field $\eta^i = \delta \phi^i$ should transform as a vector under a change of coordinates. However, under a change of coordinates,
\begin{align}
\phi^{\prime\, i}\left( \phi+\eta \right) & = \phi^{\prime\, i}\left( \phi\right) + \left( \frac{\partial \phi^{\prime\, i}}{\partial \phi^j}\right) \eta^j + \frac 12 \left(\frac{\partial^2 \phi^{\prime\, i}}{\partial \phi^j \partial \phi^k }\right) \eta^j \eta^k + \ldots \equiv \phi^{\prime\, i}\left(\phi\right) + \eta^{\prime\, i} ,
\label{4.7}
\end{align}
implies that
\begin{align}
\eta^{\prime\, i} &= \left(\frac{\partial \phi^{\prime\, i}}{\partial \phi^j}\right) \eta^j + \frac 12 \left( \frac{\partial^2 \phi^{\prime\, i}}{\partial \phi^j
\partial \phi^k }\right) \eta^j \eta^k + \ldots ,
\label{4.13a}
\end{align}
which is the correct transformation law for a vector at first order in $\eta$, but not at second order.  The solution to this problem is to use geodesic coordinates to parametrize fluctuations in $\phi$, as shown in Ref.~\cite{Honerkamp:1971sh}.  The equation for a  geodesic on $\m$ parameterized by $\lambda$ is
\begin{align}
\frac{\rd^2 \phi^i}{\rd \lambda^2} + \Gamma^i_{jk}(\phi) \frac{\rd \phi^j}{\rd \lambda}\frac{\rd \phi^k}{\rd \lambda} &=0\,.
\end{align}
Solving this equation in perturbation theory, starting at $\phi^i=\phi_0^i$ with tangent vector $\eta^i$ gives
\begin{align}
\phi^i &= \phi^i_0 + \lambda \eta^i - \frac12 \lambda^2\, \Gamma^i_{jk}(\phi_0)\, \eta^j \eta^k + \ldots
\label{4.13z}
\end{align}
Fluctuations in $\phi$ are parameterized by picking $\eta^i$ to be tangent vector such that the geodesic reaches $\phi+\delta \phi$ at $\lambda=1$, i.e.\ using the variation 
\begin{align}
\phi^i \to \phi^i + \eta^i - \frac12 \Gamma^i_{jk} \eta^j \eta^k + \mathcal{O}(\eta^3),
\label{4.13}
\end{align}
which
suffices to restore the correct transformation law for the vector $\eta^\prime$ to second order in the expansion,
\begin{align}
\eta^{\prime\, i} &= \left( \frac{\partial \phi^{\prime\, i}}{\partial \phi^j}\right) \eta^j \, .
\label{4.12}
\end{align}

Expanding the action in the geodesic fluctuation $\eta$ to quadratic order in $\eta$ yields
\begin{align}
S[\phi + \eta]= S[\phi]+ \frac{\delta S}{\delta \phi^i} \left(\eta^i-\frac{1}{2}\Gamma^i_{jk}\eta^j\eta^k \right)+\frac{\delta^2S}{\delta\phi^i\delta\phi^j}\eta^i\eta^j+\mathcal{O}(\eta^3)\,,
\end{align}
which shows that there is a quadratic in $\eta$ term proportional to the equation of motion operator $E_i \equiv \left(\delta S/\delta \phi^i\right)$.  This contribution exactly cancels the non-covariant term of Eq.~(\ref{4.6}), yielding a second variation of the action which transforms covariantly
\begin{align}
\delta^2 S &= \frac 12 \int d^4x \biggl[ g_{ij}   \left(\mathcal{D}_\mu \eta\right)^i \left(\mathcal{D}^\mu \eta \right)^j - R_{ijkl} \ \eta^i (\partial_\mu \phi)^j \eta^k (\partial^\mu \phi)^l  + \left(\nabla_i \nabla_j \mathcal{I} \right)\, \eta^i \eta^j \biggr]\,.
\label{4.14}
\end{align}
An equivalent way to implement the covariant expansion is to promote ordinary functional derivatives to covariant functional derivatives~\cite{Gaillard:1985uh},
\begin{align}
\nabla_i S &=  \frac{\delta S}{\delta \phi^i} \,,  &
\nabla_i \nabla_j S &=  \frac{\delta^2 S}{\delta \phi^i \delta \phi^j} - \Gamma_{ij}^k \frac{\delta S}{\delta \phi^k}\,.
\end{align}

The second variation of the action enters the one-loop correction to the functional integral,
\begin{align}
\Gamma_{\text{one-loop}} &= \frac{i}{2} \log \det \left( - g^{ik} \frac{\delta^2 S}{\delta \eta^k \delta \eta^j} \right)\,.
\label{4.11}
\end{align}
The one-loop corrections computed using Eq.~(\ref{4.14}) are covariant, since $\delta^2 S$ is covariant. The two forms for $\delta^2 S$, Eq.~(\ref{4.6}) and Eq.~(\ref{4.14}), differ in the form for $\phi^\prime$, Eq.~(\ref{4.13a}) and Eq.~(\ref{4.13}), i.e.\ by a field redefinition as discussed in Sec.~\ref{sec:pipi}. Thus the two formulations have the same $S$-matrix, but different Green's functions. The covariant form Eq.~(\ref{4.14}) has covariant Green's functions and $S$-matrix elements, so the non-covariant version Eq.~(\ref{4.6}) has covariant $S$-matrix elements (since they are not changed by field redefinitions) but non-covariant Green's functions.

The one-loop radiative correction can be computed from Eq.~(\ref{4.11}). For renormalization of the theory at one-loop in dimensional reqularization, we only require the divergent one-loop contribution to the Lagrangian.  This contribution can be extracted using the covariant derivative formalism in Refs.~\cite{Chan:1985ny,Chan:1986jq,Gaillard:1985uh,Drozd:2015rsp,Henning:2014wua}, which gives the same result as an earlier explicit computation by 't~Hooft~\cite{tHooft:1973us}. The results are given in Eq.~(\ref{4.33a}), after we have discussed the gauged version of Eq.~(\ref{4.11}). Since $\delta^2 S$ is covariant, the radiative corrections are also covariant when computed this way.

\subsection{Global Symmetry on ${\cal M}$}\label{sec:global}

We now consider the global symmetries of the ungauged action Eq.~(\ref{p2Lag}). The global symmetries of the scalar kinetic energy term are the isometries of $\mathcal{M}$.  These isometries are specified by a set of vector fields $t^i_\alpha$, the Killing vectors of ${\cal M}$, where the different isometries are labelled by $\alpha$.  The Killing vectors generate the infinitesimal field transformations
\begin{align}
\delta_\theta \phi^i &= \theta^\alpha \, t^i_\alpha\left( \phi \right) ,
\label{4.16}
\end{align}
where $\theta^\alpha$ are infinitesimal parameters. The gradient of $\phi$ transforms as
\begin{align}
\delta_\theta \left( \partial_\mu \phi^i \right) &= \theta^\alpha \left( \frac{\partial t^i_\alpha}{\partial \phi^j}\right) \left( \partial_\mu \phi^j \right)\,.
\label{4.26}
\end{align}

For the $O(N)$ sigma model, the global symmetries of the scalar kinetic energy term are ${\cal G} = O(N)$ transformations on the $N$-component real scalar field $\bm{\phi}$. The $N(N-1)/2$ Killing vectors of ${\cal M}$ are
\begin{align}
t^i_{ab}\left(\phi \right)=i \, {\left[ M_{ab}\right]^i}_j \, \phi^j=i \, \left( M_{ab} \, \phi \right)^i\,\,, 
\label{ONEx}
\end{align} 
where $M_{ab}$, $1 \le a < b \le N$, are the $N\times N$ anti-symmetric Hermitian matrices of Eq.~(\ref{3.4}), and the label $\alpha$ has been replaced by the bi-index $ab$. The Killing vectors in Eq.~(\ref{ONEx}) are linear in the  $N$ Cartesian components of the field $\bm{\phi}$, but not in the $N$ polar components. The $O(N)$ Killing vectors can be divided into the $(N-1)(N-2)/2$ Killing vectors of the unbroken subgroup ${\cal H}= O(N-1)$ and the $(N-1)$ Killing vectors which are spontaneously broken,
\begin{eqnarray}
t^i_{ab}\left( \phi \right) &=& i \left( M_{ab} \, \phi\right)^i = \left( {\delta^i}_a \phi_b - {\delta^i}_b \phi_a \right), \qquad 1 \le a<b < N\,, \nn
t^i_{aN}\left( \phi \right) &=& i \left( M_{aN} \, \phi\right)^i = \left( {\delta^i}_a \phi_N - {\delta^i}_N \phi_a \right) ,\qquad 1 \le a < N\, .
\label{SNKillV}
\end{eqnarray}
Restricting to the scalar submanifold $S^{N-1}$ such that $\langle \bm{\phi} \cdot \bm{\phi} \rangle = v^2$ with $h=0$, yields $N_\varphi= (N-1)$ independent real scalar fields $\varphi^a$.  The Killing vectors of $S^{N-1}$ on the first line of Eq.~(\ref{SNKillV}) act linearly on the $\varphi^a$ in both Cartesian and polar coordinates.  Those on the second line act non-linearly, since $\phi^N = \sqrt{v^2 - \bm{\varphi \cdot \varphi}}$. Explicitly,
\begin{eqnarray}
t^i_{ab}\left( \varphi \right) &=& i \left( M_{ab} \, \varphi\right)^i = \left( {\delta^i}_a \varphi_b - {\delta^i}_b \varphi_a \right), \qquad 1 \le a<b < N\,, \nn
t^i_{aN}\left( \varphi \right) &=& i \left( M_{aN} \, \phi\right)^i =  {\delta^i}_a \sqrt{v^2 - \bm{\varphi \cdot \varphi}} ,\qquad 1 \le a < N\, ,
\label{4.23}
\end{eqnarray}
for $i=1,\ldots,N-1$.

The infinitesimal field transformations generated by the Killing vectors in Eq.~(\ref{4.16}) leave the action Eq.~(\ref{p2Lag}) invariant, provided that
\begin{align}
\Lie_{t_\alpha} g &=0\,,  & \Lie_{t_\alpha} \mathcal I &=0 \,,
\label{4.17}
\end{align}
where $\Lie_{t_\alpha}$ is the Lie derivative for Killing vector $t^i_\alpha$.  The first condition in Eq.~(\ref{4.17}) is the definition of a Killing vector; it is an isometry of the metric. The second condition is that the potential is invariant.

The Lie bracket $[t_\alpha,t_\beta]$ of two isometries is also an isometry since
\begin{align}
\left[\Lie_{t_\alpha},\Lie_{t_\beta} \right] &= \Lie_{\left[t_\alpha, t_\beta\right]},
\end{align}
so the Killing vectors form the symmetry algebra
\begin{align}
\left[t_\alpha, t_\beta\right]^i &= \fcs[\alpha\beta]{\gamma}\, t_\gamma^i\,.
\end{align}
Evaluating the Lie bracket gives
\begin{align}
\left[t_\alpha, t_\beta\right]^i =t^k_\alpha t^i_{\beta,\,k}-t^k_\beta t^i_{\alpha,\,k}= \fcs[\alpha\beta]{\gamma}t_\gamma^i\,.    \label{LieBrkt}
\end{align}
Note that the above equation also holds with the ordinary derivative $t^i_{\alpha,k}$ replaced by the  covariant derivative
\begin{eqnarray}
t^i_{\alpha \, ;k} &=&t^i_{\alpha \, ,j} + \Gamma^i_{kj} \, t^j_\alpha =  \frac{\partial t^i_\alpha}{\partial \phi^k} + \Gamma^i_{kj} \, t^j_\alpha ,
\end{eqnarray}
since the Christoffel symbol is symmetric in lower indices, and cancels in the antisymmetric derivative of Eq.~(\ref{LieBrkt}). The Killing vectors in Eq.~(\ref{4.23}) are a non-trivial example of Killing vectors which form a closed set under the Lie bracket.

As noted at the beginning of the section, a covariant treatment guarantees  that vectors $\eta^i$ transform the same way as $\partial_\mu \phi^i$ under isometries, e.g.
\begin{align}
\delta_\theta \eta^i &= \theta^\alpha \left( \frac{\partial t^i_\alpha}{\partial \phi^j}\right) \eta^j\,,
\label{4.26a}
\end{align}
which is a \emph{linear} transformation law.

\subsection{Local Symmetry on ${\cal M}$}\label{sec:local}

The global symmetries Eq.~(\ref{4.16}) can be promoted to local symmetries by replacing the global symmetry parameters $\theta^\alpha$ by functions of spacetime $\theta^\alpha(x)$,
\begin{align}
\delta_\theta \phi^i(x) &= \theta^\alpha(x) \, t^i_\alpha\left( \phi(x) \right) ,
\label{4.16l}
\end{align}
and introducing gauge fields.

The gauge covariant derivative of the scalar field on the curved manifold ${\cal M}$ is defined by
\begin{align}
\left( D_\mu \phi(x) \right)^i &\equiv \partial_\mu \phi^i(x) + A_\mu^\beta (x) \, t^i_\beta\! \left(\phi(x) \right)\, ,
\label{4.25}
\end{align}
where $A_\mu^\beta(x)$ is the gauge field associated with the Killing vector $t^i_\beta\!\left(\phi\right)$, and the gauge coupling constant and a factor of $i$ has been absorbed into the gauge field. The gauge covariant derivative of the scalar field should transform the same way as $\partial_\mu \phi^i$ in Eq.~(\ref{4.26}) under the {\it local} symmetry, which implies the transformation rule
\begin{align}
\delta_\theta \left( D_\mu \phi \right)^i &=  \theta^\alpha(x) \left( \frac{\partial t^i_\alpha}{\partial \phi^j}\right) \left( D_\mu \phi \right)^j\, .
\label{4.33z}
\end{align}
Consequently, the transformation law of $A^\beta_\mu(x)$ under the local symmetry is
\begin{align}
\left(\delta_\theta A_\mu^\beta \right) t^i_\beta 
&= -\left( \partial_\mu \theta^\beta \right) t^i_\beta + \theta^\beta A^\gamma_\mu \left( t^j_\gamma \frac{\partial t^i_\beta}{\partial \phi^j} -t^j_\beta \frac{\partial t^i_\gamma}{\partial \phi^j} \right) .
\end{align}
Using the definition of the Lie bracket in Eq.~(\ref{LieBrkt}), this equation yields the usual transformation law for the gauge field
\begin{align}
\delta_\theta A_\mu^\alpha &= -\partial_\mu \theta^\alpha  -\fcs[\beta\gamma]{\alpha} \theta^\beta A^\gamma_\mu\, .
\end{align}

The gauged version of the Lagrangian Eq.~(\ref{p2Lag}) is
\begin{align}
\mathcal L=\frac12 g_{ij}(\phi) \left( D_\mu\phi \right)^i \left( D^\mu  \phi \right)^j +\mathcal I(\phi)\, ,
\label{4.35}
\end{align}
where the partial derivatives of the scalar field have been replaced by gauge covariant derivatives Eq.~(\ref{4.25}).  The first variation of the Lagrangian gives the gauged generalization of the equation of motion Eq.~(\ref{EOMphi}),
\begin{align}
E_i  &= g_{ij}\left(\partial_\mu \delta^j_k+A^\beta_\mu t^j_{\beta,\,k}\right) \left(D^\mu \phi\right)^k+g_{il}\Gamma^l_{jk}\left( D^\mu\phi\right)^j \left( D_\mu\phi \right)^k-\mathcal I_{,\, i} \nn
 &\equiv g_{ij} \left(\mathscr{D}_\mu\left( D^\mu \phi \right) \right)^j - \mathcal{I}_{,\,i}\,.
 \label{4.37}
\end{align}
The gauge covariant derivative $D_\mu \phi$ of coordinates $\phi^i$ is given in Eq.~(\ref{4.25}), and the gauged covariant derivative $\mathscr{D}_\mu$ on a vector field $\eta^i$ is
\begin{align}
\left(\mathscr{D}_\mu \eta \right)^i &= \left(\partial_\mu \eta^i + \Gamma^i_{kj} \partial_\mu \phi^k \eta^j\right)+A^\beta_\mu \left({t^i_\beta}_{,j} \,   + \Gamma^i_{jk}   \, t^k_\beta
\right) \eta^j  
\label{4.33}
\end{align}
which is the gauged generalization of Eq.~(\ref{4.4}).  Eq.~(\ref{4.33}) is the appropriate definition for covariant derivatives acting on vector fields. It arises in our calculation by a direct calculation to obtain the equations of motion Eq.~(\ref{4.37}) by varying the Lagrangian Eq.~(\ref{4.35}). One can show that Eq.~(\ref{4.33}) transforms as
\begin{align}
\delta_\theta \left( \mathscr{D}_\mu \eta \right)^i &=  \theta^\alpha(x) \left( \frac{\partial t^i_\alpha}{\partial \phi^j}\right) \left( \mathscr{D}_\mu \eta \right)^j\, .
\label{4.33b}
\end{align}
The derivation of Eq.~(\ref{4.33b}) relies on two useful identities. The first is obtained by differentiating Eq.~(\ref{LieBrkt}),
\begin{align}
{f_{\alpha\beta}}^\gamma \left( \frac{\partial t^i_\gamma}{\partial \phi^k} \right)
&=  \left[ \left( \frac{\partial^2 t^i_\beta}{\partial \phi^j \partial \phi^k}\right) t_\alpha^j  -  \left( \frac{\partial^2 t^i_\alpha}{\partial \phi^j \partial \phi^k}\right)  t_\beta^j  \right]
+\left[ \left( \frac{\partial t^i_\beta}{\partial \phi^j}\right)  \left(\frac{\partial t^j_\alpha}{\partial \phi^k}\right)  -  \left(\frac{\partial t^i_\alpha}{\partial \phi^j}\right) \left(\frac{\partial t^j_\beta}{\partial \phi^k} \right) \right]\,.
\label{id4.30}
\end{align}
The second relation is that the Lie derivative of the Levi-Civita connection vanishes because $t_\alpha$ is a Killing vector.  The explicit formula is
\begin{align}
0 &=\mathcal{L}_{t_\alpha}\Gamma^i_{kj}  = t^{\ell}_\alpha  \frac{\partial \Gamma^i_{kj}}{\partial \phi^\ell} +\frac{\partial t_\alpha^\ell}{\partial \phi^k} \Gamma^i_{\ell j}
+\frac{\partial t_\alpha^\ell}{\partial \phi^j}\Gamma^i_{k \ell} -\frac{\partial t_\alpha^i}{\partial \phi^\ell}\Gamma^\ell_{kj} + \frac{\partial^2 t^i_\alpha}{\partial \phi^k \partial \phi^j}\,.
\label{id4.31}
\end{align}

The first and second variations of the gauged action up to second order give the gauged versions of Eqs.~(\ref{4.3}) and~(\ref{4.14}),
\begin{align}
\delta S &= \int d^4x \ \left[ - g_{ij} \left( \mathscr{D}_\mu \left(D_\mu \phi \right) \right)^i \eta^j +  \mathcal{I}_{,i} \eta^i \right]\,, \nn
\delta^2 S &= \frac 12 \int d^4x \left[ g_{ij}   \left(\mathscr{D}_\mu \eta \right)^i \left(\mathscr{D}^\mu \eta \right)^j - R_{ijkl} \left(D_\mu \phi \right)^j \left(D^\mu \phi \right)^l \eta^i \eta^k + \left(\nabla_i \nabla_j \mathcal{I} \right)\, \eta^i \eta^j \right]\,.
\label{4.36}
\end{align}
The gauge field now appears implicitly in every term except for those involving the potential $\mathcal I$. The second variation $\delta^2 S$ depends on the curvature $R_{ijkl}$ of $\m$, but it does not have a term that depends on the gauge curvature (i.e.\ field-strength) $F_{\mu \nu}$.

The divergent one-loop contribution in $4-2\epsilon$ dimensions for quadratic actions such as Eq.~(\ref{4.14})
was derived by 't~Hooft in Ref.~\cite{tHooft:1973us}, 
\begin{align}
\Delta \mathcal{L}^{\rm 1-loop}=\frac{1}{32\pi^2\epsilon}\left( \frac{1}{12}\tr\left[Y_{\mu\nu}Y^{\mu\nu}\right] +\frac12\tr\left[ X^2\right]\right)\,,
\label{1loope}
\end{align}
where  
\begin{align}
{\left[ Y_{\mu\nu} \right]^i}_j &\equiv {\left[\mathscr{D}_\mu\,,\mathscr{D}_\nu \right]^i}_j\,,& {\left[X\right]^i}_{k}&\equiv 
-R^{i}_{\,\,jkl} (D_\mu \phi)^j (D^\mu \phi)^l + g^{ij} \, \mathcal{I}_{;jk} \, .
\label{XY}
\end{align}
't~Hooft's original derivation is valid when the scalar metric is $\delta_{ij}$.  Our form Eqs.~(\ref{1loope}) with $Y_{\mu \nu}$ and $X$ given by Eq.~(\ref{XY}) applies for any metric $g_{ij}$.

The matrix $X$ is the mass squared term for the fluctuations $\eta$ in Eq.~(\ref{4.36}), and $Y_{\mu \nu}$ is a field strength tensor constructed from the covariant derivative $\mathscr{D}$.  An explicit computation using the identities (\ref{id4.30}) and (\ref{id4.31}) shows that $Y_{\mu \nu}$ is equal to the sum of the curvature of $\m$ and the curvature of the gauge field,
\begin{align}
\left[Y_{\mu \nu}\right]^i{}_j  &= \left[ \mathscr{D}_\mu, \mathscr{D}_\nu \right]^i {}_j = R^i {}_{jkl} \left(D_\mu \phi \right)^k \left(D_\nu \phi\right)^l + F^\alpha_{\mu \nu} \, {t^i _\alpha }_{;j} \, .
\label{4.33a}
\end{align}

For Goldstone bosons, where $\mathcal{I}$ is a constant, $X$ and $Y_{\mu \nu}$ are both proportional to two derivatives of $\phi$ times the curvature $R_{ijkl}$, i.e. they are order ${\cal O}(R\, p^2)$, where $R$ is a typical curvature and $p$ is a typical momentum.  Thus, the one-loop correction, which is proportional to the traces of $X^2$ and $Y_{\mu \nu}^2$, is order $\mathcal{O}(R^2 p^4)$, and is $\mathcal{O}(p^4)$ as one expects in chiral perturbation theory.  The $\mathcal{O}(p^4)$ correction is proportional to the square of the curvature, and vanishes if the manifold is flat, i.e.\ in a theory such as the SM.  Thus, the SM is renormalizable even in non-linear coordinates; one-loop graphs do not require four-derivative counterterms.  The $F_{\mu \nu}$ term in $Y_{\mu \nu}$ gives the Goldstone boson contribution to the gauge coupling $\beta$-function of order $\mathcal{O}(F_{\mu \nu}^2)$, and the running of operators involving field strengths of order $\mathcal{O}(R F_{\mu \nu} p^2)$.

The quadratic invariants that enter Eq.~(\ref{1loope}) are
\begin{align}
\tr \left[ X^2 \right]  &= (\nabla^i \nabla_j \mathcal{I})(\nabla^j \nabla_i \mathcal{I}) + R^i_{(d_\mu \phi)\, j\, (d_\mu\phi)} R^j_{(d_\nu\phi)\, i\, (d_\nu\phi)}  -2  (\nabla^i \nabla^j \mathcal{I})R_{i \, (d_\mu \phi)\, j\, (d_\mu\phi)}\,,
\end{align}
and
\begin{align}
\tr \left[ Y_{\mu \nu} Y^{\mu \nu} \right]  &= R^i{}_{j\,(d_\mu \phi)\, (d_\nu\phi)}R^j{}_{i\,(d_\mu \phi)\, (d_\nu\phi)}
+ 2 R^j{}_{i\,(d_\mu \phi)\, (d_\nu\phi)} F_{\mu \nu}^\alpha (t^i_\alpha)_{;j} + F_{\mu \nu}^\alpha F_{\mu \nu}^\beta (t^i_\alpha)_{;j} (t^j_\beta)_{;i} \,.
\label{4.45}
\end{align}
Eq.~(\ref{4.45}) is universal and applies to many theories. The one-loop correction in HEFT, which is complicated, and was given previously in Refs.~\cite{Guo:2015isa,Alonso:2015fsp}, is simply an expansion of Eq.~(\ref{4.45}) into component fields.  More details about the expansion are given in Appendix~\ref{app:oneloop}.  As explained in Ref.~\cite{Alonso:2015fsp}, the same formula Eq.~(\ref{4.45}) applies to HEFT, the SM scalar sector, dilaton theories, and chiral perturbation theory.

To close this section, we consider spontaneous symmetry breaking in a theory with an invariant potential $\Lie_{t_\alpha}\mathcal{I}=0$, so that we have exact Goldstone bosons. The fields $\phi^i$ have vacuum expectation values $\vev{\phi^i}$, and transform as $\delta \phi^i = \theta^\alpha \, t^i_\alpha\! \left(\vev{\phi} \right)$. Thus, broken symmetries $t^i_A$ satisfy $t^i_A \! \left( \vev{\phi} \right) \not = 0$, and $t^i_A\! \left(\vev{\phi} \right)$ is a vector in the Goldstone boson direction --- i.e. motion along the vector field $t^i_A$ (for each broken generator) is motion between different vacuum states with the same value of the potential $\mathcal{I}$.

In the gauged case, the Goldstone bosons are eaten, giving a mass term for the gauge bosons. The Lagrangian of Eq.~(\ref{4.35}) gives the gauge boson mass term
\begin{align}
\mathcal{L}& =\frac12 M_{BC}^2 A_\mu^B {A^C}^\mu, \,   & M_{BC}^2 \left(\vev{\phi}\right) &\equiv g_{ij}\!\left(\vev{\phi} \right) \, t^i_B\!\left(\vev{\phi}\right) \,  t_C^j\! \left( \vev{\phi}\right). 
\end{align}
The rank of $M_{BC}^2$ determines the number of massive gauge bosons, which cannot exceed the dimension of the manifold $\m$. If the number of isometries exceeds $\dim \m$, then there are unbroken symmetries.  This is true in the $\m=S^{N}$ theory, where there are $N(N+1)/2$ isometries which form the group $\g=O(N+1)$, and the unbroken subgroup  $\h=O(N)$ has $N(N-1)/2$ generators. The number of broken generators is $N$, which is equal to the dimension of $S^{N}$.

\section{CCWZ and Non-compact Groups}\label{sec:CCWZ}

In this section, we connect the geometric formalism with the explicit formul\ae\ of CCWZ~\cite{Coleman:1969sm,Callan:1969sn} for Goldstone boson Lagrangians with symmetry breaking pattern $\g \to \h$.  We are interested in applying the formalism to non-compact groups, and to sigma models with non-trivial metrics on $\g/\h$. Our presentation thus parallels the discussion in the original work, while pointing out differences which arise for the case of non-compact groups.

Consider a group $\cal G$ with generators $t_\alpha$, $\alpha=1, \cdots, {\rm dim}\,{\cal G}$, satisfying the Lie algebra $\mathfrak{g}$
\begin{align}
\left[t_\alpha,t_\beta\right]&=i \fcs[\alpha\beta]{\gamma} t_\gamma\,,
\label{liealgebrag}
\end{align}
and the Jacobi identity
\begin{align}
\fcs[\alpha \beta]{\lambda} \fcs[\gamma\lambda]{\sigma} + \fcs[\gamma \alpha]{\lambda} \fcs[\beta\lambda]{ \sigma }+ \fcs[\beta \gamma]{\lambda} \fcs[\alpha\lambda]{ \sigma}
&=0 \,.
\label{jacobi}
\end{align}
To allow for negatively curved spaces~\cite{Alonso:2016btr}, we do not assume that the group ${\cal G}$ is compact.   Consequently, the Lie algebra Eq.~(\ref{liealgebrag}) implies that the structure constants $\fcs[\alpha\beta]{\gamma}$ are antisymmetric in their first two indices, $\fcs[\alpha\beta]{\gamma}=-\fcs[\beta\alpha]{\gamma}$, but total antisymmetry of the structure constants in all three indices, which is true for compact groups, is not assumed.

The group $\cal G $ is spontaneously broken to the subgroup $\cal H$ with generators $T_a$, $a=1, \cdots, {\rm dim}\, {\cal H}$,  satisfying the Lie algebra $\mathfrak{h}$,
\begin{align}
\left[T_a,T_b\right]&=i \fcs[ab]{c} T_c\, .
\label{LieAlgh}
\end{align}
The remaining broken generators of the coset ${\cal G}/{\cal H}$ needed to span $\mathfrak{g}$ are given by $X_A$, $A=1, \cdots, {\rm dim}\,{\cal G}/{\cal H}$.  The choice of the broken generators $X_A$ is not unique. In the familiar example of broken chiral symmetry in QCD, different choices of broken generators lead to different parameterizations of the chiral Lagrangian, e.g. by $\xi(x)$ which transforms as $\xi \to L \xi h^\dagger = h \xi R^\dagger$, or by $U(x)$ which transforms as $L U R^\dagger$~\cite{Manohar:1996cq}.

The $\mathfrak{g}$ commutation relations of the generators $t_\alpha= \{ T_a, X_A \}$ in Eq.~(\ref{liealgebrag}) decompose into the following commutation relations for the unbroken and broken generators
\begin{subequations}
\begin{align}
\left[T_a,T_b\right]&=i \fcs[ab]{c} \,T_c \,,  \label{LieAlga} \\
\left[T_a, X_B \right] &= i \fcs[aB]{C}\, X_C + i \fcs[aB]{c} T_c\,, \label{LieAlgb}\\
\left[X_A,X_B\right]&=i \fcs[AB]{C} \, X_C +i \fcs[AB]{c}\, T_c\, . \label{LieAlgc} 
\end{align}
\end{subequations}
The first line Eq.~(\ref{LieAlga}) is the Lie algebra $\mathfrak{h}$ of the subgroup $\cal H$ in Eq.~(\ref{LieAlgh}), which is closed under commutation, so the commutator $\left[ T_a, T_b \right]$ has no term proportional to the broken generators $X_C$, which implies that the structure constants $\fcs[ab]{C}=0$.

For compact groups, complete antisymmetry of the structure constants then implies that $\fcs[aB]{c}=0$, so Eq.~(\ref{LieAlgb}) simplifies to
\begin{align}
\left[T_a, X_B \right] &= i \fcs[aB]{C}\, X_C\,,
\label{lie2}
\end{align}
which implies that the broken generators $X_A$ form a (possibly reducible) representation $\rpi$ of the unbroken subgroup $\cal H$. The generators $T_a$ of $\cal H$ in the $\rpi$ representation are determined by the structure constants $\fcs[aB]{C}$,
\begin{align}
\left[ T_a^{\rpi} \right]_B{}^C &= - i \fcs[aB]{C}\,.
\label{lie3}
\end{align}
The $\mathfrak{h}$ commutation relations Eq.~(\ref{LieAlga}) in representation $\rpi$,
\begin{align}
\left[T_a^{\rpi} , T_b^{\rpi} \right]&=i \fcs[ab]{c} \,T_c^{\rpi} \,,
\end{align}
follow from the Jacobi identity Eq.~(\ref{jacobi}).

For non-compact groups, Eq.~(\ref{lie2}) need not be satisfied.  For now, we restrict our attention to symmetry breaking patterns where Eq.~(\ref{lie2}) holds, so $\fcs[aB]{c}=0$.  Such cosets are called reductive cosets.  Non-reductive cosets are discussed in Appendix~\ref{app:nonred}.   An example of a reductive coset is the breaking of the Lorentz group down to its rotation subgroup.  For reductive cosets, the broken generators transform as a representation $\rpi$ of the unbroken symmetry group ${\cal H}$, just as in the compact case. The coset is reductive if $\h$ is compact, even if $\g$ is non-compact.

Often, there is a discrete symmetry of the Lie algebra $X_A \to -X_A$ under which the broken generators change sign.  The presence of such a discrete symmetry implies that the structure constants $\fcs[aB]{c}$ and $\fcs[AB]{C}$ vanish, so the Lie algebra reduces to
\begin{align}
\left[T_a,T_b\right]&=i \fcs[ab]{c} \,T_c \,,  \nn
\left[T_a, X_B \right] &= i \fcs[aB]{C}\, X_C , \\
\left[X_A,X_B\right]&=i \fcs[AB]{c}\, T_c\, . \nonumber
\end{align}
An example is chiral symmetry breaking in the strong interactions, where the broken generators are odd under parity.  Cosets with such a discrete symmetry are referred to as symmetric cosets.  Symmetric cosets are automatically reductive.

The CCWZ formalism picks elements of ${\cal G}/{\cal H}$ cosets  using  the exponential map of the broken generators $\{X_A\}$
\begin{align}
\xi(x)&\equiv e^{i  \bpi \cdot X}\,, \qquad & \bpi \cdot X \equiv  \bpi^A(x) X_A = \left(\frac{\pi^A(x)}{F_\pi}\right) X_A,
\label{4.46}
\end{align}
where $ \bpi^A(x)$ are the dimensionless spacetime-dependent parameters describing the Goldstone boson directions on the vacuum  coset ${\cal G}/{\cal H}$.  This exponential map gives a unique association between a point in the coset $\cal G/\cal H$ and $\bpi^A(x)$ in a neighborhood of the identity element $e$.  An arbitrary group element $g \in {\cal G}$ in the neighborhood of the identity element $e$ can be written uniquely as 
\begin{eqnarray}
g &=& e^{i  \bpi \cdot X} e^{i \alpha \cdot T } ,
\end{eqnarray}
where $\alpha \cdot T \equiv \alpha^a(x) T_a$. 
Left action by an arbitrary group element $g \in {\cal G}$ on ${\cal G}/{\cal H}$ is given by
\begin{align}
T_{g}: \xi(x) \to g \ \xi(x),
\end{align}
which maps a point in coset space to a new point in coset space. The transformation law for $\xi(x)$ is 
\begin{align}
g\, \xi(x)&= \xi^\prime(x)\ h\left(\xi(x),g \right), \qquad g\in {\cal G},\ h \in {\cal H}\,,
\label{4.47}
\end{align}
where $\xi^\prime(x)$ is a new coset and $h \in \h$ is an implicit function of $g \in \g$ and the original coset $\xi(x)$. Using the identity
\begin{align}
g \, \xi(x) &= \left( g \, \xi(x) \, g^{-1}\right)  \,g \,,  &  g \, \xi(x) \, g^{-1} &= \exp\left(i \, \bpi(x) \, \cdot  \left( g\, X \,  g^{-1}\right)\right)\,,
\label{1.11}
\end{align}
one sees that if $g = h_0 \in \h$ is an unbroken symmetry transformation, then $h\left(\xi(x),h_0 \right)=h_0$.  In addition, $\xi^\prime = h_0 \xi h_0^{-1}$, which implies that (since the coset is assumed reductive)
\begin{align}
{ \bpi^\prime}^A(x) &= {\left[D^{\rpi}(h_0)\right]^A}_B\,  \bpi^B(x)\,,
\label{5.14a}
\end{align}
where $D^{\rpi}(h_0)$ is the $\h$ transformation matrix in the $\rpi$ representation.  Note that for reductive cosets, if $g = h_0\in H$, then $h\left(\xi(x),h_0 \right)=h_0$ is a constant (i.e.\ it does not depend on $x$ through $\xi(x)$).

The CCWZ procedure for building a $\g$-invariant Lagrangian is to map all fields to the origin of coset space $\xi=1$ with $\bpi(x)=0$ by left-action by $g=\xi^{-1}$, and to define covariant derivatives in terms of this map.  Explicitly, one starts with
\begin{align}
\xi^{-1} D_\mu \xi &= \xi^{-1} (\partial_\mu + i A^{\alpha}_{\mu} \, t_\alpha ) \xi
\end{align}
where the gauge coupling constant has been absorbed into the normalization of the gauge field $A^\alpha_\mu$.  If only a subgroup ${\cal G}_{\rm gauge} \subset {\cal G}$ is gauged, then only gauge bosons of ${\cal G}_{\rm gauge}$ appear in the above equation, or equivalently, the gauge bosons corresponding to global symmetry directions are set equal to zero.  In addition, different factor gauge groups in ${\cal G}_{\rm gauge}$ can have distinct gauge coupling constants. Power series expansion of $\xi^{-1} D_\mu \xi$ shows that it can be expressed in terms of multiple commutators, so it is an element of the Lie algebra $\mathfrak{g}$ which can be decomposed in terms of unbroken and broken generators,
\begin{align}
\xi^{-1} D_\mu \xi &= \left. \xi^{-1} D_\mu \xi \right|_T +  \left. \xi^{-1} D_\mu \xi \right|_X = i \, V_\mu + i \, \left( D_\mu  \bpi  \right)\,,\nn
\left. \xi^{-1} D_\mu \xi \right|_T &\equiv i \ V_\mu = i \ V_\mu^a T_a   , \nn
\left. \xi^{-1} D_\mu \xi \right|_X  &\equiv i \ \left( D_\mu  \bpi \right) =  i \left( D_\mu  \bpi \right)^A X_A  .
\label{5.14}
\end{align}
The above equations define $V_\mu$ and $(D_\mu  \bpi)$. Usually, one normalizes the generators so that $\tr t_\alpha t_\beta = \delta_{\alpha \beta}/2$, and projects out the broken and unbroken pieces of $\xi^{-1} D_\mu \xi$ by taking the appropriate traces.  The decomposition of a vector into a linear combination of basis vectors does not require an inner product on the vector space, so Eq.~(\ref{5.14}) is well-defined even without this normalization of generators.  An orthogonal normalization of generators is not possible for non-compact ${\cal G}$, but Eq.~(\ref{5.14}) is well-defined. Under an unbroken symmetry transformation $h \in {\cal H}$, $V_\mu$ transforms like a gauge field
\begin{align}
V_\mu & \to h \, V_\mu \, h^{-1} - \left( \partial_\mu h \right) \, h^{-1}\,, &
\label{5.15}
\end{align}
and $(D_\mu  \bpi)$ transforms by adjoint action by $\h$ in the representation $\rpi$,
\begin{align}
\left( D_\mu  \bpi \right) & \to h \, \left( D_\mu  \bpi \right) \, h^{-1}\, .
\label{5.16}
\end{align}
These last two equations require the reductive coset condition $\fcs[aB]{c}=0$.  The generalization to non-reductive cosets is discussed in Appendix~\ref{app:nonred}.

The pion covariant derivative can be decomposed into a purely pionic piece and a gauge field piece,
\begin{align}
\left( D_\mu  \bpi \right)^A &\equiv {\left[ e( \bpi)\right]^A}_B \, \left( \partial_\mu  \bpi^B \right) + F^A_\alpha( \bpi) A_\mu^\alpha
\label{split}
\end{align}
where ${\left[e( \bpi)\right]^A}_B$ are vierbeins of the ${\cal G}/{\cal H}$ vacuum manifold, and  $F^A_\alpha ( \bpi)$ are related to the Killing vectors of  ${\cal G}/{\cal H}$.

For groups where $(D_\mu  \bpi)^A$ transforms as a single irreducible representation $\rpi$, as in QCD, the simplest invariant Lagrangian is the $\mathcal{O}(p^2)$ term
\begin{align}
L &= \frac12 F_\pi^2 \sum_A  \left(D_\mu  \bpi \right)^A \left(D^\mu  \bpi \right)^A   \nn
&= \frac12 F_\pi^2 \sum_A  {\left[e( \bpi)\right]^A}_B\, {\left[e( \bpi)\right]^A}_C \left( \partial_\mu  \bpi \right)^B \left( \partial^\mu  \bpi \right)^C 
+ \cdots \, \nn
&\equiv  \frac12  \ g_{BC}( \bpi) \ \left(\partial_\mu  \bpi \right)^B \left(\partial^\mu  \bpi \right)^C + \cdots , 
\label{5.17}
\end{align}
where $F_\pi$ is the Goldstone boson decay constant, and the ellipsis denotes terms depending on the gauge fields. The Lagrangian Eq.~(\ref{5.17}) defines the scalar field metric of the ${\cal G}/{\cal H}$ vacuum manifold,
\begin{align}
g_{BC}(\bpi) &= F_\pi^2 \sum_A  {\left[e( \bpi)\right]^A}_B\, {\left[e( \bpi)\right]^A}_C\,.
\label{5.17a}
\end{align}
If the representation is reducible, the sum in Eq.~(\ref{5.17}) can be divided into sums over the individual irreducible representations, with arbitrary weights for each irreducible representation. The most general $\mathcal{O}(p^2)$ term allowed is
\begin{align}
L &= \frac12 F_\pi^2 \sum_{A,B} \eta_{AB}\, (D_\mu  \bpi)^A (D^\mu  \bpi)^B\,,
\label{5.18}
\end{align}
where $\eta_{AB}$ is a symmetric tensor invariant under the adjoint action of $\h$, Eq.~(\ref{5.16}). $\eta_{AB}$ is a positive definite matrix so that the pion kinetic energies have the correct sign. Note that $\eta_{AB}$ is a constant, i.e.\ it does not depend on $\pi$. One can always define a positive definite kinetic energy if $\h$ is a compact subgroup, e.g.\ by choosing $\eta_{AB}=\delta_{AB}$. In summary, the most general scalar metric for ${\cal G}/{\cal H}$ is
\begin{align}
g_{CD}( \bpi) &= F_\pi^2 \sum_{A,B} \eta_{AB}\, {\left[e( \bpi)\right]^A}_C\, {\left[ e( \bpi)\right]^B}_D\,,
\label{5.18a}
\end{align}
and the Killing vectors in Sec.~\ref{sec:global} are given by
\begin{align}
t^A_\alpha ( \bpi) &= {\left[e^{-1}( \bpi)\right]^A}_B\ F^B_\alpha ( \bpi) ,
\label{killing}
\end{align}
where $\left[ e^{-1} ( \bpi)  \right]$ is the inverse vierbein, which satisfies the identity
\begin{align}
\left[e^{-1}( \bpi)\right]^A{}_B\ {\left[ e( \bpi) \right]^B}_C &= {\delta^A}_C\,.
\end{align}
Eq.~(\ref{killing}) can easily be derived by looking at the shift $\bpi \to \bpi + \delta \bpi$ for an infinitesimal $\g$ transformation.

For the HEFT example, we need to evaluate the curvature tensors at the vacuum field configuration $ \bpi^A=0$, which requires knowing the metric tensor to quadratic order in $ \bpi$. The curvature at any other point can then be obtained using left-action by $\g$. Expanding Eq.~(\ref{a-3ccwz}) and using the most general Lie algebra relations Eqs.~(\ref{LieAlga}), (\ref{LieAlgb}) and~(\ref{LieAlgc}), one obtains
\begin{align}
\left( D_\mu  \bpi \right)^A &= \partial_\mu  \bpi^A +\frac12 \fcs[CB]{A}  \bpi^C \partial_\mu  \bpi^B + 
\frac16 \fcs[D\alpha]{A} \fcs[CB]{\alpha}  \bpi^D \bpi^C \partial_\mu  \bpi^B + \ldots \nn
& +  A_{\mu}^A + \fcs[B\alpha]{A}  \bpi^B A^\alpha_\mu + \frac{1}{2} \fcs[C\beta]{A} \fcs[B\alpha]{\beta}  \bpi^B  \bpi^C A_\mu^\alpha +
\ldots \,.
\end{align}
The term $A_{\mu}^A$ only involves the broken generators, and it is the square of this term in the kinetic energy which results in the broken gauge bosons acquiring a mass proportional to $F_\pi^2$.  From Eq.~(\ref{split}), the vierbein is
\begin{align}
{\left[ e( \bpi) \right]^A}_B &= {\delta^A}_B +\frac12 \fcs[CB]{A}  \bpi^C + \frac16 \fcs[D\alpha]{A} \fcs[EB]{\alpha}  \bpi^D  \bpi^E+ \ldots 
\end{align}
Using Eq.~(\ref{5.18a}), the metric $g_{CD}( \bpi)$ is
\begin{align}
\frac{1}{F_\pi^2} g_{CD}( \bpi) &=\eta_{CD} +  \frac12 \left( \eta_{AD}  \fcs[EC]{A} + \eta_{CB} \fcs[ED]{B}  \right)  \bpi^E \nn
&+   \left(  \frac16 \eta_{CB}  \fcs[E\alpha]{B} \fcs[FD]{\alpha}    + \frac16
\eta_{AD} \fcs[E\alpha]{A} \fcs[FC]{\alpha}+ \frac14 \eta_{AB} \fcs[EC]{A}  \fcs[FD]{B}  \right)  \bpi^E  \bpi^F   
+ \mathcal{O}( \bpi^3)\,.
\label{metric}
\end{align}
For a compact group, the structure constants are completely antisymmetric, so the linear term in $ \bpi$ vanishes if $\eta_{AB} \propto \delta_{AB}$.  However, in some cases, such as the SM with custodial symmetry violation, the linear term is non-zero.

The geometric quantities we need can be computed directly from the metric Eq.~(\ref{metric}). The Christoffel symbol is
\begin{align}
\Gamma^A_{BC} &=\frac12 \eta^{AG} \left(  \eta_{CE} \fcs[BG]{E} +  \eta_{BE}  \fcs[CG]{E}  \right) +\frac14 \left(  \fcs[GB]{E} \eta_{EC} +  \fcs[GC]{E} \eta_{EB}  \right)\left(\fcs[DH]{A} \eta^{HG}+\fcs[DH]{G} \eta^{AH} \right)  \bpi^D \nn
&- \frac14\left(  \fcs[HC]{G}  \fcs[DB]{E} +  \fcs[DC]{G}  \fcs[HB]{E} \right) \,\eta^{AH}\, \eta_{GE} \,   \bpi^D \nn
& + \frac1{12} \left(  \fcs[C\alpha]{A} \fcs[DB]{\alpha} + \fcs[B\alpha]{A} \fcs[DC]{\alpha}    \right)   \bpi^D  + \frac14  \left( \fcs[CG]{\alpha} \eta_{BE} + \fcs[BG]{\alpha} \eta_{CE}
   \right)  \eta^{AG} \fcs[D\alpha]{E}  \bpi^D    + \mathcal{O}( \bpi^2) ,
\label{christoffel}
\end{align}
where $\eta^{AB}$ is the inverse of $\eta_{AB}$, and the Jacobi identity has been used to simplify the final result. The Riemann curvature tensor is
\begin{align}
\frac{1}{F_\pi^2} R_{ABCD}&=\frac14\left(\fcs[AB]{\alpha}\fcs[D\alpha]{E}\eta_{CE}-\fcs[AB]{\alpha} \fcs[C\alpha]{E}\eta_{DE}+\fcs[CD]{\alpha}\fcs[B\alpha]{E}\eta_{AE}-\fcs[CD]{\alpha}\fcs[A\alpha]{E}\eta_{BE}\right) \nn
&+\frac14\left( \fcs[AD]{G}\fcs[BC]{E} - \fcs[AC]{G} \fcs[BD]{E} - 2 \fcs[AB]{G}\fcs[CD]{E}\right)\eta_{GE} \nn
&+\frac14 \eta^{GE} \bigl[ \left(\fcs[AG]{H}\eta_{DH}+\fcs[DG]{H}\eta_{AH}\right)\left(\fcs[BE]{I}\eta_{CI}+\fcs[CE]{I}\eta_{BI}\right) \nn
&-\left(\fcs[BG]{H}\eta_{DH}+\fcs[DG]{H}\eta_{BH}\right)\left(\fcs[AE]{I}\eta_{CI}+\fcs[CE]{I}\eta_{AI}\right)\bigr]\,,
\label{riemann}
\end{align}
where we recall that the sum on $\alpha$ runs over both broken and unbroken generators, whereas the sums on $E$, etc.\ are only over the broken generators. The Ricci curvature is
\begin{align}
R_{BD}&=\frac14\left(\fcs[AB]{\alpha}\fcs[D\alpha]{A}+\fcs[AD]{\alpha}\fcs[B\alpha]{A}-\fcs[AB]{\alpha} \fcs[C\alpha]{G}\eta_{DG}\eta^{AC}-\fcs[AD]{\alpha}\fcs[C\alpha]{G}\eta_{BG}\eta^{AC}\right) \nn
&-\frac34 \fcs[AB]{G}\fcs[CD]{H}\eta^{AC}\eta_{GH} +\frac14 \eta^{GH} \left(\fcs[AG]{R}\eta_{DR}+\fcs[DG]{R}\eta_{AR}\right)\left(\fcs[BH]{R}\eta_{CR}+\fcs[CH]{R}\eta_{BR}\right) \eta^{AC}\nn
&-\frac12 \left(\fcs[BG]{R}\eta_{DR}+\fcs[DG]{R}\eta_{BR}\right)\fcs[AH]{A} \eta^{GH}  \,,
\label{ricci}
\end{align}
and the scalar curvature is
\begin{align}
F_\pi^{2} R=\fcs[AB]{\alpha}\fcs[C\alpha]{A}\eta^{BC}-\frac14 \fcs[AB]{C}\fcs[GH]{D}\eta^{AG}\eta^{BH}\eta_{CD}+\frac12 \fcs[AC]{B}\fcs[BD]{A}\eta^{CD}-\fcs[AC]{A}\fcs[BD]{B}\eta^{CD}\,.
\label{scalar}
\end{align}
The scalar curvature does not have a definite sign unless the group is compact. Eqs.~(\ref{metric}), (\ref{christoffel}), (\ref{riemann}), (\ref{ricci}) and~(\ref{scalar}) are valid even for non-reductive cosets.

The results simplify considerably in a number of special cases.  For a symmetric coset, $\fcs[AB]{C}=0$, and the curvatures Eqs.~(\ref{riemann}), (\ref{ricci}) and (\ref{scalar}) reduce to
\begin{align}
\frac{1}{F_\pi^2} R_{ABCD}&=\frac14\left(\fcs[AB]{\alpha}\fcs[D\alpha]{G}\eta_{CG}-\fcs[AB]{\alpha} \fcs[C\alpha]{G}\eta_{DG}+\fcs[CD]{\alpha}\fcs[B\alpha]{G}\eta_{AG}-\fcs[CD]{\alpha}\fcs[A\alpha]{G}\eta_{BG}\right) \,, \nn
R_{BD}&=\frac14\left(\fcs[AB]{\alpha}\fcs[D\alpha]{A}+\fcs[AD]{\alpha}\fcs[B\alpha]{A}-\fcs[AB]{\alpha} \fcs[C\alpha]{G}\eta_{DG}\eta^{AC}-\fcs[AD]{\alpha}\fcs[C\alpha]{G}\eta_{BG}\eta^{AC}\right) \,,\nn
F_\pi^{2} R&=\fcs[AB]{\alpha}\fcs[C\alpha]{A}\eta^{BC}\,,
\end{align}
where the sum on $\alpha=\{ a, A\}$ can be restricted to the unbroken generator index $a$ only.

Another special case is $\g$ compact and $\eta_{AB}=\delta_{AB}$. For a compact group, the generators can be normalized so that $\tr t_\alpha t_\beta \propto \delta_{\alpha \beta}$, so the structure constants are completely antisymmetric tensors in their three indices.  Writing the structure constants with three lower indices in the usual notation for compact groups,  Eqs.~(\ref{riemann}), (\ref{ricci}) and (\ref{scalar}) simplify to
\begin{align}
\frac{1}{F_\pi^2} R_{ABCD} &= f_{AB\alpha} f_{CD \alpha}-\frac34  f_{ABG}f_{CDG} 
= f_{ABg} f_{CDg}+\frac14  f_{ABG}f_{CDG} \,, \nn
R_{BD} &= f_{ABg} f_{ADg}+\frac14  f_{ABG}f_{ADG} \,, \nn
F_\pi^2 R &=f_{ABg} f_{ABg}+\frac14  f_{ABG}f_{ABG}\,.
\label{scalar-c}
\end{align}
An interesting feature is the relative $1/4$ for the sum over broken generator index $G$ relative to the sum over unbroken generator index $g$.

If one adds the additional restriction that the coset of the compact group ${\cal G}$ is symmetric, so $f_{ABC}=0$, the formul\ae\ Eqs.~(\ref{scalar-c}) simplify further to
\begin{align}
\frac{1}{F_\pi^2} R_{ABCD} &= f_{ABg} f_{CDg}, \nn
R_{BD} &= f_{ABg} f_{ADg} = \frac12 C_A(\g) \delta_{BD}, \nn
F_\pi^2 R &=\frac12 C_A(\g) N_\pi,
\label{scalar-s}
\end{align}
where $C_A(\g)$ is the Casimir in the adjoint representation of $\g$, and $N_\pi= {\rm dim}{\cal G}/{\cal H}$ is the number of broken generators.

Finally, if the gauge group is compact and completely broken, so that $\g/\h=\g$, and $\eta_{AB}=\delta_{AB}$, Eqs.~(\ref{scalar-c}) become
\begin{align}
\frac{1}{F_\pi^2} R_{ABCD}
&=\frac14  f_{ABG}f_{CDG}, \nn
R_{BD} &= \frac14  C_A(\g) \delta_{BD}, \nn
F_\pi^2 R &=\frac14  C_A(\g)  N_\pi\,.
\label{scalar-g}
\end{align}

\subsection{Matter Fields} \label{sec:matter}

We refer to all non-Goldstone boson or gauge fields generically as matter fields. The CCWZ transformation for matter fields $\psi$ under the group transformation law Eq.~(\ref{4.47}) is
\begin{align}
\psi &\to D^{(\psi)}(h) \, \psi\,,
\label{6.41}
\end{align}
where $D^{(\psi)}(h)$ are the $\h$ representation matrices for $\psi$.   Note that $D^{(\psi)}(h)$ is assumed to be an irreducible representation, so if it is reducible, one must first decompose it into its irreducible representations.  The different irreducible representation components are then treated as separate matter fields.  One can define a chiral covariant derivative for matter field $\psi$ by
\begin{align}
D_\mu \psi &\to \left(\partial_\mu +i T_a^{(\psi)}V^a_\mu \right) \psi\,,
\label{6.42}
\end{align}
where $T^{(\psi)}_a$ are the generators of the unbroken subgroup ${\cal H}$ in the representation $D^{(\psi)}(h)$ of $\h$.  The chiral covariant derivative transforms as
\begin{align}
\left( D_\mu \psi \right) &\to D^{(\psi)}(h) \, \left( D_\mu \psi \right)\,.
\label{6.43}
\end{align}
The covariant derivative Eq.~(\ref{6.42}) is derived in CCWZ.  The argument relies on defining it as the ordinary derivative at $\xi=1$, and then using $\g$ action to define it for arbitrary $\xi$.  The key point (which is not true for non-reductive cosets) is that if $g \in H$, then $h$ in Eq.~(\ref{4.47}) is a constant, so the ordinary derivative transforms the same way as the field, Eq.~(\ref{6.41}).  Using this result at $\xi=1$, the transformation Eq.~(\ref{6.43}) for arbitrary $\xi$ follows.

The covariant derivative Eq.~(\ref{6.42}) is based on Eq.~(\ref{5.14}), and hence on the Maurer-Cartan form $g^{-1} \rd g$. This is the canonical connection on the principal $\h$-bundle $\g \to \g/\h$, and makes no reference to a metric, i.e.\ to $\eta_{AB}$. One can also define covariant derivatives based on the metric (Christoffel) connection Eq.~(\ref{christoffel}), which does depend on $\eta_{AB}$. The two are equivalent if $\eta_{AB}=\delta_{AB}$, i.e.\ if the $\g$-invariant metric on $\g/\h$ is obtained from a $\g$-invariant metric on $\g$. The difference in the connections transforms as a $\h$-invariant tensor~\cite{10.2307/2372398}, so that the change in connection can be compensated by a change in coefficients of invariant terms in the sigma model Lagrangian. The exponential map $\xi(\lambda)=\exp (X \lambda)$ is geodesic for the Maurer-Cartan connection, but not for a general $\eta_{AB}$ metric connection.

\subsection{Sectional Curvature}\label{sec:K}

The sectional curvature $K(Y,Z)$ in the plane spanned by tangent vectors $Y$ and $Z$ is defined by
\begin{align}
K(Y,Z) &=  \frac{R_{ABCD} Y^A Z^B Y^C Z^D}{\ip{Y}{Y}\ip{Z}{Z}-\ip{Y}{Z}^2}
\end{align}
where the inner product $\ip{*}{*}$ is w.r.t.\ the metric $g_{AB}$. The Cauchy-Schwartz inequality implies the denominator is positive, so the sign of the sectional curvature depends on the sign of the numerator. The sign of the sectional curvature is important, because, as shown in Refs.~\cite{Alonso:2015fsp,Alonso:2016btr}, the sign of deviations in Higgs-gauge boson scattering amplitudes from SM amplitudes is determined by the sign of the sectional curvatures of the HEFT sigma model.

From Eq.~(\ref{riemann}),
\begin{align}
&\frac{1}{F_\pi^2} R_{ABCD} Y^A Z^B Y^C Z^D = \frac12 \fcs[YZ]{\alpha} \left(\fcs[Z\alpha]{A} Y^B \eta_{AB} -\fcs[Y\alpha]{A} Z^B \eta_{AB}  \right) -\frac34\  \fcs[YZ]{A}\fcs[YZ]{B} \eta_{AB} \nn
& + \frac14 \left( \fcs[YG]{A} Z^B \eta_{AB} + \fcs[ZG]{A} Y^B \eta_{AB} \right)\left( \fcs[YH]{C} Z^D \eta_{CD} + \fcs[ZH]{C} Y^D \eta_{CD} \right) \eta^{GH} \nn
&-\fcs[YG]{A}  \fcs[ZH]{C} Y^B   Z^D \eta_{AB} \eta_{CD} \eta^{GH}
\label{riemannK}
\end{align}
and we have used the definition
\begin{align}
\left[Y,Z\right]=\left[Y^AT_A,Z^BT_B\right] &\equiv \fcs[YZ]{\alpha} t_\alpha
\label{contract}
\end{align}
for $\fcs[YZ]{\alpha}$. 
The general form Eq.~(\ref{riemannK}) does not have a definite sign. 

For compact groups with $\eta_{AB} \propto \delta_{AB}$, antisymmetry of the structure constants implies
\begin{align}
\frac{1}{F_\pi^2}  R_{ABCD} Y^A Z^B Y^C Z^D
&=  \fcs[YZ]{g}\fcs[YZ]{g}     + \frac{1}{4}  \fcs[YZ]{G}\fcs[YZ]{G}     \ge 0\,
\label{compact}
\end{align}
is positive definite of any pair of vectors $Y,Z$. For compact groups with $\eta_{AB} \not = \delta_{AB}$, the sectional curvatures need not be positive.  A simple example is $\g=SU(2)$ completely broken, with $\eta_{AB}=\text{diag}(\eta_1,\eta_2,\eta_3)$, and $Y=(1,0,0)$, $Z=(0,1,0)$, in which case
\begin{align}
K(Y,Z) &=\frac{ 2(\eta_1+\eta_2)\eta_3 + (\eta_1-\eta_2)^2-3 \eta_3^2}{4 F_\pi^2 \eta_1 \eta_2 \eta_3}
\label{example}
\end{align}
which is negative for $\eta_3 \gg \eta_{1,2}$. 

In  HEFT applications where there is only a single $h$ field, the possible sectional curvatures are:
\begin{itemize}

\item[(a)] Both $Y$ and $Z$ are in the Goldstone boson directions. Since the Goldstone boson manifold $S^3$ is a maximally symmetric space, $K(Y_\pi,Z_\pi)$ is independent of the choice $Y_\pi,Z_\pi$, and is the quantity $K(Y_\pi,Z_\pi)=\mathfrak{R}_4$ in Ref.~\cite{Alonso:2015fsp}.

\item[(b)] $Y$ is in the Goldstone boson direction, and $Z$ is in the $h$ direction. In this case $K(Y_\pi,Z_h)$ is independent of the choice $Y_\pi$ and $Z_h$ (since there is only one direction $Z_h$) and is $K(Y_\pi,Z_h)=\mathfrak{R}_{2h}$ in Ref.~\cite{Alonso:2015fsp}.

\end{itemize}

As shown in Ref.~\cite{Alonso:2015fsp}, deviations in $W_L W_L \to W_L W_L $ were proportional to $\mathfrak{r}_{4}=\mathfrak{R}_{4}(h=0)$, the sectional curvature where $Y$ and $Z$ are in Goldstone boson directions. The longitudinal gauge bosons at high energies are related to the Goldstone bosons, and so probe the Goldstone boson directions in $\m$. The $W_L W_L \to h h$ scattering amplitudes is proportional to $\mathfrak{r}_{2h}=\mathfrak{R}_{2h}(h=0)$, and probes the sectional curvature where $Y$ is in a Goldstone boson direction, and $Z$ in the Higgs direction. If the HEFT is based on a composite Higgs theory~\cite{Kaplan:1983fs}, where $h$ is itself a (pseudo) Goldstone boson of some strong dynamics at a scale $f > v$, then we see from Eq.~(\ref{compact}) that $\mathfrak{R}_{4}$ and $\mathfrak{R}_{2h}$ are both positive if the composite Higgs model is based on a compact group. On the other hand, if the sigma-model group is non-compact, it is possible to get negative values~\cite{Alonso:2016btr} for these curvatures because Eq.~(\ref{riemannK}) has no definite sign.

We also consider multi-Higgs theories in Sec.~\ref{sec:multiHiggs}. In such theories, the possible sectional curvatures are $\mathfrak{R}_4=K(Y_\pi,Z_\pi)$, $\mathfrak{R}_{2h,I}=K(Y_\pi,Z_I)$, where $Z_I$ runs over the possible Higgs directions, and $K(Y_I,Z_J)$ over distinct pairs of Higgs directions  $I \not=J$.

\section{The Standard Model and Custodial Symmetry Violation}\label{sec:custodial}

The SM sigma model for the custodial symmetric breaking pattern $SU(2)_L \times SU(2)_R \to SU(2)_V$ can be written in the CCWZ formalism, choosing the broken generators to be $T_L$. Let
\begin{align}
U(x) &= e^{i \bvphi^A(x) T_A}
\label{7.1}
\end{align}
be a $2 \times 2$ matrix, where $T_A$ are $SU(2)_L$ generators, and $\bvphi^A$ are dimensionless.

The $\xi$ field of the CCWZ formalism given by exponentiating the broken generators is
\begin{align}
\xi(x) &= \left( \begin{array}{cc}
U(x) & \\
0 & 1_{2\times2}
\end{array}\right), 
\end{align}
where the first $2\times 2$ block is the $SU(2)_L$ transformation, and the second is the $SU(2)_R$ transformation. From this $\xi$ field, one finds
\begin{align}
\xi(x)^{-1} D_\mu \xi &= \left( \begin{array}{cc}
U(x)^{-1} & \\
0 & 1_{2\times2}
\end{array}\right) \left[ \left( \begin{array}{cc}
\partial_\mu U(x) & \\
0 & 0
\end{array} \right) + \left( \begin{array}{cc}
i g_2 W_\mu^\alpha T_\alpha U(x) & \\
0 & ig_1 B_\mu T_3
\end{array}\right) \right] \nn
&=  \left( \begin{array}{cc}
U(x)^{-1} \partial_\mu U(x) + U(x)^{-1}  i g_2 W_\mu^\alpha T_\alpha U(x) & \\
0 & ig_1 B_\mu T_3
\end{array} \right) \nn
&=  \left( \begin{array}{cc}
i g_1 B_\mu T_3 & \\
0 & ig_1 B_\mu T_3
\end{array} \right) +   \left( \begin{array}{cc}
U(x)^{-1} \partial_\mu U(x) + U(x)^{-1}  i g_2 W_\mu^\alpha T_\alpha U(x) - ig_1 B_\mu T_3 & \\
0 & 0
\end{array} \right),
\end{align}
where the last line projects onto the unbroken and broken spaces, respectively.  Thus, we obtain
\begin{align}
i (D_\mu \bvphi)^A T_A &= U(x)^{-1} \partial_\mu U(x) + U(x)^{-1}  i g_2 W_\mu^\alpha T_\alpha U(x) - ig_1 B_\mu T_3, 
\label{6.4}
\end{align}
and, using the results in Appendix~\ref{app:full},
\begin{align}
(D_\mu \bvphi)^A
&=   \left( \frac{\sin \avphi}{\avphi} \right) \rd \bvphi^A + \left( \frac{1-\cos \avphi}{\avphi^2} \right)  \epsilon_{ABC} \bvphi^B \rd \bvphi^C +\left( \frac{\avphi - \sin \avphi}{\avphi ^3}\right) \bvphi^A  (\bm{\bvphi \cdot \rd \bvphi}) \nn
&+ g_2 W^A_\mu \cos \avphi + g_2 \left( \frac{\sin \avphi}{\avphi}\right) \epsilon_{ABC} \bvphi^B W_\mu^C
+ g_2 \left( \frac{1- \cos \avphi}{ \avphi^2} \right) (\bm{\bvphi \cdot W}_\mu) \bvphi^A
-  g_1 B_\mu \delta_{A3} 
\end{align}
with $\avphi^2 = \bm{\bvphi \cdot \bvphi}$.  Decomposing $(D_\mu \bvphi)^A$ into gauge and non-gauge pieces as in Eq.~(\ref{split}) yields
\begin{align}
(D_\mu \bvphi)^A &= {e^A}_B \partial_\mu \bvphi^B + F^A_\beta W^\beta_\mu + F^A_Z Z_\mu + F^A_\gamma A_\mu\,,
\label{6.6} 
\end{align}
where
\begin{align}
{e^A}_B
&=    \left( \frac{\sin \avphi}{\avphi}\right)  \delta^A_B - \left( \frac{1-\cos \avphi}{\avphi^2}\right)   \epsilon_{ABC} \bvphi^C  +\left( \frac{\avphi - \sin \avphi}{\avphi ^3}\right) \bvphi^A  \bvphi^B \,,\nn
F^A_\beta &=  \frac{e}{s_W} \left[\delta^A_\beta \cos \avphi +   \left( \frac{\sin \avphi}{\avphi}\right) \epsilon_{AD\beta} \bvphi^D
+ \left( \frac{1- \cos \avphi}{\avphi^2}\right) \bvphi^\beta \bvphi^A\right] ,\qquad \beta=1,2 \nn
F^A_Z &=   \frac{e}{s_W c_W}\left[\delta^A_3 \left( s_W^2+c_W^2  \cos \avphi \right) +  c_W^2 \left( \frac{\sin \avphi}{\avphi}\right) \epsilon_{AB3} \bvphi^B
+ c_W^2   \left( \frac{1- \cos \avphi}{ \avphi^2}\right)  \bvphi^3 \bvphi^A \right] \,,  \nn
F^A_\gamma &=  e\left[ - \delta^A_3 \left(1- \cos \avphi\right) +   \left( \frac{\sin \avphi}{\avphi}\right)  \epsilon_{AB3} \bvphi^B 
+  \left( \frac{1- \cos \avphi}{ \avphi^2}\right)  \bvphi^3 \bvphi^A\right]\,,
\label{7.7}
\end{align}
with $c_W=\cos\theta_W$ and $s_W=\sin\theta_W$. The $F^A_\alpha$ can be used to construct the Killing vectors using Eq.~(\ref{killing}). Expanding these equations gives
\begin{align}
{e^A}_B &= \delta^A_B - \frac12 \epsilon_{ABC} \bvphi^C 
+\frac16 \left[  \bvphi^A \bvphi^B - \avphi^2\delta^A_B \right] + \ldots \nn
F^A_\beta &=  \frac{e}{s_W} \left[\delta^A_\beta \left(1 - \frac12 \avphi^2\right)+  \epsilon_{AD\beta} \bvphi^D
+ \frac12 \bvphi^\beta \bvphi^A\right] + \ldots,\qquad \beta=1,2\nn
F^A_Z &=   \frac{e}{s_W c_W}\left[\delta^A_3 \left( 1  -  \frac12c_W^2 \avphi^2 \right)  + c_W^2  \epsilon_{AB3} \bvphi^B 
+  \frac12  c_W^2\bvphi^3 \bvphi^A \right] + \ldots \,, \nn
F^A_\gamma &=  e\left[ -  \frac12 \avphi^2  \delta^A_3+   \epsilon_{AB3} \bvphi^B
+ \frac12 \bvphi^3 \bvphi^A\right]+ \ldots\,.
\label{7.8}
\end{align}
In unitary gauge, $\bvphi=0$ and
\begin{align}
F^A_\beta &=  \frac{e}{s_W}\delta^A_\beta,\quad \beta=1,2, &
F^A_Z &=   \frac{e}{s_W c_W} \delta^A_3, &
F^A_\gamma &=  0,
\end{align}
so the photon is massless, and $W,Z$ acquire mass.

The most general $\mathcal{O}(p^2)$ Lagrangian is
\begin{align}
L=\frac12 \sum_{AB} \eta_{AB} (D_\mu\bvphi)^A (D_\mu\bvphi)^B
\label{6.16}
\end{align}
where $\eta_{AB}$ is a $\h$-invariant tensor.  For the SM with custodial $SU(2)$ symmetry, the breaking pattern is $SU(2)_L \times SU(2)_R \to SU(2)_V$.  The tensor 
$\eta_{AB}$ must be invariant under the unbroken $\h = SU(2)_V$ symmetry, so
\begin{align}
\eta_{AB} = \frac{v^2}{8} \delta_{AB},
\end{align}
where $v \sim 246$~GeV is chosen to give the correct gauge boson masses. 

If custodial symmetry is not exact, the breaking pattern is $SU(2)_L \times U(1)_Y \to U(1)_{\text{em}}$, and $\eta_{AB}$ must be invariant under the unbroken $\h=U(1)_{\text{em}}$ symmetry. In this case,
\begin{align}
\eta_{AB} = \frac{v^2}{8} \left( \begin{array}{ccc} 1 & 0 & 0 \\ 0 & 1 & 0 \\ 0 & 0 & \rho \end{array}\right),
\label{6.12}
\end{align}
where $\rho$ is the $\rho$-parameter
\begin{align}
\rho &=  \frac{M_Z^2 c_W^2}{M_W^2} \,,
\end{align}
which is no longer equal to one. The experimental constraint on the $\rho$ parameter is an extremely stringent constraint on custodial symmetry violation, since it requires $\abs{\rho-1}\lesssim 0.01$. A simple example of custodial symmetry violating is the SM with an additional triplet scalar field~\cite{Gelmini:1980re}
\begin{align}
\chi &=   \left[ \begin{array}{cccc} 
\frac{1}{\sqrt 2} \chi^+ & - \chi^{++} \\
\chi^0 & - \frac{1}{\sqrt 2} \chi^+
\end{array} \right] \,.
\end{align}
If the doublet and triplet vacuum expectation values are
\begin{align}
\vev{H} &= \left[ \begin{array}{cccc} 
0 \\ \frac{v_D}{\sqrt 2} \end{array} \right]\,,  &
\vev{\chi} &= \left[ \begin{array}{cccc} 
0 & 0 \\
\frac{v_T}{\sqrt 2} & 0
\end{array} \right] \,,
\label{6.20}
\end{align}
then the values of the $\eta_{AB}$ parameters in Eq.~(\ref{6.12}) are
\begin{align}
v^2 &= v_D^2 + 2 v_T^2, &  \rho &=\frac{v_D^2 +4 v_T^2}{v_D^2 + 2  v_T^2}\,.
\end{align}

The geometry of the scalar manifold with metric Eq.~(\ref{6.16}) has been studied in other contexts~\cite{Eguchi:1980jx}. The configuration space of a rigid body with one point fixed is given by the rotation matrix $R(\theta,\phi,\psi) \in SO(3)$ parameterized by three Euler angles, and, up to $\mathbb{Z}_2$ factors, is the same as the Goldstone boson manifold of the SM. Rotations of the body about space-fixed axes correspond to $SO(3)_L$ rotations $R \to g_L R$, $g_L \in SO(3)$, and rotations about the body-fixed axes correspond to $SO(3)_R$ rotations $R \to R g_R^{-1}$, $g_R \in SO(3)$. The body-axis angular momenta are given by $\omega^A T_A = R^{-1} \dot R$. The kinetic energy for a rigid body is then given by the analog of Eq.~(\ref{6.16}), 
\begin{align}
L=\frac12 \sum_{A} I_A \left(\omega^A\right)^2,
\label{6.16a}
\end{align}
where $\eta_{AB}$ can be chosen to be diagonal by picking the body axes to coincide with the principal axes of the body. The kinetic energy for a spherical top with all three principal moments of inertia equal, $I_1=I_2=I_3$ is the analog of the SM with custodial symmetry. The configuration space of the top is the (undeformed) three-sphere $S^3$. The custodial symmetry violating case is analogous to $I_1=I_2 \not = I_3$, which is the configuration space of a symmetric top. This space is known as the squashed three-sphere, and also occurs in the metric for the Taub universe~\cite{Eguchi:1980jx}. The asymmetric top with all $I_i$ different would correspond to the SM with electromagnetism broken.

\section{HEFT with Multiple Singlet Scalar Bosons}\label{sec:multiHiggs}

The HEFT formalism can be extended to the case of multiple singlet (under custodial $SU(2)$) Higgs fields $h^I$, $I=1,2, \cdots$, which involves adding additional singlet scalars to the SM field content.  The generalization of the HEFT Lagrangian Eq.~(\ref{heft}) to multiple singlet scalar fields is
\begin{align}
L &= \frac12 v^2 F(h)^2 \left(\partial_\mu \bm{n} \right)^2  + \frac12  g_{IJ} (h) \left(\partial_\mu h^I \right) \left(\partial_\mu h^J \right) -V(h) + \ldots
\label{heft2}
\end{align}
where $F(h)$ is an arbitrary function of the dimensionless singlet scalar fields $h^I/v$. The coordinates $\{h^I\}$ are chosen so that $h=(0,0,\ldots,0)$ is the ground state, and the HEFT function  $F(h)$ is normalized so that
\begin{align}
F(0,\ldots,0)=1\,
\label{f0a}
\end{align}
since the radius of $S^3$ in the vacuum is fixed to be $v$ by the gauge boson masses.

Consider the $O(4) \to O(3)$ symmetry breaking pattern of the SM, with multiple scalar fields $h^I$ which are \emph{singlets} under the unbroken custodial $O(3)$ symmetry.  The most general metric of the scalar fields $\Phi^i \equiv \{ \pi^A, h^I \}$ has the form
\begin{align}
g_{ij}\left( \Phi \right) &= \left[ \begin{array}{cc} F(h)^2 g_{AB}(\pi) & 0 \\  0 & g_{IJ}(h)
\end{array}\right] ,
\label{6.1}
\end{align}
where $\pi^A/v$ are coordinates on the coset space ${\cal G}/{\cal H}=O(4)/O(3)=S^3$, and $g_{AB}(\pi)$ is the metric on the unit 3-sphere. $O(4)$ invariance implies that  the off-diagonal metric terms $g_{AI}$ and $g_{IA}$ vanish, and that $g_{IJ}(h)$ has no dependence on the $\pi$ fields. An easy way to prove that the general metric takes the form Eq.~(\ref{6.1}) is to note that a point on $S^3$ is given by a four-component unit vector $\bm{n}$. The entry $g_{IJ}(h)$ can depend on $\bm{n}$, but not on its derivatives; $O(4)$ invariance then requires it to be function of $\bm{n \cdot n}=1$, and therefore independent of $\bm{\pi}$.  Similarly, $g_{IA} \partial_\mu \pi^A$ is an $O(4)$ invariant function of $\bm{n}$ and $\partial_\mu \bm{n}$ with one derivative; the only invariant object is  $\partial_\mu\bm{n \cdot n}=0$, so the off-diagonal entries vanish. The $11$ entry has the form $F(h)^2 g_{AB}(\pi)$ because $\cal G$-invariance requires that $h$ dependence is an overall multiplicative factor, since there is only one $\cal G$-invariant metric on $S^3$.  We will consider the geometry of the metric Eq.~(\ref{6.1}), with a general metric $g_{AB}(\pi)$, so the results are valid for a general ${\cal G}/{\cal H}$ manifold as long as the off-diagonal terms of $g_{ij}(\Phi)$ vanish as in Eq.~(\ref{6.1}).

Using the metric Eq.~(\ref{6.1}), the Christoffel symbols are
\begin{align}
\Gamma^A_{BC} &= \gamma^A_{BC} , &
\Gamma^A_{BK} &= \frac{F_{,K}}{F} {\delta^A}_B , &
\Gamma^A_{JK} &= 0, \nn
\Gamma^I_{BC} &= - F F_{,M} g^{IM} g_{BC}, &
\Gamma^I_{BK} &= 0 ,&
\Gamma^I_{JK} &= \gamma^I_{JK},
\end{align}
where $ \gamma^A_{BC}$ and $\gamma^I_{JK}$ are the Christoffel symbols computed from the metrics $g_{AB}(\pi)$ and $g_{IJ}(h)$, respectively. Similarly, in the expressions below, $r^A{}_{BCD}$, $r_{BD}$ and $r_\pi$ are the curvatures computed from the metric $g_{AB}(\pi)$, whereas $r^I{}_{JKL}$, $r_{JL}$ and $r_h$ are the curvatures computed from the metric $g_{IJ}(h)$. The Riemann curvature tensor is
\begin{align}
{R^A}_{BCD} &= {r^A}_{BCD} - g^{MN} F_{,M} F_{,N}\left( {\delta^A}_C \, g_{DB}-{\delta^A}_D \, g_{BC} \right) ,&
{R^A}_{BCL} &=0, \nn
{R^A}_{BKL} &= 0, &
{R^I}_{JCD} &= 0 , \nn
{R^I}_{JKD} &= 0, &
{R^I}_{JKL} &= {r^I}_{JKL}, \nn
{R^A}_{JCD} &= 0 ,&
{R^A}_{JKL} &=0  , \nn
{R^I}_{BCD} &=0 , &
{R^I}_{BKD} &=  -g_{DB} \, g^{IM} \, F_{;MK}, \nn
{R^I}_{BKL} &= 0, &
{R^A}_{JCL} &=- {\delta^A}_C  \, F_{;JL} \,.
\end{align}
The covariant derivatives of $F$ are w.r.t.\ $\gamma^I_{JK}$. The Ricci tensor is
\begin{align}
R_{BD} & = r_{BD} - g^{RS} F_{,R} F_{,S}(N_\pi-1) g_{BD}  - g_{BD} g^{RS} F F_{;RS}, \nn
R_{BL} & = 0 ,\nn
R_{JL} &= - N_\pi F_{;JL} + r_{JL},
\end{align}
and the curvature scalar is
\begin{align}
R &=  \frac{1}{F^2} r_\pi  -N_\pi (N_\pi-1) \frac{1}{F^2} g^{RS} F_{,R} F_{,S}  -2 N_\pi \frac{1}{F} g^{RS}  F_{;RS}
+ r_h .
\end{align}
If $\cal G/H$ is a maximally symmetric space,
\begin{align}
{r^A}_{BCD} &= \frac{1}{F_\pi^2} \left({\delta^A}_C \, g_{BD}-{\delta^A}_D \, g_{BC} \right), &
r_{BD} &= \frac{1}{F_\pi^2} (N_\pi-1) g_{BD}, &
r_\pi &=  \frac{1}{F_\pi^2} N_\pi (N_\pi-1).
\end{align}
The above expressions reduce to the formul\ae\  given in Ref.~\cite{Alonso:2015fsp} for one Higgs singlet field $h$ and $\cal G/H$ a symmetric space, which used 
\begin{align}
g_{IJ}(h) &= 1, & F(h) &= 1 + c_1 \left(\frac hv\right) + \frac12 c_2  \left(\frac hv\right)^2 + \ldots
\end{align}
with $F_\pi=v$. 

The above expressions can be further simplified if one picks one $h$ field to be the radius of $S^3$, $F(\{h\})=h^1$, in which case $F$ does not depend on $h^I$, $I \not =1$. The radial direction is in general not a mass-eigenstate direction. Letting $\rho$ be the radial direction, with $\rho=1$ in the vacuum, and letting the remaining directions still be called $h^I$ (there is one less $h$ now), with $I,J,K$ running over $\rho, \{h\}$, one gets a simpler version of the above equations, where $F_{,K}=1$ if $K=\rho$, and zero otherwise. For example,
\begin{align}
F &\to \rho, & G^{RS} F_{,R} F_{,S} &\to G^{\rho\rho}, & F_{;RS} & \to - \gamma^\rho_{RS},
\end{align}
etc.

\section{Conclusions}\label{sec:conclusions}

In this paper, we have discussed the relation between the SM and two of its generalizations, SMEFT and HEFT, and have shown that HEFT can be written in SMEFT form if and only if there is an $O(4)$ invariant fixed point of the scalar manifold in a neighborhood of which the scalar fields transform as a vector of $O(4)$. We have shown that the SM can be written using scalar fields transforming either linearly or non-linearly under $SU(2)_L \times U(1)_Y$, and is renormalizable with either choice. Whether ``the Higgs transforms linearly or non-linearly'' is not observable; the correct question, which can be resolved experimentally, is whether the SM scalar manifold $\cal{M}$ is flat or curved.

We have discussed the formulation of scalar fields on a curved manifold, including the case with gauge symmetry, reviewed the computation of  one-loop corrections in terms of the curvature, and applied these known results to the case where the manifold is a coset. The general expressions were used to obtain the one-loop renormalization of HEFT~\cite{Guo:2015isa,Alonso:2015fsp},
and details of the computation are given here.

Deviations of Higgs and longitudinal gauge boson scattering amplitudes from their SM values are given by sectional curvatures of the  scalar manifold. In simple examples based on $\g/\h$ symmetry breaking with compact groups, the sectional curvatures are positive, which fixes the signs of deviations from the SM. We are investigating examples where sectional curvatures can be negative, and have given the generalization of the CCWZ formalism to non-compact groups.

\section{Acknowledgments}

AM would like to thank Luis Alvarez-Gaum\'e, Luca Merlo, and Vyacheslav Rychkov for for helpful discussions.
This work was partially supported by grants from the Simons Foundation (\#340282 to Elizabeth Jenkins and
\#340281 to Aneesh Manohar) and by DOE grant DE-SC0009919. RA would like to thank the CERN theory group for hospitality and funding.

\newpage
\begin{appendix}

\section{Exponential Parametrization of the $O(N)$ Model}\label{app:full}

The real antisymmetric Goldstone boson matrix is given by
\begin{align}
\bm{\Pi} &\equiv i  \left( \bm{ \bpi} \cdot \bm{X} \right)
=   \left[ \begin{array}{cc} 0 & \bm{ \bpi} \\ -\bm{ \bpi}^T & 0 \end{array}\right]
=  \left[ \begin{array}{cccc} 0 & \ldots & 0 &  \bpi^1 \\ 
0 & \ldots & 0 &  \bpi^2 \\
\vdots & & \vdots & \vdots \\ 
0  & \ldots & 0 &  \bpi^{\n_G} \\ 
- \bpi^1 & \ldots & - \bpi^{\n_G} & 0 \end{array}\right] ,
\end{align}
where $ \bpi^A \equiv \pi^A/F_\pi$.
$\xi$ is
\begin{align}
\xi &\equiv e^{\bm{\Pi}} = \bm{1} + \left( \frac{ \sin  \api}{\api}\right) \bm{\Pi}+ \left( \frac{1-\cos  \api }{ \api^2}\right) \bm{\Pi}^2  ,  
& { \api}^2 & \equiv   \bpi^A  \bpi^A.
\label{a-3}
\end{align}
The Mauer-Cartan form is
\begin{align}
\xi^{-1} \partial_\mu \xi 
&= \left( \frac{\sin  \api}{ \api}\right)  i \left( \partial_\mu \bm{ \bpi} \right) \cdot \bm{X} +\left( \frac{ \api - \sin  \api}{ \api^3}\right)  \left(  \bpi^B \partial_\mu  \bpi^B  \right) i \bm{ \bpi} \cdot \bm{X}  \nn
& +\left( \frac{1 - \cos  \api  }{ \api^2} \right)  \left[ \begin{array}{cc}  -\left( \partial_\mu  \bpi^A \right)  \bpi^B +  \bpi^A \left(\partial_\mu  \bpi^B \right) & 0 \\ 0 & 0 \end{array} \right] ,
\label{a-2}
\end{align}
where the first two terms are linear combinations of the broken generators, and the last term is a linear combination of the unbroken generators. The indices $A,B$ in the last term are the row and column indices of the $N_\varphi \times N_\varphi $ submatrix in the upper $11$ block. 
Using Eq.~(\ref{5.14}),
\begin{align}
(D_\mu  \bpi)^A &= \left( \frac{\sin  \bpi}{ \bpi}\right)  \left( \partial_\mu { \bpi} \right)^A +\left( \frac{ \bpi - \sin  \bpi}{\bpi^3}\right)  \left(  \bpi^B \partial_\mu  \bpi^B  \right) { \bpi}^A ,
\label{a-3ccwz}
\end{align}
and
\begin{align}
 \left . \xi^{-1} \partial_\mu \xi \right|_T &=i \bm{V}_\mu \cdot \bm{T} = \left(\frac{1- \cos  \bpi}{ \bpi^2} \right)  \left[ \begin{array}{cc}  -\left( \partial_\mu  \bpi^A \right)  \bpi^B +  \bpi^A \left(\partial_\mu  \bpi^B \right) & 0 \\ 0 & 0 \end{array} \right] .
\label{a-4ccwz}
\end{align}

\section{One-Loop Renormalization of HEFT} \label{app:oneloop}

In this appendix, we provide some intermediate results in the computation of the one-loop renormalization of HEFT~\cite{Guo:2015isa,Alonso:2015fsp}. 

The metric for the scalar manifold ${\cal M}$ in HEFT is
\begin{align}
g_{ij}(\phi) &= \left[ \begin{array}{cc}  v^2 F(h)^2 g_{ab}(\bvphi ) & 0 \\ 0 & 1 \end{array} \right],
\label{d1}
\end{align}
where $F(h)$ is a dimensionless function with a power series expansion in $h/v$, and $g_{ab}(\bvphi)$ is the metric on the Goldstone boson manifold ${\cal G}/{\cal H}=S^3$.  The field $h$ has mass dimension one, $\bvphi$ is dimensionless, and $i$ runs over indices $a$, $h$.
The scalar kinetic term in HEFT is given by
\begin{eqnarray}
L &=& \frac 12 g_{ij}(\phi) \partial_\mu \phi^i \partial^\mu \phi^j = \frac 12 v^2 F(h)^2 \, g_{ab} ( \bvphi ) \, \partial_\mu \bvphi^a \partial^\mu \bvphi^b+ \frac 12 \partial_\mu h \,\partial^\mu h \nn
&\equiv& \frac 12 F(h)^2 \, v^2 \, \partial_\mu \bm{n} \cdot \partial^\mu \bm{n} + \frac 12 \partial_\mu h \, \partial^\mu h,
\label{d2}
\end{eqnarray} 
where the unit vector $\bm{n}(\bvphi)$ is a dimensionless function of the the three independent coordinates $\overline \pi^a =\pi^a/v$ on $S^3$.
Note that we have chosen to normalize $\bvphi^a$ to be dimensionless coordinates, which differs from Ref.~\cite{Alonso:2015fsp} by a rescaling by $v$.
Eq.~(\ref{d2}) implies that the $S^3$ metric $g_{ab}(\bvphi)$ is given in terms of the unit vector $\bm{n}(\bvphi)$ by
\begin{align}
g_{ab}(\bvphi) &\equiv  \frac{\partial \bm{n}(\bvphi)}{\partial \bvphi^a} \cdot \frac{\partial \bm{n}(\bvphi)}{\partial \bvphi^b}.
\end{align}

The Riemann curvature tensor $R_{ijkl}(\phi)$ obtained from the scalar metric $g_{ij}(\phi)$ consists of the non-vanishing components
\begin{align}
R_{abcd}(\phi) &= \left[1 - v^2 (F^\prime(h))^2\right] v^2 F(h)^2 \left( g_{ac}(\bvphi) g_{bd}(\bvphi)-g_{ad}(\bvphi) g_{bc}(\bvphi) \right), \nn
R_{ah b h}(\phi) &=  - v^2 F(h) F^{\prime \prime} (h) g_{ab}(\bvphi),
\label{d3}
\end{align}
and components related to these by the permutation symmetry of the Riemann tensor. $R_{abcd}(\phi)$ is proportional to the tensor 
$(g_{ac} g_{bd}-g_{ad} g_{bc})$ because $S^3$ is a maximally symmetric space.

The quantities $X$ and $Y_{\mu\nu}$ from Eqs.~(\ref{XY}) and~(\ref{4.33a}) that appear in the one-loop correction Eq.~(\ref{1loope}) contain terms depending on the Riemann curvature tensor.  The Riemann curvature tensor components contributing to ${\left[X\right]^i}_k$ and ${\left[ Y_{\mu \nu} \right]^i}_j$, respectively, are
\renewcommand{\arraycolsep}{0.3cm}
\begin{align}
&{R^i}_{jkl} (D_\mu \phi)^j (D^\mu \phi)^l  \nn
&= \left[ \begin{array}{cc} 
\left[1-v^2 (F^\prime)^2\right] \left[ (D_\mu \bvphi)^2 \ \delta^a_c - (D_\mu \bvphi)^a (D_\mu \bvphi)_c\right] -\frac{F^{\prime \prime}}{F}  (\partial_\mu h)(\partial^\mu h) \delta^a_c 
&
 \frac{F^{\prime \prime}}{F} (D_\mu \bvphi)^a ( \partial_\mu h) 
  \\[10pt]
v^2 F F^{\prime \prime}(\partial_\mu h) (D_\mu \bvphi)_c
&
- v^2 F F^{\prime \prime}  (D_\mu \bvphi)^2  
\end{array}\right] , \nn
&{R^i}_{jkl} (D_\mu \phi)^k (D^\mu \phi)^l \nn
&= \left[ \begin{array}{cc} 
\left[1-v^2 (F^\prime)^2\right] \left[ (D_\mu \bvphi)^a (D_\nu \bvphi)_b - (D_\nu \bvphi)^a (D_\nu \bvphi)_b\right]
& 
\frac{F^{\prime \prime}}{F}  \left[ (D_\nu \bvphi)^a ( \partial_\mu h)- (D_\mu \bvphi)^a (\partial_\nu h )\right] 
 \\[10pt]
- v^2 F F^{\prime \prime} \left[ (\partial_\mu h) (D_\nu \bvphi)_b -( \partial_\nu h) (D_\mu \bvphi)_b \right] 
& 0
\end{array}\right].
\end{align}

The Lagrangian term ${\cal I}(\phi)$ containing the potential and Yukawa couplings is
\begin{align}
\mathcal{I}(\phi) &=-V(h) + K(h) \, \bm{n} \cdot \bm{W}
\end{align}
where $\bm{W}$ is a constant, in the notation of Ref.~\cite{Alonso:2015fsp}.
\begin{align}
\nabla_i \nabla_j \mathcal{I}
&= \left[ \begin{array}{cc} 
g_{ab} \left[ v^2 F F^\prime \left( \bm{W \cdot n} \, K^\prime - V^\prime \right)- \bm{W \cdot n}\, K \right] & F \left( \frac{K}{F} \right)^\prime 
\bm{W \cdot n}_{,a}  \\
 F \left( \frac{K}{F} \right)^\prime \bm{W \cdot n}_{,b} & -V^{\prime \prime} + K^{\prime \prime}\,  \bm{n \cdot W}
\end{array}\right]
\end{align}
where $\bm{n}_{,a}=\partial \bm{n}/\partial \bvphi^a$.

The field strength $Y_{\mu \nu}$ is
\begin{align}
\left[Y_{\mu \nu}\right]^i{}_j
&= \left[ \begin{array}{cc} 
\left[1-v^2(F^\prime)^2\right] \left[ (D_\mu \bvphi)^a (D_\nu \bvphi)_b - (D_\nu \bvphi)^a (D_\nu \bvphi)_b\right]
& \frac{F^{\prime \prime}}{F}  \left[( \partial_\mu h) (D_\nu \bvphi)^a - (\partial_\nu h )(D_\mu \bvphi)^a \right] 
\\[10pt]
- v^2 F F^{\prime \prime} \left[ (\partial_\mu h) (D_\nu \bvphi)_b -( \partial_\nu) h (D_\mu \bvphi)_b \right] 
& 0
\end{array}\right]\nn
& +A^\beta_{\mu \nu} (t^i_\beta)_{;j}  
\end{align}
with
\begin{align}
A^\beta_{\mu \nu} (t^i_\beta)_{;j} 
&= \left[ \begin{array}{cc} 
0 & - F F^{\prime} (\partial_b n)^T \mathscr{A}_{\mu \nu} n \\[10pt]
v^2\frac{F^{\prime}}{F} g^{ac} (\partial_c n)^T \mathscr{A}_{\mu \nu} n
& g^{ac} (\partial_c n)^T \mathscr{A}_{\mu \nu} (\partial_b n)
\end{array}\right] 
\end{align}
and
\begin{align}
\mathscr{A}_\mu
&= \left[ \begin{array}{cccc} 
0&gW_\mu^3+g^\prime B_\mu & -gW_\mu^2&gW_\mu^1\\
-gW_\mu^3-g^\prime B_\mu &0& gW_\mu^1&gW_\mu^2\\
gW_\mu^2&-gW_\mu^1&0& gW_\mu^3-g^\prime B_\mu \\
-gW_\mu^1&-gW_\mu^2&-gW_\mu^3 +g^\prime B_\mu&0\\
\end{array}\right] 
\label{amu}
\end{align}
in terms of the electroweak gauge bosons. The field strength tensor $\mathscr{A}_{\mu \nu}$ is given by Eq.~(\ref{amu})
with the replacements $W_{\mu}^\alpha \to W_{\mu \nu}^\alpha$, $B_\mu \to B_{\mu \nu}$. The covariant derivative $D_\mu \bm{n}$ is given by
\begin{align}
D_\mu \bm{n} &= \partial_\mu \bm{n} +  \mathscr{A}_\mu \bm{n}
\end{align}
treating $\bm{n}$ as a four-component column vector, and using matrix multiplication. The covariant derivative on $\bvphi$ is defined implicitly through
\begin{align}
D_\mu \bm{n} \cdot D^\mu \bm{n} &= g_{ab}(\bvphi) (D_\mu \bvphi)^a(D_\mu \bvphi)^b
\end{align}

Substituting the above equations into Eq.~(\ref{1loope}) gives Eq.~(59) in Ref.~\cite{Alonso:2015fsp}.

\section{Non-reductive Cosets}\label{app:nonred}

In this appendix, we comment briefly on the CCWZ formalism when $\left[T_a,X_B\right]$ contains a piece proportional to the unbroken generators, so that the coset is non-reductive. Such examples are relevant for constructing $\g/\h$ theories with negative sectional curvature~\cite{Alonso:2015fsp}.

One can still define the CCWZ $\xi$ field as in Eq.~(\ref{4.46}) which transforms as in Eq.~(\ref{4.47}).
The complication for the non-reductive case is in Eq.~(\ref{1.11}). For $g \in H$,
\begin{align}
g \left( \pi^A(x) X_A\right) g^{-1}
\end{align}
is no longer a linear combination of the broken generators, but also has a component along the unbroken generators,
\begin{align}
g \, \pi^A(x) X_A\, g^{-1} &=  X_A \left[D^{\rpi}(g)\right]^A{}_B \pi^B +  T_a M^a_B \pi^B \,,
\label{a2}
\end{align}
where $D^{\rpi}$ is the $\rpi$ transformation matrix constructed out of $\fcs[aB]{C}$, as in Eq.~(\ref{lie3}), and $M^a_B \pi^B$ is the component in the unbroken direction. The exponential
of Eq.~(\ref{a2}) can be schematically written as
\begin{align}
e^{X+T} &= e^{X^\prime}e^{T^\prime}
\end{align}
where $X,X^\prime$ are linear combinations of broken generators, and $T,T^\prime$ are linear combinations of unbroken generators, and the primed and unprimed quantities are connected by the Baker-Campbell-Hausdorff formula. Thus one gets Eq.~(\ref{4.47}) with some important changes even in if $g \in H$: (a) The relation between $\pi$ and $\pi^\prime$ is non-linear. Eq.~(\ref{5.14a}) only holds for the linear term, i.e.\ for the transformation of the tangent vector to the Goldstone boson manifold at the origin, and (b) $h^\prime(\xi(x),g)$ depends on $\xi$ and hence $x$, even if $g \in H$. 

The transformation of $(D_\mu \pi)$ and $V_\mu$ in Eqs.~(\ref{5.15}) and~(\ref{5.16}) is also changed,
\begin{align}
(D_\mu\pi) & \to \left. h (D_\mu \pi) h^{-1} \right|_X
\label{a4}
\end{align}
\begin{align}
V_\mu & \to h V_\mu h^{-1} - \partial_\mu h \, h^{-1} +  \left. h (D_\mu \pi) h^{-1} \right|_T
\label{a5}
\end{align}
$(D_\mu \pi)$ transforms by adjoint action by $\h$ in the representation $\rpi$, as before. However, $V_\mu$ picks up an additional piece and no longer transforms as a  gauge field under $\h$. One can still define Goldstone boson kinetic terms as before, Eq.~(\ref{5.17}). However, since $V_\mu$ does not transform as a gauge field, it is not possible to define covariant derivatives on matter fields $\psi$ which transform as arbitrary irreducible representations of $\h$, as was done in CCWZ.
Nevertheless, some matter fields are allowed in the EFT. For example, if $\psi$ transforms as a representation $\mathbf{R}_G$ of the full group $\g$,
\begin{align}
\psi  \to D(g) \psi \,,
\end{align}
then
\begin{align}
(\partial_\mu + i t_\alpha A^\alpha_\mu)\,\psi 
\label{a7}
\end{align}
is a covariant derivative, where the generators $t_\alpha$ are in the $\mathbf{R}_G$ representation. Following CCWZ, we can define new fields $\chi$ by
\begin{align}
\chi &= D(\xi^\dagger) \psi
\end{align}
which transform as
\begin{align}
\chi &\to D(h) \chi \,,
\end{align}
where $h$ is given by Eq.~(\ref{4.47}). The covariant derivative Eq.~(\ref{a7}) turns into
\begin{align}
(\partial_\mu + \xi^{-1} D_\mu \xi )\chi = \left[ \partial_\mu +i (D_\mu \pi)^A X_A + i V_\mu^a\, T_a  \right] \chi
\label{a10}
\end{align}
on $\chi$ using Eq.~(\ref{5.15}). The sum $\left( D_\mu \pi + V_\mu \right)$ in the covariant derivative transforms as a gauge field
\begin{align}
\left( D_\mu \pi + V_\mu \right) & \to h \left(D_\mu \pi + V_\mu \right) h^{-1}- \partial_\mu h \, h^{-1}  \,,
\label{a11}
\end{align}
and the covariant derivative Eq.~(\ref{a10}) is well-defined. For compact groups, where $(D_\mu \pi)$ transforms as
\begin{align}
(D_\mu \pi) & \to h \left(D_\mu \pi\right) h^{-1} \,,
\label{a12}
\end{align}
and does not mix with $V_\mu$,
one can omit $(D_\mu \pi)$ in Eq.~(\ref{a10}) to get the CCWZ covariant derivative. In this case, for the covariant derivative on $\chi$ to make sense, it is only necessary to define the action of the unbroken generators $T_a$ on $\chi$, i.e.\ one can restrict $\chi$ to only be in an irreducible representation of $\h$; it does not have to form a representation of $\g$. Baryons in QCD are an example --- they form a representation of the unbroken $SU(3)_V$ symmetry, but not of chiral $SU(3)_L \times SU(3)_R$.
However, for the non-reductive case, it is necessary to retain the $(D_\mu \pi)$ term in the covariant derivative, to cancel the extra piece in the transformation of $V_\mu$, the last term in Eq.~(\ref{a5}). In this case, we need to define the action of $T_a$ \emph{and} $X_A$, which requires $\chi$ to form a representation of the full symmetry $\g$, not just its unbroken subgroup.

The main difficulty for sigma models with non-compact $\h$ is unitarity. The $\psi$ kinetic energy term for compact groups $\h$ is
\begin{align}
\sum_a \left( D_\mu \psi \right)^\dagger_a\, \left( D^\mu \psi \right)^a
\label{6.44}
\end{align}
if $\psi$ is a complex scalar. If $\h$ is non-compact, then the unitary representations are infinite dimensional. For a finite dimensional non-unitary representation, the kinetic term Eq.~(\ref{6.44}) is not an invariant, since $\psi^\dagger$ does not transform as the inverse of $\psi$. One can construct invariant terms. For example, if $\h$ is $SO(3,1)$, and $\psi$ transforms as the (real) vector representation,
\begin{align}
\sum_{i=1,2,3} \left( D_\mu \psi_i \right)\, \left( D^\mu \psi_i \right)-\left( D_\mu \psi_4 \right)\, \left( D^\mu \psi_4 \right)
\label{6.45}
\end{align}
is invariant, as should be familiar from the Lorentz group. Eq.~(\ref{6.45}) has a wrong sign kinetic term, and leads to ghosts.
We do not know, in general, whether there are finite dimensional representations for a non-compact group $\h$ with a positive definite
$\h$-invariant kinetic energy term. This is possible for a trivial example: if $\h$ is a non-compact $U(1)$, i.e.\ of the form $h=\exp \alpha T$, $-\infty \le \alpha \le \infty$, one can pick the fermion to transform as $\exp i q \alpha$, and the kinetic energy Eq.~(\ref{6.44}) is $\h$-invariant.

One can construct a suitable kinetic energy term if $\h$ is compact even if $\g$ is non-compact, since $\psi$ transforms under $\h$, not $\g$. An example of this type based on $SO(4,1) \to SO(4)$ was studied in Ref.~\cite{Alonso:2016btr}. In this case, the low energy EFT is unitary. However, implementing a unitary UV theory in which $\g$ invariance is manifest is problematic, and we do not know of any examples where this is possible.\footnote{A simple argument due to S.~Rychkov is to look at $\g$-current correlators $\vev{J^\mu_\alpha J^\nu_\beta}$ in the UV theory. $\g$ invariance requires the correlator to be proportional to the Killing form $B_{\alpha \beta}$, which is not positive definite if $\g$ is non-compact, so that unitarity is violated. However, the low-energy EFT correlators are unitary, so it might be possible to construct theories where the $\g$ symmetry of $\g/\h$ arises only in the low energy limit.}

\subsection{Example of a Non-reductive Coset.}

A simple example of a non-reductive coset is the 2-parameter group of matrices
\renewcommand{\arraycolsep}{0.2cm}
\begin{align}
\left[ \begin{array}{cc} 1 & 0 \\ x & y \end{array}\right] , \qquad y>0,
\end{align}
under multiplication. The generators (absorbing a factor of $i$) can be chosen as 
\begin{align}
T &= \left[ \begin{array}{cc} 0 & 0 \\ 1 & 0 \end{array}\right], &
X &= \left[ \begin{array}{cc} 0 & 0 \\ 0 & -1 \end{array}\right] ,
\end{align}
with the commutation relation
\begin{align}
\left[T,X\right] &= T\,.
\end{align}

If the matrices act on a vector
\begin{align}
v &=  \left[ \begin{array}{c} 0 \\ 1 \end{array}\right],
\end{align}
then $Tv=0$, $Xv \not=0$, so that $T$ is an unbroken generator and $X$ is a broken generator. The matrices are sufficiently simple that the CCWZ formul\ae\ can be computed explicitly. The exponential of a Lie algebra element is
\begin{align}
g=e^{a T + b X} &= \left[ \begin{array}{cc} 1 & 0 \\ \frac{a}{b} (1-e^{-b}) & e^{-b} \end{array}\right] ,
\label{e17}
\end{align}
so that
\begin{align}
\xi &= e^{\bvphi X} = \left[ \begin{array}{cc} 1 & 0 \\ 0 & e^{-\bvphi}  \end{array}\right] ,
\end{align}
and
\begin{align}
e^{u T}   &=
\left[ \begin{array}{cc} 1 & 0 \\  u & 1  \end{array}\right] .
\end{align}
The CCWZ multiplication rule
\begin{align}
g e^{\bvphi X} &= e^{\bvphi^\prime X} e^{u^\prime T}
\label{e20}
\end{align}
with $g$ in Eq.~(\ref{e17}) gives
\begin{align}
\bvphi^\prime &= \bvphi + b ,\nn
u^\prime &= \frac{a}{b} \left( e^b- 1 \right) e^{\bvphi}.
\end{align}
In the special case where $ g \in \h$, $b=0$ and
\begin{align}
\bvphi^\prime(x) &= \bvphi(x) ,\nn
u^\prime(x) &= a  e^{\bvphi(x)},
\end{align}
so that $u^\prime$ depends on $x$ through $\bvphi(x)$. Eq.~(\ref{e20}) becomes
\begin{align}
e^{aT} e^{\bvphi\, X} &= e^{\bvphi X} h, & h(x) &=e^{a e^{\bvphi}(x)\, T},
\label{e23}
\end{align}
and $h$ depends on $x$ even for an unbroken transformation.

The Maurer-Cartan form is
\begin{align}
\xi^{-1} \rd \xi &= \rd\bvphi \, X , & \omega &= \left. \xi^{-1} \rd \xi \right|_X,   & V &= \left. \xi^{-1} \rd \xi \right|_T,
\end{align}
so that
\begin{align}
\omega &= \rd\bvphi \, X, &
V &= 0.
\end{align}
Under a global unbroken transformation $g = \exp a T$,
\begin{align}
\xi^{-1} \rd \xi &\to \xi^{\prime\, -1} \rd \xi^\prime = \rd\bvphi^\prime \, X.
\end{align}
Using Eq.~(\ref{e20}),
\begin{align}
\omega^\prime &= \omega, & V^\prime = 0.
\end{align}
The transformation laws are
\begin{align}
\omega^\prime &= \left. h \omega h^{-1}\right|_X &
V^\prime &=\left. h \omega h^{-1}\right|_T + h V h^{-1}  - \rd h h^{-1}
\end{align}
with $h$ in Eq.~(\ref{e23}). These equations are satisfied because of the extra $h \omega h^{-1}$ term in the $V$ transformation.

\end{appendix}

\bibliographystyle{JHEP}
\bibliography{linear}

\end{document}